# A Type System to Ensure Non-Interference in ReScript

Benjamin Bennetzen, Daniel Vang Kleist, Emilie Sonne Steinmann, Loke Walsted, Nikolaj Rossander Kristensen, Peter Buus Steffensen

October, 2024

**Abstract**

Protecting confidential data from leaking is a critical challenge in computer systems, particularly given the growing number of observers on the internet. Therefore, limiting information flow using robust security policies becomes increasingly vital. We focus on the non-interference policy, where the goal is to ensure that confidential data can not impact public data. This paper presents a type system, for a subset of the ReScript syntax, designed to enforce non-interference. We conclude with a proof of soundness for the type system, demonstrating that if an expression is type-able, it is inherently non-interferent. In addition, we provide a brief overview of a type checker that implements the previously mentioned type system.

Aalborg University
Department of Computer Science
Selma Lagerlöfs Vej 300, Aalborg, Denmark

# 1 Introduction

Regarding the ever growing world of computers, the question of security and protection of confidential data has become highly relevant. One security policy that aims to guarantee data confidentiality is that of non-interference. Non-interference is achieved when confidential data cannot affect public data, ensuring that unauthorized observers cannot access confidential information. The property of non-interference was defined as early as 1982 by Goguen and Meseguer [1], and can be analyzed through information flow analysis. In their work, they present a formal notion for describing security policies and models.

Ensuring non-interference is especially important in web-based services, as these can easily be accessed by unauthorized observers. The most commonly used language throughout the last couple of years for development is JavaScript [2, 3, 4], and thus several solutions have been developed for performing information flow analysis on JavaScript code, as seen in, e.g., the work of Just et al. from 2011 [5] and Hedin, Bello, and Sabelfeld's work from 2016 [6]. Both articles suggest mechanisms for dynamic (run-time) information flow analysis, though recognize that static (compile-time) analysis is required to analyze implicit information flows. Information flow is generally categorized into two types: explicit and implicit [6]. Explicit information flow covers situations, where data is copied directly, e.g., via an assignment. An example of this is given in Listing 1. We denote confidential data as `h`, and public data as `l`, for the code examples, in this paper.

```
let h = 2
let l = h
```

Listing 1: Explicitly leaking the value of `h` to `l` using an assignment.

Implicit information flow is slightly more complicated, as it describes situations where information is used to control program flow. An example of this is seen in Listing 2. Here, information about confidential data is indirectly revealed, as confidential data controls public aspects of the program behavior.

```
let h = true
let l = if (h) {
    2
    } else { 3 }
```

Listing 2: Implicitly revealing the value of `h` through a conditional assignment to `l`.

Listing 1 and Listing 2 showcase examples of information flow which capture exactly what we want to avoid in order to achieve non-interference.

In this paper we will look at a more recently developed programming language called ReScript where a non-interferent information flow could be relevant. ReScript is a functional programming language that compiles to JavaScript, and its developers claim ReScript to be the fastest build system on the web [7]. As the language is still quite new, information flow analysis for ReScript is yet to be developed.

An interesting aspect of ReScript is that some imperative features are included in the otherwise functional programming language. These complicate the analysis of non-interference, e.g., through the possibility to use `ref` to create mutable bindings. This allows for other ways to create information flow as seen in Listing 3 and Listing 4.

```
1  let h = ref(2)
2  let l = h
```

Listing 3: Leaking confidential information from variable `h` to public variable `l` by copying the reference to the location of the value of `h`.

```
1  let l = ref(2)
2  let h = l
3  h := 4
```

Listing 4: Leaking confidential data by copying the reference of a public variable, `l`, to a confidential variable, `h`, and then updating the value of h.

In both cases, the confidential and public variables end up pointing to the same memory, which means that the confidential information in `h` is accessible through the public variable `l`. This clearly breaks with non-interference. To deal with this, a solution must be able to control references.

In [8], Sabelfeld and Myers describe how a type system is a natural way to implement information flow analysis, as existing type systems can easily be extended with security levels. As with many properties, non-interference is an undecidable problem. Therefore, any type system aiming to capture non-interference will be an approximation.

In this paper, we introduce a type system for ReScript similar to the security type system in [8], with the aim of ensuring non-interference for a subset of ReScript. To prove that this type system ensures non-interference, we give a proof of soundness that shows if an expression is typeable it is also non-interferent. This method of proving non-interference is similar to the method by Volpano et al. in [9].

The rest of the paper is organized as follows. In section 2, we introduce the abstract syntax as well as the semantics of ReScript. In section 3, we introduce the type system to ensure non-interference in ReScript. In section 4, we prove the soundness of the type system. In section 5, we give an overview of a type checker implementing our type system, and how we used this as a tool to estimate the efficacy of the type rules. In section 6 we conclude this paper and discuss future work.

## 2 Preliminaries

We will start by introducing an abstract syntax based on a reduced ReScript syntax. This is followed by the semantics for the reduced syntax. The abstract syntax is based on [7] and [10]. The given semantics are based on our own testing of ReScript programs. This is due to the fact that we were unable to find any official documentation containing the semantic specification. We conclude this section by defining non-interference in the context of this project.

### 2.1 Reduced ReScript Syntax

We will not be covering the full syntax of the ReScript language. Instead, we have selected a subset of the language, encompassing the most essential and commonly used features from both the imperative and functional paradigms. This reduction allows for a focused exploration aligned with the project's objectives and time constraints. The chosen abstract syntax can be viewed in Figure 1.

$$
\begin{array}{rcll}
\textit{program pgm} & ::= & e & \\
\textit{expression } e & ::= & n & \text{(number)} \\
& | & b & \text{(boolean)} \\
& | & () & \text{(unit)} \\
& | & x & \text{(variable)} \\
& | & e\ e & \text{(application)} \\
& | & \lambda\ x.e & \text{(function)} \\
& | & e; e & \text{(sequential expressions)} \\
& | & \textsf{Let } x =\ e & \text{(let binding)} \\
& | & e\ \textit{bop}\ e & \text{(binary operators)} \\
& | & \textsf{Ref } e & \text{(new location)} \\
& | & x := e & \text{(write to location)} \\
& | & !x & \text{(read from location)} \\
& | & \textsf{For } x\ e\ e\ e & \text{(for loop)} \\
& | & \textsf{While } e\ e & \text{(while loop)} \\
& | & \textsf{If } e \textsf{ then } e \textsf{ else } e & \text{(if-then-else)} \\
\textit{binary operator bop} & ::= & + & \text{(addition)} \\
& | & - & \text{(subtraction)} \\
& | & * & \text{(multiplication)} \\
& | & / & \text{(division)} \\
& | & == & \text{(equality)} \\
& | & < & \text{(less than)} \\
& | & > & \text{(greater than)} \\
\end{array}
$$

Figure 1: The abstract syntax based on a reduced ReScript-syntax.

From Figure 1, it is seen that, at its core, ReScript is a functional language, as everything is an expression. However, ReScript has many imperative elements, such as iterative control structures and references. Most noteworthy is the ability to have mutable bindings through ref, by binding a variable to a location containing a value, instead of binding directly to the value. The use of Ref also allows aliasing, in which multiple variables are bound to the same location.

## 2.2 Semantics

Our semantics of ReScript are described in Big-Step-Semantics as a transition system of the form $((\textbf{Exp} \times \textbf{Env} \times \textbf{Sto}), \rightarrow, (\textbf{Val} \times \textbf{Env} \times \textbf{Sto}))$, where:

- Expressions (**Exp**): This set represents all expressions, possible through our ReScript abstract syntax.

- Environments (**Env**): This set represents all possible environments, which are partial functions from variables to values ($\textbf{Var} \rightharpoonup \textbf{Val}$). We use $\gamma$ to denote an element in **Env**.

- Stores (**Sto**): This set represents all possible stores, which are partial functions from locations to values ($\textbf{Loc} \rightharpoonup \textbf{Val}$). We use $\sigma$ to denote an element in **Sto**.

- Values (**Val**): This set represents all possible values (**Loc** ∪ ℤ ∪ {True, False} ∪ **Func** ∪ **Unit**), where:
    - Locations (**Loc**): Is the set of possible locations. We denote a location as $\ell$.
    - Integers (ℤ): Is the set of integers. We denote an integer as $n$.
    - Booleans ({True, False}): Is the set of boolean values. We denote a boolean value as $b$.
    - Functions (**Func**): Is the set of function closures, in the form (**Exp** × **Var** × **Env**).
    - Unit (**Unit**): Is the set $\{u\}$, which is the set containing only a single value, unit.

The function ℕ is defined to map each number-literal in the syntax to the actual number it represents. Similarly, its inverse function $ℕ^{-1}$, when given a numerical value, returns the number-literal equivalent to it. This approach is also applied to booleans with the function 𝔹. Additionally, the function *new* is introduced, which, upon receiving a store, returns a new location that is not in use by the given store. This is to ensure, that when we need a to use a new location, there is not accidentally a variable that is already referencing that location.

Below, are shown the most interesting semantic rules. The full set of semantic rules can be found in Appendix A.

(S-Let) represents let bindings and works by first evaluating the sub-expression $e$. We can then add the mapping of $x$ to $v$ to our environment. (S-Let) is a statement-like expression, that is used to change the environment, and is not used to compute a value. The evaluation value will therefore always be $u$ (unit), which is also the case for other statement-like expressions.

The (S-If-Else-True) and (S-If-Else-False) rules, that are shown in Figure 2, function as one would expect an if-expression to do in a functional language. In the case where the conditional expression evaluates to True, the whole expression evaluates to the value of the expression in the `then` branch. Conversely, if the conditional expression evaluates to False, the whole expression evaluates to the value of the expression in the `else` branch. Due to the scoping rules of ReScript, the environment is the same before and after the transition, thus any bindings made in the sub-expressions do not propagate throughout the rest of the program. This will be the case for all rules, except for (S-Let) and (S-Seq) which represents let bindings and sequential expressions, where we do want bindings to propagate to the following expressions. We do, however, always propagate the store, as side-effects are always possible through the use of a reassign expression.

As ReScript is a functional language, functions are first class citizens. As such, a function evaluates to a value representing the functions own closure. Therefore, (S-Func) will evaluate to a closure, that includes the sub-expression, the formal parameter, and the environment, when the function was declared.

(S-App) represents function application and works in three steps. First, evaluate $e_1$ to its value, which needs to be a function closure. Second, evaluate $e_2$ to its value. Third, evaluate the function body with the environment from the closure with added binding

$$(\text{S-Let}) \quad \frac{\langle e, \gamma^1, \sigma^1 \rangle \to \langle v, \gamma^2, \sigma^2 \rangle}{\langle \text{Let } x = e, \gamma^1, \sigma^1 \rangle \to \langle u, \gamma^1[x \mapsto v], \sigma^2 \rangle}$$

$$(\text{S-If-Else-True}) \quad \frac{\langle e_1, \gamma^1, \sigma^1 \rangle \to \langle \text{True}, \gamma^2, \sigma^2 \rangle \quad \langle e_2, \gamma^1, \sigma^2 \rangle \to \langle v, \gamma^3, \sigma^3 \rangle}{\langle \text{If } e_1 \text{ then } e_2 \text{ else } e_3, \gamma^1, \sigma^1 \rangle \to \langle v, \gamma^1, \sigma^3 \rangle}$$

$$(\text{S-If-Else-False}) \quad \frac{\langle e_1, \gamma^1, \sigma^1 \rangle \to \langle \text{False}, \gamma^2, \sigma^2 \rangle \quad \langle e_3, \gamma^1, \sigma^2 \rangle \to \langle v, \gamma^3, \sigma^3 \rangle}{\langle \text{If } e_1 \text{ then } e_2 \text{ else } e_3, \gamma^1, \sigma^1 \rangle \to \langle v, \gamma^1, \sigma^3 \rangle}$$

$$(\text{S-Func}) \quad \frac{}{\langle \lambda x.e, \gamma, \sigma \rangle \to \langle v, \gamma, \sigma \rangle} \quad v = (e, x, \gamma)$$

$$(\text{S-App}) \quad \frac{\begin{array}{c} \langle e_1, \gamma^1, \sigma^1 \rangle \to \langle v^1, \gamma^2, \sigma^2 \rangle \quad \langle e_2, \gamma^1, \sigma^2 \rangle \to \langle v^2, \gamma^3, \sigma^3 \rangle \\ \langle e_3, \gamma^5[x \mapsto v^2], \sigma^3 \rangle \to \langle v^3, \gamma^4, \sigma^4 \rangle \end{array}}{\langle e_1 \ e_2, \gamma^1, \sigma^1 \rangle \to \langle v^3, \gamma^1, \sigma^4 \rangle} \quad v^1 = (e_3, x, \gamma^5)$$

$$(\text{S-Ref}) \quad \frac{\langle e, \gamma^1, \sigma^1 \rangle \to \langle v, \gamma^2, \sigma^2 \rangle}{\langle \text{Ref}(e), \gamma^1, \sigma^1 \rangle \to \langle \ell, \gamma^1, \sigma^2[\ell \to v] \rangle} \quad \ell = new(\sigma^2)$$

$$(\text{S-Reassign}) \quad \frac{\langle e, \gamma^1, \sigma^1 \rangle \to \langle v^1, \gamma^2, \sigma^2 \rangle}{\langle x := e, \gamma^1, \sigma^1 \rangle \to \langle u, \gamma^1, \sigma^2[\ell \to v^1] \rangle} \quad \gamma^1(x) = \ell$$

Figure 2: Subset of semantics

from the formal parameter to the actual parameter. The evaluation value of the function body is then the value of the application.

(S-Ref) creates references to values. It works by first evaluating the sub-expression. Then we find an unused location using the *new* function. The evaluation of (S-Ref) is then the unused location and an updated store, where there is a mapping from the new location to the value of the sub-expression.

The last semantic rule introduced here is (S-Reassign), which involves overwriting values in the store. To implement this rule, the sub-expression $e$ is initially evaluated. Following this, the binding $x$ is looked up in the environment to find the associated location, that is pointing at the value to be overwritten. The store is then modified by creating a new mapping from the location to the value of the sub-expression, thereby overwriting the previous value. As (S-Reassign) functions as a statement-like expression, the resulting evaluation value is therefore $u$.

## 2.3 Non-interference

In this section we will give a definition of non-interference. Prior to defining non-interference, we need to define a type environment, and then low-equivalence.

A type environment is a partial function from variables to types ($\mathbf{Var} \rightharpoonup \mathbf{T}$), and we use **TEnv** to denote the set of all type environments. We use $\Gamma$ to denote a type environment. **T** is the set of all possible types using the formation rules in Figure 3. A type can be a security-level, that being either Low or High. Confidential data should be typed as High, and public data should be typed as Low. It can also be a function from one type to another type. To handle side effects there is a side effect type $t_1 @ t_2$, read as; the expression that produces $t_1$ has a side effect of $t_2$. The main objective of the side effect type to keep track of the lowest security-level affected in the evaluation of an expression. In the case where an expression has no side effects we have the empty type, which can be use in conjunction with the side effect type. Lastly we have ref $t$ types, which represents the security level for references.

In the context of security levels, the notion of a partial order can be very useful. One introduction to partial orders can be seen in [11] where Aceto et al. defines it as a pair $(D, \sqsubseteq)$, where $D$ is a set and $\sqsubseteq$ is a relation over $D$. This relation $\sqsubseteq$ must be reflexive, antisymmetric, and transitive.

Consider $X \subseteq D$. We define $d \in D$ as an upper bound for $X$ if and only if $\forall x \in X. x \sqsubseteq d$. Furthermore, $d$ is the least upper bound if $d \sqsubseteq d'$ for every $d' \in D$ that is also an upper bound for $X$. Similarly, $d \in D$ is a lower bound for $X$ if and only if $\forall x \in X. d \sqsubseteq x$, and $d$ is the greatest lower bound if $d' \sqsubseteq d$ for every $d' \in D$ that is also a lower bound for $X$. We denote the least upper bound of $X$ as $\sqcup X$ and the greatest lower bound of $X$ as $\sqcap X$.

$$
\begin{aligned}
t \;::=\; & \mathsf{Low} && \text{(Low security level)} \\
\mid\; & \mathsf{High} && \text{(High security level)} \\
\mid\; & t \to t && \text{(abstraction)} \\
\mid\; & t @ t && \text{(effects)} \\
\mid\; & () && \text{(empty)} \\
\mid\; & \mathsf{ref}\ t && \text{(reference)}
\end{aligned}
$$

Figure 3: The formation rules for types

We operate under the assumption that anything defined and typed as Low or as a function is observable to the user. While the restriction on functions is not strictly necessary for ensuring non-interference, we have opted to impose this requirement. This choice is made in recognition that, although the reassignment of functions is prohibited, a scenario might arise where the definition of a function slips out of a conditional.

Two states are low-equivalent if the following premises hold. All Low-variables have identical values. All Ref Low variables have locations, that point to identical values and all function variables are equivalent. Equivalence of functions are defined in definition 2.2. This definition acknowledges the potential for High-variables to exhibit varying values, between the two states.

**Definition 2.1** (State). *A state is defined as a pair consisting of a store and an environment, $s = (\gamma, \sigma)$, and **States** is defined as the set of all possible states.*

**Definition 2.2** (Equivalence). *Let*

$$(v_1, \sigma_1) \sim_t (v_2, \sigma_2)$$

*defined by*

$$u \sim_{\text{High}} u$$
$$n_1 \sim_{\text{High}} n_2$$
$$b_1 \sim_{\text{High}} b_2$$
$$\ell_1 \sim_{\text{High}} \ell_2$$
$$u \sim_{\text{Low}} u$$
$$n_1 \sim_{\text{Low}} n_2 \text{ where } n_1 = n_2$$
$$b_1 \sim_{\text{Low}} b_2 \text{ where } b_1 = b_2$$
$$\ell_1 \sim_{\text{Low}} \ell_2 \text{ where } \sigma_1(\ell_1) \sim \sigma_2(\ell_2)$$

*and*

$$\langle e_1, x_1, \gamma^1 \rangle \sim_{(t_1 \to t_2)} \langle e_1, x_1, \gamma^2 \rangle$$

*where*

$$\forall v_3, v_4 \text{ where } (v_3, \sigma_1) \sim_{t_1} (v_4, \sigma_2)$$

*and we have $\Gamma_1$ is the Environment where the functions was defined and we have that*
$$\mathbf{dom}(\Gamma_1) = \mathbf{dom}(\gamma^1) = \mathbf{dom}(\gamma^2)$$
$$\Gamma_1[x \to t_1] \vdash (\gamma^1[x \to v_3], \sigma^1) =_{\text{Low}} (\gamma^2[x \to v_4], \sigma^2)$$

*then if*

$$\langle e_1, \gamma^1[x_1 \mapsto v_3], \sigma_1 \rangle \longrightarrow \langle v_5, \gamma_3, \sigma_3 \rangle$$
$$\langle e_2, \gamma^2[x_2 \mapsto v_4], \sigma_2 \rangle \longrightarrow \langle v_6, \gamma_4, \sigma_4 \rangle$$

*then*

$$(v_5, \sigma_3) \sim_{t_2} (v_6, \sigma_4) \quad (\textit{Unless } t_2 = \textsf{ref } t_3, \textit{ then } (v_5, \sigma_3) \sim_{t_3} (v_6, \sigma_4))$$
$$\Gamma_1 \vdash (\gamma_1, \sigma_3) =_{\text{Low}} (\gamma_2, \sigma_4)$$

The reason for the requirements of $\sim_t$ is that we need booleans and numbers to be strictly equal, when they are typed as Low. However, we do not care about their values, when they are typed as High. With regards to locations, then when we have 2 location that is referencing something typed as Low, we need the value stored at the locations to be equal. This is also why $\sim_t$ is a relation over pairs, containing a value and a store. The equivalence of functions need to show that regardless of input, if the inputs to the functions are $\sim_{t_1}$ with regards to the input type $t_1$, then the output of the functions are $\sim_{t_2}$, with regards to the output type $t_2$. Furthermore, we need to know that the functions maintain low-equivalence between the states.

**Definition 2.3** (Low-equivalence). *Consider $\Gamma \in \textbf{TEnv}$ and $s_1, s_2 \in \textbf{States}$ then $\Gamma \vdash s_1 =_{\text{Low}} s_2$ if for all $x \in \textbf{dom}(\Gamma)$ the following conditions apply.*

*If*

- $\Gamma(x) \in \{\text{Low}, \text{ref Low}\}$, *then* $(\gamma_{s_1}(x), \sigma_{s_1}) \sim_{\text{Low}} (\gamma_{s_2}(x), \sigma_{s_2})$.
- $\Gamma(x) = (t_1 \to t_2 @ t_3)$, *then* $(\gamma_{s_1}(x), \sigma_{s_1}) \sim_{t_1 \to t_2} (\gamma_{s_2}(x), \sigma_{s_2})$.

We define non-interference in the context of our semantics and our type environment. The definition states an expression upholds the non-interference property under a type environment, if it is the case that for any two states that are low-equivalent, then any new states reachable must also be low-equivalent. In section 4, we prove it to be the case that any well-typed expression using our type system always upholds the non-interference property.

**Definition 2.4** (Non-interference). *Consider $e \in \textbf{Exp}$ and $\Gamma, \Gamma' \in \textbf{TEnv}$. We say that $e$ upholds the non-interference property under $\Gamma$ if the following condition holds, for any given pair $s_1, s_2 \in \textbf{States}$.*

$$
\begin{aligned}
&\textit{if} \quad \Gamma \vdash s_1 =_{\text{Low}} s_2 \\
&\textit{and} \quad \langle e, s_1 \rangle \to \langle v_1, s_1' \rangle \quad (1) \\
&\textit{and} \quad \langle e, s_2 \rangle \to \langle v_2, s_2' \rangle \quad (2) \\
&\textit{then} \ \Gamma' \vdash s_1' =_{\text{Low}} s_2'
\end{aligned}
$$

# 3 Type system

For this chapter, we commence with a brief overview of type judgments, followed by an exploration of the partial order employed for our security levels. Subsequently, we delve into an explanation of the most interesting type rules within our type system.

## 3.1 Type judgments

Type judgments are of the form $\Gamma, pc \vdash e : t_1 @ t_2 \triangleright \Gamma'$, read as: Given a type environment $\Gamma$ and a security level $pc$, then the expression $e$ has the type $t_1$, with the lowest side effect being $t_2$, and produce a new type environment $\Gamma'$. Here, type environments are partial functions from variables to types ($\textbf{Var} \rightharpoonup \textbf{T}$), $pc \in \{\text{Low}, \text{High}\}$, $e \in \textbf{Exp}$ and $t \in \textbf{T}$. $pc$ represents the security level of the context, meaning that if $pc = \text{High}$, then when $e$ is evaluated in the program, there is a High value that controls the program flow. An example of when $pc$ is High is seen in line 3 in Listing 2.

## 3.2 Partial Order of Security Levels

We define the partial order $(\mathbf{T}, \sqsubseteq)$ using the inference rules seen in Figure 4. The first six rules describe the order between the types Low, High and (), while (Abs-Abs) describes the order between functions of the same arity. The partial order does not define any ordering between Low, High and () and functions, nor does it define any ordering for side-effect or reference types.

$$(\text{Low-Low}) \quad \frac{}{\mathsf{Low} \sqsubseteq \mathsf{Low}} \qquad (\text{Low-High}) \quad \frac{}{\mathsf{Low} \sqsubseteq \mathsf{High}}$$

$$(\text{Low-Empty}) \quad \frac{}{\mathsf{Low} \sqsubseteq ()} \qquad (\text{High-High}) \quad \frac{}{\mathsf{High} \sqsubseteq \mathsf{High}}$$

$$(\text{High-Empty}) \quad \frac{}{\mathsf{High} \sqsubseteq ()} \qquad (\text{Empty-Empty}) \quad \frac{}{() \sqsubseteq ()}$$

$$(\text{Abs-Abs}) \quad \frac{t_1 \sqsubseteq t_1' \quad t_2' \sqsubseteq t_2}{t_1 \to t_2 \sqsubseteq t_1' \to t_2'}$$

Figure 4: Partial order of security levels.

## 3.3 Type rules

The type rules of our type system is presented in Figure 5 and Figure 6.

$$(\text{Bop}) \quad \frac{\Gamma_1, pc_1 \vdash e_1 : t_1 @ t_2 \triangleright \Gamma_2 \quad \Gamma_1, pc_1 \vdash e_2 : t_3 @ t_4 \triangleright \Gamma_3}{\Gamma_1, pc_1 \vdash e_1 \, bop \, e_2 : t_5 @ t_6 \triangleright \Gamma_1} \quad \begin{array}{l} t_1, t_3 \in \{\mathsf{Low}, \mathsf{High}\} \\ t_5 = \sqcup\{t_1, t_3\} \\ t_6 = \sqcap\{t_2, t_4\} \end{array}$$

$$(\text{Let-n}) \quad \frac{\Gamma, pc \vdash e : t_2 @ t_3 \triangleright \Gamma_1}{\Gamma, pc \vdash \mathsf{Let} \, x_{t_1} = e : \mathsf{Low} @ t_4 \triangleright \Gamma[x \to t_1]} \quad \begin{array}{l} t_1, t_2 \in \{\mathsf{Low}, \mathsf{High}\} \\ t_1 \sqsupseteq t_2 \\ t_1 \sqsupseteq pc \\ t_4 = \sqcap\{t_3, t_1\} \end{array}$$

$$(\text{Let-Base}) \quad \frac{\Gamma, pc \vdash e : t_1 @ t_2 \triangleright \Gamma_1}{\Gamma, pc \vdash \mathsf{Let} \, x = e : \mathsf{Low} @ t_3 \triangleright \Gamma[x \to t_1]} \quad \begin{array}{l} t_1 \in \{\mathsf{Low}, \mathsf{High}\} \\ t_1 \sqsupseteq pc \\ t_3 = \sqcap\{t_1, t_2\} \end{array}$$

Figure 5: Subset of type system

$$\text{(Let-Base-Func)} \quad \frac{\Gamma, pc \vdash e : (t_1 \to t_2 @ t_3) @ t_4 \rhd \Gamma_1}{\Gamma, pc \vdash \mathsf{Let}\ x = \ e\ : \mathsf{Low@Low} \rhd \Gamma[x \to (t_1 \to t_2 @ t_3)]} \quad \mathsf{Low} \sqsupseteq pc$$

$$\text{(Let-Base-Ref)} \quad \frac{\Gamma, pc \vdash e : \mathsf{ref}\ t_1 @ t_2 \rhd \Gamma_1}{\Gamma, pc \vdash \mathsf{Let}\ x = \ e\ : \mathsf{Low} @ t_3 \rhd \Gamma[x \to \mathsf{ref}\ t_1]} \quad \begin{array}{c} t_3 = \sqcap\{t_2, t_1\} \\ t_3 \sqsupseteq pc \\ t_1 \in \{\mathsf{Low}, \mathsf{High}\} \end{array}$$

$$\text{(If-Else)} \quad \frac{\begin{array}{c}\Gamma_1, pc_1 \vdash e_1 : t_1 @ t_2 \rhd \Gamma_2 \\ \Gamma_1, pc_2 \vdash e_2 : t_3 @ t_4 \rhd \Gamma_3 \\ \Gamma_1, pc_2 \vdash e_3 : t_5 @ t_6 \rhd \Gamma_4 \end{array}}{\Gamma_1, pc_1 \vdash \mathsf{If}\ e_1\ \mathsf{then}\ e_2\ \mathsf{else}\ e_3 : t_7 @ t_8 \rhd \Gamma_1} \quad \begin{array}{c} pc_2 = \sqcup\{pc_1, t_1\} \\ t_7 = \sqcup\{t_1, t_3, t_5\} \\ t_8 = \sqcap\{t_2, t_4, t_6\} \\ t_1 \in \{\mathsf{Low}, \mathsf{High}\} \end{array}$$

$$\text{(For)} \quad \frac{\begin{array}{c}\Gamma_1, pc_1 \vdash e_1 : t_1 @ t_2 \rhd \Gamma_2 \\ \Gamma_1, pc_1 \vdash e_2 : t_3 @ t_4 \rhd \Gamma_3 \\ \Gamma_1[x \to pc_2], pc_2 \vdash e_3 : t_5 @ t_6 \rhd \Gamma_4 \end{array}}{\Gamma_1, pc_1 \vdash \mathsf{For}\ x\ e_1\ e_2\ e_3 : \mathsf{Low} @ t_8 \rhd \Gamma_1} \quad \begin{array}{c} pc_2 = \sqcup\{pc_1, t_1, t_3\} \\ t_8 = \sqcap\{t_2, t_4, t_6\} \\ t_1, t_3 \in \{\mathsf{Low}, \mathsf{High}\} \end{array}$$

$$\text{(While)} \quad \frac{\begin{array}{c}\Gamma_1, pc_1 \vdash e_1 : t_1 @ t_2 \rhd \Gamma_2 \\ \Gamma_1, pc_2 \vdash e_2 : t_3 @ t_4 \rhd \Gamma_3 \end{array}}{\Gamma_1, pc_1 \vdash \mathsf{While}\ e_1\ e_2 : \mathsf{Low} @ t_6 \rhd \Gamma_1} \quad \begin{array}{c} pc_2 = \sqcup\{pc_1, t_1\} \\ t_6 = \sqcap\{t_2, t_4\} \\ t_1 \in \{\mathsf{Low}, \mathsf{High}\} \end{array}$$

$$\text{(Seq)} \quad \frac{\begin{array}{c}\Gamma_1, pc \vdash e_1 : t_1 @ t_2 \rhd \Gamma_2 \\ \Gamma_2, pc \vdash e_2 : t_3 @ t_4 \rhd \Gamma_3 \end{array}}{\Gamma_1, pc \vdash e_1; e_2 : t_3 @ t_5 \rhd \Gamma_3} \quad t_5 = \sqcap\{t_2, t_4\}$$

$$\text{(Func)} \quad \frac{\Gamma_1[x \mapsto t_1], pc \vdash e : t_2 @ t_3 \rhd \Gamma_2}{\Gamma_1, pc \vdash \lambda x_{t_1}.e : (t_1 \to t_2 @ t_3) @ () \rhd \Gamma_1} \quad t_3 \sqsupseteq pc$$

$$\text{(App)} \quad \frac{\begin{array}{c}\Gamma_1, pc_1 \vdash e_2 : t_1 @ t_2 \rhd \Gamma_2 \\ \Gamma_1, pc_1 \vdash e_1 : (t_1 \to t_3 @ t_4) @ t_5 \rhd \Gamma_3 \end{array}}{\Gamma_1, pc_1 \vdash e_1\ e_2 : t_3 @ t_6 \rhd \Gamma_1} \quad \begin{array}{c} t_6 = \sqcap\{t_2, t_4, t_5\} \\ t_6 \sqsupseteq pc_1 \end{array}$$

$$\text{(Deref)} \quad \frac{}{\Gamma, pc \vdash !x : t_1 @ () \rhd \Gamma} \quad \Gamma(x) = \mathsf{ref}\ t_1$$

$$\text{(Reassign)} \quad \frac{\Gamma, pc \vdash e : t_1 @ t_2 \rhd \Gamma_1}{\Gamma, pc \vdash x := e : \mathsf{Low} @ t_4 \rhd \Gamma} \quad \begin{array}{c} \Gamma(x) = \mathsf{ref}\ t_3 \\ t_3 \in \{\mathsf{Low}, \mathsf{High}\} \\ t_3 \sqsupseteq t_1 \\ t_3 \sqsupseteq pc \\ t_4 = \sqcap\{t_3, t_2\} \end{array}$$

Figure 6: Subset of type system

For all our rules typing a `Let`-binding, the entire expression will be typed as Low. This is because the semantic rule (S-Let) always evaluates to $u$, which means regardless of the type of $e$, the `Let`-binding will always evaluate to the same value. However, we still need to check if the type of $e$ is allowed in the current context, which is why we check that $t_1 \sqsupseteq pc$ is in the partial order. Checking that the type of an sub-expression is allowed within its surrounding expression, is a common operation throughout the type system. The side-effect of the `Let`-binding will be the greatest lower bound of the type of $e$ and the side-effect of $e$. Think of this as either the lowest side-effect is the side-effect of $e$ or the lowest side-effect is the act of binding $e$. Lastly we add a mapping from the variable $x$ to the type of $e$ to the type environment $\Gamma$.

For (Let-Base-Func), a crucial observation is that the side-condition imposes a limitation on the assignment of a variable to a function, permitting it only in a $pc$ that is less than or equal to Low. The limitation of the $pc$ comes from our assumption, that everything, which is defined, and is typed as Low or as a function, is observable to the user. Using this assumption, the user can not know the value of h, but running the program multiple times could leak the value of h, as they would be able to see the definition of the function f change. An example of this can be seen in Listing 5.

```
let h = true
let l = Ref(4)
if(h) {
    let f = (x) => {l := 5; x}
    f 2
} else {
    let f = (x) => {l := 6; x}
    f 3
}
```
Listing 5: Implicit information flow about h, leaked by a conditional and function application

(Let-Base-Ref) ensures that the type of the value stored at the location, must be higher or equal to the $pc$, to ensure that we cannot change Low values/€, in parts of the program, where $pc = $ High.

(Let-n) works similarly to (Let-Base), but instead of inferring the type of $x$ through the type of $e$, $x$ has its own type annotation. This is also why there is a side-condition which states that the type of the expression $e$ must be typed as less than or equal to $t_1$, ensuring there is no direct information-flow. The mapping in the type environment is also changed such that it maps the variable to the type given in the type annotation. This also allows programmers to explicitly say that some values should be regarded as High, as there is no other way, for High values to appear otherwise.

We only allow defining variables typed as High or ref High, when $pc = $ High. This restriction is captured in (Let-n) by the side condition $t_1 \sqsupseteq pc$, as the type of the variable to be defined, has to be the the same as, or higher than the $pc$ it is being defined in. An example of this can be seen in Listing 6.

```
let h = true
if(h) {
    let l = 3
} else {
```

```
5        let l = 4
6   }
```
Listing 6: Implicit information flow about h, leaked by a conditional

The (Reassign) rule works in the same way as the (Let-n) rule, except it is used for variables that are already declared. So instead of having the type of the variable in a type annotation, we have to retrieve it from the type environment. Furthermore, the type of the variable has to be a reference type.

It is possible to have an implicit information-flow through the use of control structures, such as in the case of the if-expression in Listing 7. In the example it can be seen that it is possible to deduce the value of $h$ through the value of $l$, even though $l$ was never explicitly assigned the value of $h$. The type system should therefore keep track of the contexts in which it is permissible to reassign to variables of varying security-levels.

```
1   let h = ref(true)
2   let l = ref(false)
3   if(!h) {
4        l := true
5   } else {
6        l := false
7   }
```
Listing 7: Implicit flow of information from High to Low through a conditional

(If-Else) works by first typing the conditional expression. To prevent implicit information-flow as seen in Listing 7, we need to type the expressions for branches under a new *pc*, which will be the lowest upper bound of the previous *pc* and the type of the conditional expression. The type of the if-expression will then be the lowest upper bound of the types for all the sub-expressions, while the side-effect will be the greatest lower bound of the side-effects of all the sub-expressions.

In the (Seq) rule, we ensure that the bindings carry over through sequential expressions. When typing sequential expressions, our focus is solely on the type of the last expression. Since the two expressions are executed separately, the final value of the entire sequential expression is determined by the type of the last expression.

As was the case with if-expressions, it is also possible to have implicit information-flow using a while loop as seen in Listing 8.

```
1   let h = ref(true)
2   let l = ref(false)
3   while(!h) {
4        l := true
5   }
```
Listing 8: Implicit information flow in while loop

To prevent any implicit flow stemming from the while loop, we follow the same procedure as in (If-Else). We set the program counter of the sub-expression, to the least upper bound of the current *pc* and the type of the conditional expression.

The (For) type rule follows the same principles as (If-Else) and (While), that being to prevent implicit information-flow in the body. This is done by typing the body under a new *pc* which will be the lowest upper bound of the types of both $e_1$ and $e_2$, and the previous *pc*, as both $e_1$ and $e_2$ are used to determine how many times the `For`-loop will be executed. The body will also be typed under a type environment in which the mapping from the variable $x$ to the new *pc* is added.

ReScript, being a functional language with imperative tools, facilitates the utilization of variables from external scopes within functions. In the context of Listing 9, the function `f` is invoked within an if-expression marked by a High program counter. However the function `f` makes an alteration to the Low-variable `l`, necessitating the implementation of measures to prohibit such behavior.

```
let h = true;
let l = 2;
let f = (x) => {l := 3; x}

if (h) {
   f(2)
} else {
   3
}
```

Listing 9: Leaking through function call in High program counter

When typing a function with (Func), we first need to type the function body under a type environment where the parameter $x$ is mapped to its type annotation. The type of the function will then be a function type, from the type of the parameter to the type of the body. The inner side-effect will be the same as the side-effect of the body, as it corresponds to the side-effect caused upon an application of the function. The outer side-effect will be empty as it corresponds to the side-effect of simply evaluating to the function definition. The side-condition $t_3 \sqsubseteq pc$ ensures that it is not permissible to create functions in contexts where the body of the function would cause side-effects not permissible in the current context. With this side-condition we catch unusable functions, as functions with Low side-effects where $pc =$ High will never be applied.

The (App) rule demonstrates that an application can be typed if $e_1$ can be typed as a function, with the parameter type aligning with the type of $e_2$. Taking into consideration what we learned from Listing 9, the side-effects embedded in $e_1$, $e_2$, and the function body, should not induce implicit information flow if executed in the current context.

In the binary operation rule (Bop) the type of the whole operation will be the lowest upper bound of types of the two sub-expressions. This is because, even though we dont directly leak a High value, it still affects the resulting value. If we had an expression such as $h + 3$, then the value of the High variable $h$ could be deduced by subtracting 3 from the resulting value. The side-effect will the the greatest lower bound of the side-effects of the two sub-expressions.

When creating a reference, the type of the whole expression will be typed with (Ref), and the type of the sub-expression is wrapped into a reference type. We also limit the types that can be referenced to, such that a reference type can only point to a type of High or Low. This is to limit the scope of the subset of ReScript, that we have choosen to create a type-system for.

In (Deref), we can use the type environment to look up the type of *x*. Given that it is a reference-variable, we can unwrap the type being referenced and adopt this type as the type for the Deref operation.

# 4 Soundness of the type system

## 4.1 Soundness Theorem

We will now outline how we have proven, that our type system ensures, that all well-typed expressions maintain non-interference. For the full proof of every type-rule go to Appendix C.

Firstly, we need to define precisely what well-typed means:

**Definition 4.1** (Well-typed). *An expression e is well-typed in regards to a type environment $\Gamma$ and a security level pc, if:* $\Gamma, pc \vdash e : t_1 @ t_2 \triangleright \Gamma'$

We want to show that any given well-typed expression, will maintain non-interference. This means, that if an expression can be typed with our type rules, you are guaranteed that there is no information flow from High variables to Low variables. This can be described with the following theorem:

**Theorem 1** (Soundness). *Let*

- $\Gamma \in \mathbf{Env}$
- $s_1, s_2 \in \mathbf{States}$
- $pc \in \{\mathsf{High}, \mathsf{Low}\}$

*Assume that*

$$\Gamma, pc \models e : t_1 @ t_2 \triangleright \Gamma'$$
$$\Gamma \models s_1 =_{\mathsf{Low}} s_2$$

*if*

$$\langle e, s_1 \rangle \to \langle v^1, s_1' \rangle \tag{1}$$
$$\langle e, s_2 \rangle \to \langle v^1, s_2' \rangle \tag{2}$$

*then*

$$\Gamma' \models s_1' =_{\mathsf{Low}} s_2'$$

If Theorem 1 holds for all type environments, states, expressions and security levels, we can be sure that all well typed expressions will maintain non-interference. It is this theorem, that we will prove holds for our type system and semantics of ReScript.

To prove Theorem 1, we will use induction on the rules used to conclude (1). This means that we assume that Theorem 1 holds for the rules in the premise, and from that, we will show that it also holds for the actual rule. We will need 3 lemmas, that are proven in the appendixes Appendix D, Appendix E and Appendix F to do this. The first is that we require Low-evaluation, meaning that, if an expression *e* is typed as Low, it evaluates to the same value, if evaluated in two different low-equivalent states. Furthermore, if *e* is typed as $t_1 \rightarrow t_2@t_3$, then the function that the expression evaluates to, must maintain non-interference. Lastly, if *e* evaluates to a $l_{\mathsf{Low}}$, the value stored at this location must also be the same.

**Lemma 1** (Low Evaluation). *Let*

- $\Gamma, pc \vdash e : t_1@t_2$
- $t_1 \in \{\mathsf{Low}, \mathbf{Func}, l_{\mathsf{Low}}\}$
- $\forall s_1, s_2 \in \mathbf{States}$ *where* $s_1 =_{\mathsf{Low}} s_2$

*if*

$$\langle e, \gamma_{s_1}, \sigma_{s_1} \rangle \rightarrow \langle v_{s_1}, \gamma'_{s_1}, \sigma'_{s_1} \rangle \qquad (3)$$
$$\langle e, \gamma_{s_2}, \sigma_{s_2} \rangle \rightarrow \langle v_{s_2}, \gamma'_{s_2}, \sigma'_{s_2} \rangle \qquad (4)$$

*then if* $t_1 \in \{\mathsf{Low}, l_{\mathsf{Low}}\}$

$$v_{s_1} \sim_{\mathsf{Low}} v_{s_2}$$

*else if* $t_1 = (t_1 \rightarrow t_2@t_3)$

$$v_{s_1} \sim_{t_1 \rightarrow t_2} v_{s_2}$$

One of the reasons this lemma is needed, is because when we have constructs such as `if`-, `while`- and `for`-expressions, then if the type of the condition is Low, we need to be sure that the same rule is used for the derivation tree of the expression.

We also need to be sure, that if the condition of an conditional-expression, or a while-expression is typed as High, then we cannot guarantee that the derivation-tree of an expression is the same in two different states. Therefore, we need to be sure that the side-effects of an expression evaluated in such an environment, where $pc = \mathsf{High}$, does not change the non-interference property, meaning that there must not be assigned to any low-variables. Therefore, we have Lemma 2.

**Lemma 2** (Side-effect Equivalent High PC). *Given*

$$\Gamma, \mathsf{High} \vdash e : t_1@t_2 \rhd \Gamma_1$$

*if*

$$\langle e, s \rangle \longrightarrow \langle v, s' \rangle \qquad (5)$$

*then*

$$\Gamma_1 \models s =_{\mathsf{Low}} s'$$

Furthermore, we need to be sure, that when a local scope ends, going back to the earlier type environment and environments do not change non-interference. Therefore, we need Lemma 3.

**Lemma 3** (Exiting scope non-interference). *If*

$$\Gamma \models (\gamma_{s_1}^1, \sigma_{s_1}^1) =_{\text{Low}} (\gamma_{s_2}^1, \sigma_{s_2}^1)$$

*and*

$$\langle e, \gamma_{s_1}^1, \sigma_{s_1}^1 \rangle \rightarrow \langle v_{s_1}, \gamma_{s_1}^2, \sigma_{s_1}^2 \rangle \quad (6)$$
$$\langle e, \gamma_{s_2}^1, \sigma_{s_2}^1 \rangle \rightarrow \langle v_{s_2}, \gamma_{s_2}^2, \sigma_{s_2}^2 \rangle \quad (7)$$

$$\Gamma, pc \vdash e : t \triangleright \Gamma'$$

*then*

$$\Gamma \models (\gamma_{s_1}^1, \sigma_{s_1}^2) =_{\text{Low}} (\gamma_{s_2}^1, \sigma_{s_2}^2)$$

The last lemma, Lemma 4 ensures that if an expression does not change non-interference, meaning that the state after the evaluation is low-equivalent to the state before the evaluation, then we can go back to the type-environment and environment from before the evaluation.

**Lemma 4** (Exiting scope non-interference extended). *If*

$$\Gamma' \models (\gamma_{s_1}^1, \sigma_{s_1}^1) =_{\text{Low}} (\gamma_{s_1}^2, \sigma_{s1}^2)$$

*and*

$$\langle e, \gamma_{s_1}^1, \sigma_{s_1}^1 \rangle \rightarrow \langle v_{s_1}, \gamma_{s_1}^2, \sigma_{s_1}^2 \rangle \quad (8)$$
$$\Gamma, pc \vdash e : t \triangleright \Gamma'$$

*then*

$$\Gamma \models (\gamma_{s_1}^1, \sigma_{s_1}^1) =_{\text{Low}} (\gamma_{s_1}^1, \sigma_{s_1}^2)$$

We will begin with the proving Theorem 1 holds for (S-If-Else-True).

## 4.2 If-Else

In the case of the If-construct, we cannot be sure that (1) and (2) use the same rule. This comes from the fact that $\Gamma \models s_1 =_{\text{Low}} s_2$ does not guarantee that High variables have the same value. So if, for an example, the conditional expression $e_1$ was a High variable, then it might in one state evaluate to True, and in another state evaluate to False. For an example, the transition for the expression "if $h$ then 2 else 3", where $\Gamma(h) = $ High could be derived using the following 2 derivation trees:

$$\text{(S-If-Else-True)} \frac{\text{(S-Var)} \frac{}{\langle h, \gamma_{s_1}, \sigma_{s_1} \rangle \rightarrow \langle \text{True}, \gamma_{s_1}, \sigma_{s_1} \rangle} \quad \gamma_{s_1}(h) = \text{True} \quad \text{(S-Num)} \frac{}{\langle 2, \gamma_{s_1}, \sigma_{s_1} \rangle \rightarrow \langle v_1, \gamma_{s_1}, \sigma_{s_1} \rangle} \quad \mathbb{N}(2) = v_1}{\langle \text{if } h \text{ then } 2 \text{ else } 3, \gamma_{s_1}, \sigma_{s_1} \rangle \rightarrow \langle v_1, \gamma_{s_1}, \sigma_{s_1} \rangle}$$

and

$$\text{(S-If-Else-False)} \frac{\text{(S-Var)} \frac{}{\langle h, \gamma_{s_2}, \sigma_{s_2} \rangle \rightarrow \langle \text{False}, \gamma_{s_2}, \sigma_{s_2} \rangle} \quad \gamma_{s_2}(h) = \text{False} \quad \text{(S-Num)} \frac{}{\langle 3, \gamma_{s_2}, \sigma_{s_2} \rangle \rightarrow \langle v_2, \gamma_{s_2}, \sigma_{s_2} \rangle} \quad \mathbb{N}(3) = v_2}{\langle \text{if } h \text{ then } 2 \text{ else } 3, \gamma_{s_1}, \sigma_{s_1} \rangle \rightarrow \langle v_2, \gamma_{s_1}, \sigma_{s_1} \rangle}$$

Therefore, we need to show that regardless of which of the two rules are chosen for (2), non-interference still holds.

Therefore, (1) is concluded with:

(S-If-Else-True) $\dfrac{\langle e_1, \gamma_{s_1}^1, \sigma_{s_1}^1 \rangle \to \langle \mathsf{True}, \gamma_{s_1}^2, \sigma_{s_1}^2 \rangle \quad \langle e_2, \gamma_{s_1}^1, \sigma_{s_1}^2 \rangle \to \langle v_{s_1}, \gamma_{s_1}^3, \sigma_{s_1}^3 \rangle}{\langle \mathsf{If}\ e_1\ \mathsf{then}\ e_2\ \mathsf{else}\ e_3, \gamma_{s_1}^1, \sigma_{s_1}^1 \rangle \to \langle v_{s_1}, \gamma_{s_1}^1, \sigma_{s_1}^3 \rangle}$

and (2) is concluded with either

(S-If-Else-True) $\dfrac{\langle e_1, \gamma_{s_2}^1, \sigma_{s_2}^1 \rangle \to \langle \mathsf{True}, \gamma_{s_2}^2, \sigma_{s_2}^2 \rangle \quad \langle e_2, \gamma_{s_2}^1, \sigma_{s_2}^2 \rangle \to \langle v_{s_2}, \gamma_{s_2}^3, \sigma_{s_2}^3 \rangle}{\langle \mathsf{If}\ e_1\ \mathsf{then}\ e_2\ \mathsf{else}\ e_3, \gamma_{s_2}^1, \sigma_{s_2}^1 \rangle \to \langle v_{s_2}, \gamma_{s_2}^1, \sigma_{s_2}^3 \rangle}$

or

(S-If-Else-False) $\dfrac{\langle e_1, \gamma_{s_2}^1, \sigma_{s_2}^1 \rangle \to \langle \mathsf{False}, \gamma_{s_2}^2, \sigma_{s_2}^2 \rangle \quad \langle e_3, \gamma_{s_2}^1, \sigma_{s_2}^2 \rangle \to \langle v_{s_2}, \gamma_{s_2}^3, \sigma_{s_2}^3 \rangle}{\langle \mathsf{If}\ e_1\ \mathsf{then}\ e_2\ \mathsf{else}\ e_3, \gamma_{s_2}^1, \sigma_{s_2}^1 \rangle \to \langle v_{s_2}, \gamma_{s_2}^1, \sigma_{s_2}^3 \rangle}$

However, regardless of what rule is chosen, a conditional-expression is always typed with:

(If-Else) $\dfrac{\begin{array}{c}\Gamma_1, pc_1 \vdash e_1 : t_1 @ t_2 \triangleright \Gamma_2 \\ \Gamma_1, pc_2 \vdash e_2 : t_3 @ t_4 \triangleright \Gamma_3 \\ \Gamma_1, pc_2 \vdash e_3 : t_5 @ t_6 \triangleright \Gamma_4\end{array}}{\Gamma_1, pc_1 \vdash \mathsf{If}\ e_1\ \mathsf{then}\ e_2\ \mathsf{else}\ e_3 : t_7 @ t_8 \triangleright \Gamma_1}$ $\begin{array}{l}pc_2 = \sqcup\{pc_1, t_1\} \\ t_7 = \sqcup\{t_1, t_3, t_5\} \\ t_8 = \sqcap\{t_2, t_4, t_6\} \\ t_1 \in \{\mathsf{Low}, \mathsf{High}\}\end{array}$

Now, from the fact that $t_1 \in \{\mathsf{Low}, \mathsf{High}\}$ we see that there are two cases, that we need to handle separately, namely that $t_1 = \mathsf{Low}$ and $t_1 = \mathsf{High}$.

We will begin with the case where $t_1 = \mathsf{Low}$. We will now use Lemma 1, which states that since $\Gamma_1, pc \vdash e_1 : \mathsf{Low} @ t_2 \triangleright \Gamma_2$, and

$$\langle e_1, \gamma_{s_1}^1, \sigma_{s_1}^1 \rangle \to \langle v_{s_1}, \gamma_{s_1}^2, \sigma_{s_1}^2 \rangle$$

*and*

$$\langle e_1, \gamma_{s_2}^1, \sigma_{s_2}^1 \rangle \to \langle v_{s_2}, \gamma_{s_2}^2, \sigma_{s_2}^2 \rangle$$

then we have that $v_{s_1} = v_{s_2}$.

Therefore, the same transition rule is used, as if $v_{s_1} = v_{s_2} = \mathsf{True}$, then both (1) and (2) is concluded with (S-If-Else-True).

We will now use the inductive hypothesis, meaning that we assume the expressions in the precondition uphold Theorem 1. Then, since $e_1$ must be well-typed, for the expression "if $e_1$ then $e_2$ else $e_3$" to be well-typed, and the states it is evaluated in is $=_{\mathsf{Low}}$, we then know that $\Gamma_2 \models (\gamma_{s_1}^2, \sigma_{s_1}^2) =_{\mathsf{Low}} (\gamma_{s_2}^2, \sigma_{s_2}^2)$.

We can then use lemma 4, to switch out the type environment and environments with earlier ones, so that we get $\Gamma_1 \models \langle \gamma^1_{s_1}, \sigma^2_{s_1} \rangle =_{\mathsf{Low}} \langle \gamma^1_{s_2}, \sigma^2_{s_2} \rangle$

Then, since the the states that the next expression is evaluated in is $=_{\mathsf{Low}}$, and $e_2$ and $e_3$ must also be well-typed, we have that

$$\langle e_2, \gamma^1_{s_1}, \sigma^2_{s_1} \rangle \to \langle v^1_{s_1}, \gamma^3_{s_1}, \sigma^3_{s_1} \rangle$$

*and*

$$\langle e_2, \gamma^1_{s_2}, \sigma^2_{s_2} \rangle \to \langle v^2_{s_2}, \gamma^3_{s_2}, \sigma^3_{s_2} \rangle$$

We then from the inductive hypothesis know, that $\Gamma_3 \models (\gamma^3_{s_1}, \sigma^3_{s_1}) =_{\mathsf{Low}} (\gamma^3_{s_2}, \sigma^3_{s_2})$

We can then use lemma 4, to once again, exit the local scope, and go back to $\Gamma_1$, $\gamma^1_1$ and $\gamma^1_2$, so that we have $\Gamma_1 \models \langle \gamma^1_{s_1}, \sigma^3_{s_1} \rangle =_{\mathsf{Low}} \langle \gamma^1_{s_2}, \sigma^3_{s_2} \rangle$.

As this is the final states and type environment, after the evaluation of an conditional-expression, this means that Theorem 1 holds for (S-If-Else-True) if $t_1 = \mathsf{Low}$. We will now look at the case where $t_1 = \mathsf{High}$

Since $t_1 = \mathsf{High}$, we then know that since $pc_2 = \sqcup \{pc_1, \mathsf{High}\}$, that $pc_2 = \mathsf{High}$. This can be seen by inspection of our inference rules found in Section 3.2, and the fact that $pc_2 \in \{\mathsf{Low}, \mathsf{High}\}$ by definition of our type system. We therefore have that:

$$\Gamma_1, \mathsf{High} \vdash e_2 : t_3 @ t_4 \triangleright \Gamma_2$$

*and*

$$\Gamma_1, \mathsf{High} \vdash e_3 : t_5 @ t_6 \triangleright \Gamma_2$$

We can then use Lemma 2, to say that the state before evaluation of $e_2$ or $e_3$ in (2), will be low-equivalent to the new state after evaluation, meaning that

$$\Gamma_3 \models (\gamma^1_{s_1}, \sigma^2_{s_1}) =_{\mathsf{Low}} (\gamma^3_{s_1}, \sigma^3_{s_1})$$

*and*

$$\Gamma_3 \models (\gamma^1_{s_2}, \sigma^2_{s_2}) =_{\mathsf{Low}} (\gamma^3_{s_2}, \sigma^3_{s_2})$$

Here, we can now use Lemma 4, to exchange back to $\Gamma_1$, $\gamma^1_{s_1}$ and $\gamma^1_{s_2}$.

$$\Gamma_1 \models (\gamma^1_{s_1}, \sigma^2_{s_1}) =_{\mathsf{Low}} (\gamma^1_{s_1}, \sigma^3_{s_1})$$

and

$$\Gamma_1 \models (\gamma^1_{s_1}, \sigma^2_{s_1}) =_{\mathsf{Low}} (\gamma^1_{s_1}, \sigma^3_{s_1})$$

Now, we make an observation. The relation of $=_{\mathsf{Low}}$ with regards to a type environment $\Gamma$ is a transitive relation. This can be seen from the definition of Definition 2.2. Therefore, since we know that

$$\Gamma_1 \models (\gamma^1_{s_1}, \sigma^2_{s_1}) =_{\mathsf{Low}} (\gamma^1_{s_2}, \sigma^2_{s_2})$$

and

$$\Gamma_1 \models (\gamma_{s_1}^1, \sigma_{s_1}^2) =_{\text{Low}} (\gamma_{s_1}^1, \sigma_{s_1}^3)$$

and

$$\Gamma_1 \models (\gamma_{s_1}^1, \sigma_{s_1}^2) =_{\text{Low}} (\gamma_{s_1}^1, \sigma_{s_1}^3)$$

We can then conclude that:

$$\Gamma_1 \models (\gamma_{s_1}^1, \sigma_{s_1}^3) =_{\text{Low}} (\gamma_{s_2}^1, \sigma_{s_2}^3)$$

Since this is the final state and type environment after the evaluation, this means that Theorem 1 holds for (S-If-Else-True) if $t_1 = \text{High}$. Since it hold for all possible typing's of $t_1$, where the conditional expression is well-typed, we know that Theorem 1 holds for (S-If-Else-True).

If (1) was concluded with

$$(\text{S-If-Else-False}) \quad \frac{\langle e_1, \gamma_{s_2}^1, \sigma_{s_2}^1 \rangle \to \langle \text{False}, \gamma_{s_2}^2, \sigma_{s_2}^2 \rangle \quad \langle e_3, \gamma_{s_2}^1, \sigma_{s_2}^2 \rangle \to \langle v_{s_2}, \gamma_{s_2}^3, \sigma_{s_2}^3 \rangle}{\langle \text{If } e_1 \text{ then } e_2 \text{ else } e_3, \gamma_{s_2}^1, \sigma_{s_2}^1 \rangle \to \langle v_{s_2}, \gamma_{s_2}^3, \sigma_{s_2}^3 \rangle}$$

Then the proof for it is also upholding Theorem 1, follows the exact same logic as for (S-If-Else-True), except that $e_2$ must be replaced with $e_3$.

## 4.3 Let-construct

The Let-construct is interesting to look at, as there is needed many different typing rules for it, as we need to secure different things when assigning to different types of variables. If (1) is found with

$$(\text{S-Let}) \quad \frac{\langle e, \gamma_{s_1}^1, \sigma_{s_1}^1 \rangle \to \langle v_{s_1}, \gamma_{s_1}^2, \sigma_{s_1}^2 \rangle}{\langle \text{Let } x = e, \gamma_{s_1}^1, \sigma_{s_1}^1 \rangle \to \langle u, \gamma_{s_1}^1[x \mapsto v_{s_1}], \sigma_{s_1}^2 \rangle}$$

Then (2) is also found with

$$(\text{S-Let}) \quad \frac{\langle e, \gamma_{s_2}^1, \sigma_{s_2}^1 \rangle \to \langle v_{s_2}, \gamma_{s_2}^2, \sigma_{s_2}^2 \rangle}{\langle \text{Let } x = e, \gamma_{s_2}^1, \sigma_{s_2}^1 \rangle \to \langle u, \gamma_{s_2}^1[x \mapsto v_{s_2}], \sigma_{s_2}^2 \rangle}$$

And both are typed with either:

$$(\text{Let-n}) \quad \frac{\Gamma, pc \vdash e : t_2@t_3 \rhd \Gamma_1}{\Gamma, pc \vdash \text{Let } x_{t_1} = e : \text{Low}@t_4 \rhd \Gamma[x \to t_1]} \quad \begin{array}{c} t_1, t_2 \in \{\text{Low}, \text{High}\} \\ t_1 \sqsupseteq t_2 \\ t_1 \sqsupseteq pc \\ t_4 = \sqcap\{t_3, t_1\} \end{array}$$

or

$$(\text{Let-Base}) \quad \frac{\Gamma, pc \vdash e : t_1@t_2 \rhd \Gamma_1}{\Gamma, pc \vdash \text{Let } x = e : \text{Low}@t_3 \rhd \Gamma[x \to t_1]} \quad \begin{array}{c} t_1 \in \{\text{Low}, \text{High}\} \\ t_1 \sqsupseteq pc \\ t_3 = \sqcap\{t_1, t_2\} \end{array}$$

or

$$(\text{Let-Base-Func}) \quad \frac{\Gamma, pc \vdash e : (t_1 \to t_2@t_3)@t_4 \rhd \Gamma_1}{\Gamma, pc \vdash \text{Let } x = e : \text{Low}@\text{Low} \rhd \Gamma[x \to (t_1 \to t_2@t_3)]} \quad \text{Low} \sqsupseteq pc$$

or

$$\text{(Let-Base-Ref)} \quad \frac{\Gamma, pc \vdash e : \text{ref } t_1@t_2 \triangleright \Gamma_1}{\Gamma, pc \vdash \text{Let } x = \ e \ : \text{Low}@t_3 \triangleright \Gamma[x \to \text{ref } t_1]} \quad \begin{array}{c} t_3 = \sqcap\{t_2, t_1\} \\ t_3 \sqsupseteq pc \\ t_1 \in \{\text{Low}, \text{High}\} \end{array}$$

For all typing rules, we can use the inductive hypothesis, since we must have that $e$ is welltyped, for the whole let-expression to be well-typed. Furthermore, we assume that $\Gamma \models (\gamma^1_{s_1}, \sigma^1_{s_1}) =_{\text{Low}} (\gamma^1_{s_2}, \sigma^1_{s_2})$.

It must also be that $\langle e, \gamma^1_{s_1}, \sigma^1_{s_1} \rangle \to \langle v_{s_1}, \gamma^2_{s_1}, \sigma^2_{s_1} \rangle$ and $\langle e, \gamma^1_{s_2}, \sigma^1_{s_2} \rangle \to \langle v_{s_2}, \gamma^2_{s_2}, \sigma^2_{s_2} \rangle$

and therefore, since we assume that $e$ upholds Theorem 1 we can say that

$\Gamma_1 \models (\gamma^2_{s_1}, \sigma^2_{s_1}) =_{\text{Low}} (\gamma^2_{s_2}, \sigma^2_{s_2})$.

We can then use Lemma 3, to exchange $\Gamma_1$ and $\gamma^2_{s_1}$ $\gamma^2_{s_2}$ with the earlier environments and initial type environment and still retain non-interference. Therefore, we have that $\Gamma \models (\gamma^1_{s_1}, \sigma^2_{s_1}) =_{\text{Low}} (\gamma^1_{s_2}, \sigma^2_{s_2})$.

We will now look at each typing-rule separately.

### 4.3.1 Let-n

The (Let-n) rule is used to force a variable to be High or Low. For (Let-n) we have two cases, either $t_1 = \text{Low}$ or $t_1 = \text{High}$. We look at each case separately.

We will begin with the case where ($t_1 = \text{Low}$). In this case, we know that $t_1 = \text{Low}$, and since $t_1 \sqsupseteq t_2$, therefore $t_2 = \text{Low}$. from Lemma 1, we then know that $v_{s_1} = v_{s_2}$. Therefore, the value stored in the Low variable is the same in both states, and therefore $\Gamma[x \to \text{Low}] \models (\gamma^1_{s_1}[x \to v_{s_1}], \sigma^2_{s_1}) =_{\text{Low}} (\gamma^1_{s_2}[x \to v_{s_2}], \sigma^2_{s_2})$.

Now we will look at when ($t_1 = \text{High}$). In this case, we know that the resulting type environment is $\Gamma[x \to \text{High}]$, meaning that it does add any bindings to low variables, it therefore still upholds non-interference. Therefore, we have that $\Gamma[x \to \text{High}] \models (\gamma^1_{s_1}[x \to v_{s_1}], \sigma^2_{s_1}) =_{\text{Low}} (\gamma^1_{s_2}[x \to v_{s_2}], \sigma^2_{s_2})$.

Therefore, Theorem 1 holds for (S-Let), when it is typed with (Let-n)

### 4.3.2 Let-Base

For (Let-Base) we have two cases, either $t_1 = \text{Low}$ or $t_1 = \text{High}$. We look at each case separately.

We will begin with the case where $t_1 = \text{Low}$. In this case, we know from Lemma 1, that $v_{s_1} = v_{s_2}$. Therefore, the value stored in the Low variable $x$ is the same in both states, and therefore $\Gamma[x \to \text{Low}] \models (\gamma^1_{s_1}[x \to v_{s_1}], \sigma^2_{s_1}) =_{\text{Low}} (\gamma^1_{s_2}[x \to v_{s_2}], \sigma^2_{s_2})$.

We will now look at the case where $t_1 = \text{High}$. In this case, we know that the resulting type environment is $\Gamma[x \to \text{High}]$, meaning that it does not add any bindings to Low variables, and we therefore still upholds non-interference. Therefore, we know that $\Gamma[x \to \mathsf{High}] \models (\gamma^1_{s_1}[x \to v_{s_1}], \sigma^2_{s_1}) =_{\mathsf{Low}} (\gamma^1_{s_2}[x \to v_{s_2}], \sigma^2_{s_2})$.

Therefore, Theorem 1 holds for (S-Let), when it is typed with (Let-Base)

### 4.3.3 Let-Base-Func

From Lemma 1, since $\Gamma, pc \vdash e : (t_1 \to t_2@t_3)t_4 \triangleright \Gamma_1$ we know that the functions will be $\sim_{t_1 \to t_2}$. This is also the requirement for function variables in the definition of $=_{\mathsf{Low}}$ (Definition 2.3). Therefore, $\Gamma[x \to t_1 \to t_2@t_3] \vdash (\gamma^1_{s_1}[x \to v_{s_1}], \sigma^2_{s_1}) =_{\mathsf{Low}} (\gamma^1_{s_2}[x \to v_{s_2}], \sigma^2_{s_2})$.

Therefore, Theorem 1 holds for (S-Let), when it is typed with (Let-Base-Func)

### 4.3.4 Let-Base-Ref

For (Let-Base-Ref) we have two cases, either $t_1 = \mathsf{Low}$ or $t_1 = \mathsf{High}$. We look at each case separately. An interesting observation is that we cannot here directly choose that a reference must be ref High, and then assign a reference of type ref Low, as we could with variables typed High or Low. This comes from the fact, that we must keep pointers to High and Low variables separate, otherwise changes to High variables might accidental change values that a variable of type ref Low also was pointing to.

We will begin with the case when $t_1 = \mathsf{Low}$. In this case, we know that $e$ is typed as ref Low, and therefore from Lemma 1, we know that the same value is stored in the locations in both states, meaning that $\sigma^2_{s_1}(v_{s_1}) = \sigma^2_{s_2}(v_{s_2})$

Therefore $\Gamma[x \to l_{\mathsf{Low}}] \models (\gamma^1_{s_1}[x \to v_{s_1}], \sigma^2_{s_1}) =_{\mathsf{Low}} (\gamma^1_{s_2}[x \to v_{s_2}], \sigma^2_{s_2})$, as it is the same value, that is stored in the location of the variable $x$.

Now we will look at the case where $t_1 = \mathsf{High}$. In this case, we know that the resulting type environment is $\Gamma[x \to \mathsf{ref\ High}]$, meaning that it does not add a binding to a variable with type ref Low, and we therefore still uphold non-interference, as we do not care about values stored in High references. Therefore, we know that $\Gamma[x \to \mathsf{High}] \models (\gamma^1_{s_1}[x \to v_{s_1}], \sigma^2_{s_1}) =_{\mathsf{Low}} (\gamma^1_{s_2}[x \to v_{s_2}], \sigma^2_{s_2})$.

Therefore, Theorem 1 holds for (S-Let), when it is typed with (Let-Base-Ref). Since Theorem 1 holds, regardless of which typing rule is used, it holds for (S-Let)

## 4.4 Application

The (App) rule is interesting to look at, as it requires that functions work similarly in different states, in order to prove that we maintain non-interference. Essentially, we need to know that if an expression can evaluate to a function, then applying that function will not change the non-interference property of the states. This is why it is required in Definition 2.3, that the functions in the states are $\sim_{t_1 \to t_2}$

Now if (1) is found with

$$\text{(S-App)} \quad \frac{\langle e_1, \gamma^1_{s_1}, \sigma^1_{s_1}\rangle \to \langle v^1_{s_1}, \gamma^2_{s_1}, \sigma^2_{s_1}\rangle \quad \langle e_2, \gamma^1_{s_1}, \sigma^2_{s_1}\rangle \to \langle v^2_{s_1}, \gamma^3_{s_1}, \sigma^3_{s_1}\rangle \quad \langle e_3, \gamma^5_{s_1}[x \mapsto v^2_{s_1}], \sigma^3_{s_1}\rangle \to \langle v^3_{s_1}, \gamma^4_{s_1}, \sigma^4_{s_1}\rangle}{\langle e_1\ e_2, \gamma^1_{s_1}, \sigma^1_{s_1}\rangle \to \langle v^3_{s_1}, \gamma^1_{s_1}, \sigma^4_{s_1}\rangle} \quad v^1_{s_1} = (e_3, x, \gamma^5_{s_1})$$

Then (2) is found with

(S-App) $$\dfrac{\langle e_1, \gamma_{s_2}^1, \sigma_{s_2}^1 \rangle \to \langle v_{s_2}^1, \gamma_{s_2}^2, \sigma_{s_2}^2 \rangle \quad \langle e_2, \gamma_{s_2}^1, \sigma_{s_2}^2 \rangle \to \langle v_{s_2}^2, \gamma_{s_2}^3, \sigma_{s_2}^3 \rangle \quad \langle e_3, \gamma_{s_2}^5[x \mapsto v_{s_2}^2], \sigma_{s_2}^3 \rangle \to \langle v_{s_2}^3, \gamma_{s_2}^4, \sigma_{s_2}^4 \rangle}{\langle e_1\ e_2, \gamma_{s_2}^1, \sigma_{s_2}^1 \rangle \to \langle v_{s_2}^3, \gamma_{s_2}^1, \sigma_{s_2}^4 \rangle} \quad v_{s_2}^1 = (e_3, x, \gamma_{s_2}^5)$$

And both are typed with:

(App) $$\dfrac{\Gamma_1, pc_1 \vdash e_2 : t_1@t_2 \triangleright \Gamma_2 \quad \Gamma_1, pc_1 \vdash e_1 : (t_1 \to t_3@t_4)@t_5 \triangleright \Gamma_3}{\Gamma_1, pc_1 \vdash e_1\ e_2 : t_3@t_6 \triangleright \Gamma_1} \quad \begin{array}{l} t_6 = \sqcap\{t_2, t_4, t_5\} \\ t_6 \sqsupseteq pc_1 \end{array}$$

Since we know that $\langle e_1, \gamma_{s_1}^1, \sigma_{s_1}^1 \rangle \to \langle v_{s_1}^1, \gamma_{s_1}^2, \sigma_{s_1}^2 \rangle$, and $\langle e_1, \gamma_{s_2}^1, \sigma_{s_2}^1 \rangle \to \langle v_{s_2}^1, \gamma_{s_2}^2, \sigma_{s_2}^2 \rangle$, we can use the inductive hypothesis, to say that:
$\Gamma_2 \models (\gamma_{s_2}^2, \sigma_{s_2}^2) =_{\mathsf{Low}} (\gamma_{s_1}^2, \sigma_{s_1}^2)$.

We can then use Lemma 3, to exit the local scope, and we then have that:
$\Gamma_1 \models (\gamma_{s_2}^1, \sigma_{s_2}^2) =_{\mathsf{Low}} (\gamma_{s_1}^1, \sigma_{s_1}^2)$.

Since we then know that $\langle e_2, \gamma_{s_1}^1, \sigma_{s_1}^2 \rangle \to \langle v_{s_1}^2, \gamma_{s_1}^3, \sigma_{s_1}^3 \rangle$, and $\langle e_2, \gamma_{s_2}^1, \sigma_{s_2}^2 \rangle \to \langle v_{s_2}^2, \gamma_{s_2}^3, \sigma_{s_2}^3 \rangle$, and their states are $=_{\mathsf{Low}}$, we can use the inductive hypothesis, to say that $\Gamma_3 \models (\gamma_{s_1}^3, \sigma_{s_1}^3) =_{\mathsf{Low}} (\gamma_{s_2}^3, \sigma_{s_2}^3)$.

We can then again use Lemma 3, to exit the local scope, and we say that $\Gamma_1 \models (\gamma_{s_2}^1, \sigma_{s_2}^3) =_{\mathsf{Low}} (\gamma_{s_1}^1, \sigma_{s_1}^3)$.

From Lemma 1, we have that since $e_1$ is typed as a $(t_1 \to t_2@t_3)$, we then know that $(v_{s_1}^1, \sigma_{s_1}^2) \sim_{t_1 \to t_2} (v_{s_2}^1, \sigma_{s_2}^2)$. Since locations always are new with regards to the store, we cannot overwrite old locations, and therefore, we can exchange $\sigma_{s_1}^2$ and $\sigma_{s_2}^2$, with the newer stores $\sigma_{s_1}^3$ and $\sigma_{s_2}^3$, and we have that $(v_{s_1}^1, \sigma_{s_1}^3) \sim_{t_1 \to t_2} (v_{s_2}^1, \sigma_{s_2}^3)$. Therefore, we know that regardless of what value the functions are applied to, as long as the values are $\sim_{t_1}$ they will maintain non-interference of the states. This we know, since the type of $e_2$ must also be $t_1$, and therefore, from Lemma 1 we know that the values are $\sim_{t_1}$ if $t_1 \in \mathbf{Func} \cup \{\mathsf{Low}, \mathsf{ref}\ \mathsf{Low}\}$. We also know that if $t_1 \in \{\mathsf{High}, \mathsf{ref}\ \mathsf{High}\}$, we do not care about the values, or the values stored at the location.

Therefore since,
$\langle e_3, \gamma_{s_1}^5[x \to v_{s_1}^2], \sigma_{s_1}^3 \rangle \to \langle v_{s_1}^3, \gamma_{s_1}^4, \sigma_{s_1}^4 \rangle$
and
$\langle e_3, \gamma_{s_2}^5[x \to v_{s_2}^2], \sigma_{s_2}^3 \rangle \to \langle v_{s_2}^3, \gamma_{s_2}^4, \sigma_{s_2}^4 \rangle$
Then we know that, for some $\Gamma'$, that was the type-environment when the functions was defined, then $\Gamma' \models (\gamma_{s_2}^4, \sigma_{s_2}^4) =_{\mathsf{Low}} (\gamma_{s_1}^4, \sigma_{s_1}^4)$.

To show how to find $\Gamma'$, we simply inspect the function typing rule. This typing rule must have been used where the function was initially defined:

(Func) $$\dfrac{\Gamma_1[x \mapsto t_1], pc \vdash e : t_2@t_3 \triangleright \Gamma_2}{\Gamma_1, pc \vdash \lambda x_{t_1}.e : (t_1 \to t_2@t_3)@() \triangleright \Gamma_1} \quad t_3 \sqsupseteq pc$$

It is the typing environment $\Gamma_1$ in the (Func)-rule, we refer to when we say "the type-environment $\Gamma'$ when the functions was defined".

The expression $e$ in the (Func) rule, is the inner expression found in a function value, as can be seen by the (S-Func) rule:

$$\text{(S-Func)} \quad \frac{}{\langle \lambda x.e, \gamma, \sigma \rangle \to \langle v, \gamma, \sigma \rangle} \quad v = (e, x, \gamma)$$

Now, we switch back to the type-environment $\Gamma_1$, and environments $\gamma_{s_1}^1$ and $\gamma_{s_2}^1$, using Lemma 3, and we get that:

$\Gamma_1 \models (\gamma_{s_1}^1, \sigma_{s_1}^4) =_{\textsf{Low}} (\gamma_{s_2}^1, \sigma_{s_2}^4)$

We are aware, that this is a little different from the usual use of Lemma 3, to exit a scope, since the environment we go back to is an earlier one, instead of a modified version of our current one. However, from inspection of our type system, we know that there cannot be removed bindings from our type environment and environments, and therefore, this exchange does not change non-interference properties. But since the final states are $=_{\textsf{Low}}$, we know that Theorem 1 holds for (S-App).

These were the most advanced rules to prove, that Theorem 1 holds for. For the proof that all of the semantic rules uphold it, we refer to Appendix C. We will now show how our type checker has actually been implemented, such that it could ensure that programs maintained non-interference.

## 5  Type Checker

During the initial phase of writing our type system, we also implemented a type checker in Haskell. While the type checker did not prove anything about our type system, we could use it to programmatically test new iterations of rules against a lot of different ReScript programs. If any of the programs were not typed correctly, we could easily change the rule and try again. Haskell was chosen as its type system is more than powerful enough to encompass our type system and most of our ReScript specification, and the type system could be translated rather easily.

We will now give an overview of the most important implementation details. However, the source code, which is freely available on GitHub [12], also includes the following:

- Example ReScript programs, written within our subset of the ReScript syntax presented in Section 2.1.
- A parser written using the Parsec library [13].
- Datatypes for the ReScript abstract syntax and our type formation rules.
- A Lattice instance using the Lattice library [14].
- The type checker.

On Listing 10, the data types for ReScript expressions, type system types, and type environment are depicted. These all come rather naturally and are almost completely identical to the formation rules. However, there are two things of note. Firstly, the type environment in the implementation is a map data structure and not a partial function. Secondly, we have tried to mimic the same notation as seen in this paper, which is why we have the *LevelTEnv* type instead of using the pair (*LevelT, Env*).

```
data Expr
    = N Int
    | B Bool
```

```
4        | Unit
5        | Var Variable
6        | App Expr Expr
7        | Abs Variable LevelT Expr
8        | Seq Expr Expr
9        | Let Variable LevelT Expr
10       | LetInf Variable Expr
11       | BO BinOper Expr Expr
12       | Rec (NE.NonEmpty (Label, Expr))
13       | Proj Expr Label
14       | IfThenElse Expr Expr Expr
15       | IfThen Expr Expr
16       | While Expr Expr
17       | For Variable Expr Expr Expr
18       | Loc Location
19       | Ref Expr
20       | Deref Variable
21       | Assign Variable Expr
22       deriving (Eq)
23
24  data LowHigh = Low | High
25  data LevelT
26      = LH LowHigh
27      | RefLH LowHigh
28      | LevelT :@ LevelT
29      | LevelT :→ LevelT
30      | Empty
31
32  type Env = M.Map Variable LevelT
33  data LevelTEnv = LevelT :|> Env
```

Listing 10: Relevant datatypes in type checker implentation

On Listing 11, a snippet of the type checker is shown. The type for the type checker is a function, which, when given a type environment, a type context, and an expression, produces a `StateEither [String] LevelTEnv`.

The result type is very closely related to the type `StateT (Either String) [String] LevelTEnv`, which is a stateful computation which can fail with a value of type `String` or produce a tuple containing a value of type `LevelTEnv` and the state of type `[String]`. The difference comes in the case of failure because `StateEither [String] LevelTEnv` will then result in a tuple containing a value of type `String` and the state of type `[String]`, thus with `StateEither [String] LevelTEnv` the state is part of the result even if the computation fails. This is very useful, because in our case the state represents the current path in the derivation tree, meaning that if the type checking fails we can show the complete path to the rule that failed and the sub-expression the rule was used on. In the function definition, we start by doing pattern matching on the expression. We have chosen only to show the definition where `expr` matches the pattern for a binary operation, but most other cases follow the same structure, and in the cases that they do not, they only differ very slightly. When we match the pattern, we call the trace function, which is a helper function, that makes sure that our state always contains the current path.

The actual type checking happens on the following lines where we start by type checking the sub-expressions `e1` and `e2`. To check the side conditions for the binary operation type rule we use another helper function, namely `sat`. `sat` does nothing in the case that the first parameter evaluates to `True`, and in the case that it evaluates to `False`, `sat` results in a monad fail. When we have type checked sub-expressions, and checked the side conditions, we can use operators from the Lattice type class to find the least upper bound of `l1` and `l2`, and the greatest lower bound of `eff1` and `eff2`.

```
check :: Env → LevelT → Expr → StateEither [String] LevelTEnv
check env pc expr = case expr of
    ...
    (BO _ e1 e2) → trace ("BO: " ++ show expr) $ do
        l1 :@ eff1 :|> _ <- check env pc e1
        l2 :@ eff2 :|> _ <- check env pc e2
        sat (l1 `elem` [LH Low, LH High])
            "NotSat: l1 `elem` [LH Low, LH High]"
        sat (l2 `elem` [LH Low, LH High])
            "NotSat: l2 `elem` [LH Low, LH High]"
        return $ (l1 \/ l2) :@ (eff1 /\ eff2) :|> env
```

Listing 11: Code snippet of type checker implementation

We believe that our type-checker implements our type system correctly, and therefore, the resulting type and side effect types should match the corresponding types in our type-system. This means, that given any program, if our type-checker returns a type, and a side-effect type, then you know that the program is non-interferent.

# 6 Conclusion

## 6.1 Project Results

In this paper, we have explored the integration of a type system, that ensures the non-interference property, within the programming language ReScript. Our approach involved the development of a type system, designed for a subset of ReScript, aimed at enforcing non-interference. A proof of soundness was used to validate that the type system ensures that every well-typed expression in our type system is non-interferent. While our type system does enhance the security of programs, it also imposes certain limitations in the way you can use certain language constructs. An example of this is the inability of if-expressions to evaluate to anything other than **Low** or **High**. As an addition, we developed a type checker in Haskell to quickly estimate the validity of new or revised typing rules as well as ensuring that the shown ReScript examples will not be typed.

## 6.2 Future Works

As the type system only works for a subset of ReScript, the obvious extension would be a type system that encompasses all of ReScript. Most notably, we are missing records, lists, arrays, destructuring, switches and "try-catch" constructs. Additionally, there is a significant number of binary and unary operations currently missing, although their inclusion should be a relatively straightforward extension. It's worth mentioning that an attempt was made to integrate records into the type system. However, this effort faced challenges in developing a satisfying solution that allowed record fields to have different types. Consequently, records were ultimately removed from the type system, highlighting an area for future research and development.

Building on this, another place that merits attention is the simplicity of our security levels. One could easily imagine scenarios where more complex security levels would be needed. Consider as an example the management of exam materials in a university. In such a scenario, students should not have access to exams before the scheduled date. Department heads should have access exclusively to their department's exams, while the university head should be able to access all exams. A very useful extension would therefore be one in which the security levels is a lattice that can be extended both laterally and vertically.

Our initial thoughts on how this may be done can be seen on Figure 7, where we have made formation rules to construct arbitrary security levels. Using these formation rules, we can make security levels for a university consisting of two departments, $Head\ of\ university \downarrow [Physics, Chemistry] \downarrow Students$. Our inference rules for the partial order might then be expressed as: $sl \vDash Label_1 \sqsubseteq Label_2$, where $sl$ represents predefined security levels, and $Label_1$ is ordered before $Label_2$.

$$
\begin{aligned}
sl \quad ::=& \quad [sl*] \quad &\text{(lateral extension)} \\
|& \quad sl \downarrow sl \quad &\text{(vertical extension)} \\
|& \quad Label \quad &\text{(element name)}
\end{aligned}
$$

Figure 7: The formation rules for arbitrary security levels

In parallel with these developments, we also made the type checker. While its primary function was to facilitate our own work, the type checker has demonstrated potential for broader applications. However, for it to be useful beyond our immediate needs, certain improvements are necessary, especially in how it presents information during the type checking of a program. For research-oriented applications, the type checker would benefit from being able to construct complete derivation trees for both well-typed and ill-typed programs, providing insights into the typing processes. This capability would be particularly valuable if the type checker were to be adapted to include new type rules emerging from ongoing research. Conversely, for ReScript users aiming to verify non-interference in their programs, the type checker should ideally identify specific points of non-compliance and offer potential solutions. Nonetheless, it is important to note that as the type checker is currently only capable of typing a subset of ReScript, its practicality for use in production environments is questionable. However, if they choose to program their application inside our sub-set of the ReScript language, we can guarantee that the application is non-interferent.

# A  Full semantics

This appendix contains all of the inference rules for the big-step semantics of ReScript.

$$(\text{S-Num}) \quad \frac{}{\langle n, \gamma, \sigma \rangle \to \langle v, \gamma, \sigma \rangle} \quad v = \mathbb{N}(n)$$

$$(\text{S-Bool}) \quad \frac{}{\langle b, \gamma, \sigma \rangle \to \langle v, \gamma, \sigma \rangle} \quad v = \mathbb{B}(b)$$

$$(\text{S-Unit}) \quad \frac{}{\langle (), \gamma, \sigma \rangle \to \langle u, \gamma, \sigma \rangle}$$

$$(\text{S-Var}) \quad \frac{}{\langle x, \gamma, \sigma \rangle \to \langle v, \gamma, \sigma \rangle} \quad \gamma(x) = v$$

$$(\text{S-Bop}) \quad \frac{\langle e_1, \gamma^1, \sigma^1 \rangle \to \langle v^1, \gamma^2, \sigma^2 \rangle \quad \langle e_2, \gamma^1, \sigma^2 \rangle \to \langle v^2, \gamma^3, \sigma^3 \rangle}{\langle e_1 \text{ bop } e_2, \gamma^1, \sigma^1 \rangle \to \langle v^3, \gamma^1, \sigma^3 \rangle} \quad v^3 = v^1 \text{ bop } v^2$$

$$(\text{S-Let}) \quad \frac{\langle e, \gamma^1, \sigma^1 \rangle \to \langle v, \gamma^2, \sigma^2 \rangle}{\langle \text{Let } x = e, \gamma^1, \sigma^1 \rangle \to \langle u, \gamma^1[x \mapsto v], \sigma^2 \rangle}$$

$$(\text{S-If-Else-True}) \quad \frac{\langle e_1, \gamma^1, \sigma^1 \rangle \to \langle \text{True}, \gamma^2, \sigma^2 \rangle \quad \langle e_2, \gamma^1, \sigma^2 \rangle \to \langle v, \gamma^3, \sigma^3 \rangle}{\langle \text{If } e_1 \text{ then } e_2 \text{ else } e_3, \gamma^1, \sigma^1 \rangle \to \langle v, \gamma^1, \sigma^3 \rangle}$$

$$(\text{S-If-Else-False}) \quad \frac{\langle e_1, \gamma^1, \sigma^1 \rangle \to \langle \text{False}, \gamma^2, \sigma^2 \rangle \quad \langle e_3, \gamma^1, \sigma^2 \rangle \to \langle v, \gamma^3, \sigma^3 \rangle}{\langle \text{If } e_1 \text{ then } e_2 \text{ else } e_3, \gamma^1, \sigma^1 \rangle \to \langle v, \gamma^1, \sigma^3 \rangle}$$

$$(\text{S-While-True}) \quad \frac{\begin{array}{c}\langle e_1, \gamma^1, \sigma^1 \rangle \to \langle \text{True}, \gamma^2, \sigma^2 \rangle \\ \langle e_2, \gamma^1, \sigma^2 \rangle \to \langle v^1, \gamma^3, \sigma^3 \rangle \\ \langle \text{While } e_1\ e_2, \gamma^1, \sigma^3 \rangle \to \langle v^2, \gamma^4, \sigma^4 \rangle\end{array}}{\langle \text{While } e_1\ e_2, \gamma^1, \sigma^1 \rangle \to \langle u, \gamma^1, \sigma^4 \rangle}$$

$$(\text{S-While-False}) \quad \frac{\langle e_1, \gamma^1, \sigma^1 \rangle \to \langle \text{False}, \gamma^2, \sigma^2 \rangle}{\langle \text{While } e_1\ e_2, \gamma^1, \sigma^1 \rangle \to \langle u, \gamma^1, \sigma^2 \rangle}$$

$$(\text{S-For-Base}) \quad \frac{\begin{array}{c}\langle e_1, \gamma^1, \sigma^1 \rangle \to \langle v^1, \gamma^2, \sigma^2 \rangle \quad \langle e_2, \gamma^1, \sigma^2 \rangle \to \langle v^2, \gamma^3, \sigma^3 \rangle \\ \langle e_3, \gamma^1[x \mapsto v^2], \sigma^3 \rangle \to \langle v^3, \gamma^4, \sigma^4 \rangle\end{array}}{\langle \text{For } x\ e_1\ e_2\ e_3, \gamma^1, \sigma^1 \rangle \to \langle u, \gamma^1, \sigma^4 \rangle} \quad v^1 = v^2$$

$$(\text{S-For-Rec}) \quad \frac{\begin{array}{c}\langle e_1, \gamma^1, \sigma^1\rangle \to \langle v^1, \gamma^2, \sigma^2\rangle \quad \langle e_2, \gamma^1, \sigma^2\rangle \to \langle v^2, \gamma^3, \sigma^3\rangle \\ \langle \text{For } x \ n \ m, \gamma^1, \sigma^3\rangle \to (u, \gamma^4, \sigma^4) \\ \langle e_3, \gamma^1[x \mapsto v^2], \sigma^4\rangle \to (v^4, \gamma^5, \sigma^5) \\ \hline \langle \text{For } x \ e_1 \ e_2 \ e_3, \gamma^1, \sigma^1\rangle \to \langle u, \gamma^1, \sigma^5\rangle\end{array}} \quad \begin{array}{l} n = \mathbb{N}^{-1}(v^1) \\ m = \mathbb{N}^{-1}(v^2 - 1) \\ v^1 < v^2 \end{array}$$

$$(\text{S-Seq}) \quad \frac{\langle e_1, \gamma^1, \sigma^1\rangle \to \langle v^1, \gamma^2, \sigma^2\rangle \quad \langle e_2, \gamma^2, \sigma^2\rangle \to \langle v^2, \gamma^3, \sigma^3\rangle}{\langle e_1; e_2, \gamma^1, \sigma^1\rangle \to \langle v^2, \gamma^3, \sigma^3\rangle}$$

$$(\text{S-Func}) \quad \frac{}{\langle \lambda x.e, \gamma, \sigma\rangle \to \langle v, \gamma, \sigma\rangle} \quad v = (e, x, \gamma)$$

$$(\text{S-App}) \quad \frac{\begin{array}{c}\langle e_1, \gamma^1, \sigma^1\rangle \to \langle v^1, \gamma^2, \sigma^2\rangle \quad \langle e_2, \gamma^1, \sigma^2\rangle \to \langle v^2, \gamma^3, \sigma^3\rangle \\ \langle e_3, \gamma^5[x \mapsto v^2], \sigma^3\rangle \to \langle v^3, \gamma^4, \sigma^4\rangle \\ \hline \langle e_1 \ e_2, \gamma^1, \sigma^1\rangle \to \langle v^3, \gamma^1, \sigma^4\rangle\end{array}} \quad v^1 = (e_3, x, \gamma^5)$$

$$(\text{S-Ref}) \quad \frac{\langle e, \gamma^1, \sigma^1\rangle \to \langle v, \gamma^2, \sigma^2\rangle}{\langle \text{Ref}(e), \gamma^1, \sigma^1\rangle \to \langle \ell, \gamma^1, \sigma^2[\ell \to v]\rangle} \quad \ell = new(\sigma^2)$$

$$(\text{S-Deref}) \quad \frac{}{\langle !x, \gamma^1, \sigma^1\rangle \to \langle v^2, \gamma^1, \sigma^1\rangle} \quad \begin{array}{l}\gamma^1(x) = \ell^1 \\ v^2 = \sigma^1(\ell^1)\end{array}$$

$$(\text{S-Reassign}) \quad \frac{\langle e, \gamma^1, \sigma^1\rangle \to \langle v^1, \gamma^2, \sigma^2\rangle}{\langle x := e, \gamma^1, \sigma^1\rangle \to \langle u, \gamma^1, \sigma^2[\ell \to v^1]\rangle} \quad \gamma^1(x) = \ell$$

# B  Full type system

$$(\text{Num}) \quad \overline{\Gamma, pc \vdash n : \mathsf{Low}@() \triangleright \Gamma}$$

$$(\text{Bool}) \quad \overline{\Gamma, pc \vdash b : \mathsf{Low}@() \triangleright \Gamma}$$

$$(\text{Unit}) \quad \overline{\Gamma, pc \vdash () : \mathsf{Low}@() \triangleright \Gamma}$$

$$(\text{Var}) \quad \overline{\Gamma, pc \vdash x : t@() \triangleright \Gamma} \quad \Gamma(x) = t$$

$$(\text{Bop}) \quad \frac{\Gamma_1, pc_1 \vdash e_1 : t_1@t_2 \triangleright \Gamma_2 \qquad \Gamma_1, pc_1 \vdash e_2 : t_3@t_4 \triangleright \Gamma_3}{\Gamma_1, pc_1 \vdash e_1\, bop\, e_2 : t_5@t_6 \triangleright \Gamma_1} \quad \begin{array}{l} t_1, t_3 \in \{\mathsf{Low}, \mathsf{High}\} \\ t_5 = \sqcup\{t_1, t_3\} \\ t_6 = \sqcap\{t_2, t_4\} \end{array}$$

$$(\text{Let-Base}) \quad \frac{\Gamma, pc \vdash e : t_1@t_2 \triangleright \Gamma_1}{\Gamma, pc \vdash \mathsf{Let}\ x = e : \mathsf{Low}@t_3 \triangleright \Gamma[x \to t_1]} \quad \begin{array}{l} t_1 \in \{\mathsf{Low}, \mathsf{High}\} \\ t_1 \sqsupseteq pc \\ t_3 = \sqcap\{t_1, t_2\} \end{array}$$

$$(\text{Let-Base-Func}) \quad \frac{\Gamma, pc \vdash e : (t_1 \to t_2@t_3)@t_4 \triangleright \Gamma_1}{\Gamma, pc \vdash \mathsf{Let}\ x = e : \mathsf{Low}@\mathsf{Low} \triangleright \Gamma[x \to (t_1 \to t_2@t_3)]} \quad \mathsf{Low} \sqsupseteq pc$$

$$(\text{Let-Base-Ref}) \quad \frac{\Gamma, pc \vdash e : \mathsf{ref}\ t_1@t_2 \triangleright \Gamma_1}{\Gamma, pc \vdash \mathsf{Let}\ x = e : \mathsf{Low}@t_3 \triangleright \Gamma[x \to \mathsf{ref}\ t_1]} \quad \begin{array}{l} t_3 = \sqcap\{t_2, t_1\} \\ t_3 \sqsupseteq pc \\ t_1 \in \{\mathsf{Low}, \mathsf{High}\} \end{array}$$

$$(\text{Let-n}) \quad \frac{\Gamma, pc \vdash e : t_2@t_3 \triangleright \Gamma_1}{\Gamma, pc \vdash \mathsf{Let}\ x_{t_1} = e : \mathsf{Low}@t_4 \triangleright \Gamma[x \to t_1]} \quad \begin{array}{l} t_1, t_2 \in \{\mathsf{Low}, \mathsf{High}\} \\ t_1 \sqsupseteq t_2 \\ t_1 \sqsupseteq pc \\ t_4 = \sqcap\{t_3, t_1\} \end{array}$$

$$(\text{If-Else}) \quad \frac{\begin{array}{l}\Gamma_1, pc_1 \vdash e_1 : t_1@t_2 \triangleright \Gamma_2 \\ \Gamma_1, pc_2 \vdash e_2 : t_3@t_4 \triangleright \Gamma_3 \\ \Gamma_1, pc_2 \vdash e_3 : t_5@t_6 \triangleright \Gamma_4\end{array}}{\Gamma_1, pc_1 \vdash \mathsf{If}\ e_1\ \mathsf{then}\ e_2\ \mathsf{else}\ e_3 : t_7@t_8 \triangleright \Gamma_1} \quad \begin{array}{l} pc_2 = \sqcup\{pc_1, t_1\} \\ t_7 = \sqcup\{t_1, t_3, t_5\} \\ t_8 = \sqcap\{t_2, t_4, t_6\} \\ t_1 \in \{\mathsf{Low}, \mathsf{High}\} \end{array}$$

$$(\text{While}) \quad \frac{\Gamma_1, pc_1 \vdash e_1 : t_1@t_2 \rhd \Gamma_2 \qquad \Gamma_1, pc_2 \vdash e_2 : t_3@t_4 \rhd \Gamma_3}{\Gamma_1, pc_1 \vdash \text{While } e_1 \ e_2 : \text{Low}@t_6 \rhd \Gamma_1} \quad \begin{array}{l} pc_2 = \sqcup\{pc_1, t_1\} \\ t_6 = \sqcap\{t_2, t_4\} \\ t_1 \in \{\text{Low}, \text{High}\} \end{array}$$

$$(\text{For}) \quad \frac{\begin{array}{c} \Gamma_1, pc_1 \vdash e_1 : t_1@t_2 \rhd \Gamma_2 \\ \Gamma_1, pc_1 \vdash e_2 : t_3@t_4 \rhd \Gamma_3 \\ \Gamma_1[x \to pc_2], pc_2 \vdash e_3 : t_5@t_6 \rhd \Gamma_4 \end{array}}{\Gamma_1, pc_1 \vdash \text{For } x \ e_1 \ e_2 \ e_3 : \text{Low}@t_8 \rhd \Gamma_1} \quad \begin{array}{l} pc_2 = \sqcup\{pc_1, t_1, t_3\} \\ t_8 = \sqcap\{t_2, t_4, t_6\} \\ t_1, t_3 \in \{\text{Low}, \text{High}\} \end{array}$$

$$(\text{Seq}) \quad \frac{\Gamma_1, pc \vdash e_1 : t_1@t_2 \rhd \Gamma_2 \qquad \Gamma_2, pc \vdash e_2 : t_3@t_4 \rhd \Gamma_3}{\Gamma_1, pc \vdash e_1; e_2 : t_3@t_5 \rhd \Gamma_3} \quad t_5 = \sqcap\{t_2, t_4\}$$

$$(\text{Func}) \quad \frac{\Gamma_1[x \mapsto t_1], pc \vdash e : t_2@t_3 \rhd \Gamma_2}{\Gamma_1, pc \vdash \lambda x_{t_1}.e : (t_1 \to t_2@t_3)@() \rhd \Gamma_1} \quad t_3 \sqsupseteq pc$$

$$(\text{App}) \quad \frac{\Gamma_1, pc_1 \vdash e_2 : t_1@t_2 \rhd \Gamma_2 \qquad \Gamma_1, pc_1 \vdash e_1 : (t_1 \to t_3@t_4)@t_5 \rhd \Gamma_3}{\Gamma_1, pc_1 \vdash e_1 \ e_2 : t_3@t_6 \rhd \Gamma_1} \quad \begin{array}{l} t_6 = \sqcap\{t_2, t_4, t_5\} \\ t_6 \sqsupseteq pc_1 \end{array}$$

$$(\text{Ref}) \quad \frac{\Gamma, pc \vdash e : t_1@t_2 \rhd \Gamma'}{\Gamma, pc \vdash \text{Ref}(e) : \text{ref } t_1@t_2 \rhd \Gamma} \quad t_1 \in \{\text{Low}, \text{High}\}$$

$$(\text{Deref}) \quad \frac{}{\Gamma, pc \vdash !x : t_1@() \rhd \Gamma} \quad \Gamma(x) = \text{ref } t_1$$

$$(\text{Reassign}) \quad \frac{\Gamma, pc \vdash e : t_1@t_2 \rhd \Gamma_1}{\Gamma, pc \vdash x := e : \text{Low}@t_4 \rhd \Gamma} \quad \begin{array}{l} \Gamma(x) = \text{ref } t_3 \\ t_3 \in \{\text{Low}, \text{High}\} \\ t_3 \sqsupseteq t_1 \\ t_3 \sqsupseteq pc \\ t_4 = \sqcap\{t_3, t_2\} \end{array}$$

# C  Soundness Proof

*Proof.* We will prove the following theorem:

**Theorem 1** (Soundness). *Let*

- $\Gamma \in \mathbf{Env}$
- $s_1, s_2 \in \mathbf{States}$
- $pc \in \{\mathsf{High}, \mathsf{Low}\}$

*Assume that*

$$\Gamma, pc \models e : t_1 @ t_2 \triangleright \Gamma'$$
$$\Gamma \models s_1 =_{\mathsf{Low}} s_2$$

*if*

$$\langle e, s_1 \rangle \to \langle v^1, s_1' \rangle \qquad (1)$$
$$\langle e, s_2 \rangle \to \langle v^1, s_2' \rangle \qquad (2)$$

*then*

$$\Gamma' \models s_1' =_{\mathsf{Low}} s_2'$$

We will do this, we will use induction on (1), and then show for each semantic rule that such a transition could be found with, Theorem Theorem 1 holds.

Furthermore, to prove it, we will need 5 lemmas; Lemma 1, Lemma 2, Lemma 5, Lemma 3, and Lemma 4. They are proved in the following Appendixes.

## Bool

If (1) is found with

(S-Bool) $\dfrac{}{\langle b, \gamma_{s_1}, \sigma_{s_1} \rangle \to \langle v_{s_1}, \gamma_{s_1}, \sigma_{s_1} \rangle} \quad v_{s_1} = \mathbb{B}(b)$

We then know that (2) is also found with

(S-Bool) $\dfrac{}{\langle b, \gamma_{s_2}, \sigma_{s_2} \rangle \to \langle v_{s_2}, \gamma_{s_2}, \sigma_{s_2} \rangle} \quad v_{s_2} = \mathbb{B}(b)$

And both are typed with:

(Bool) $\dfrac{}{\Gamma, pc \vdash b : \mathsf{Low}@() \triangleright \Gamma}$

Since there is no changes to the states, it is trivially true that for the final states, we have that $\Gamma \models (\gamma_{s_1}, \sigma_{s_1}) =_{\mathsf{Low}} (\gamma_{s_1}, \sigma_{s_1})$, and therefore, (S-Bool) upholds Theorem 1.

## Num

If (1) is found with

(S-Num) $\dfrac{}{\langle n, \gamma_{s_1}, \sigma_{s_1} \rangle \to \langle v_{s_1}, \gamma_{s_1}, \sigma_{s_1} \rangle} \quad v_{s_1} = \mathbb{N}(n)$

Then (2) is found with

(S-Num) $$\frac{}{\langle n, \gamma_{s_2}, \sigma_{s_2}\rangle \to \langle v_{s_2}, \gamma_{s_2}, \sigma_{s_2}\rangle} \quad v_{s_2} = \mathbb{N}(n)$$

And both are typed with:

(Num) $$\frac{}{\Gamma, pc \vdash n : \mathsf{Low}@() \triangleright \Gamma}$$

Since there is no changes to the states, it is trivially true that for the final states, we have that $\Gamma \models (\gamma_{s_1}, \sigma_{s_1}) =_{\mathsf{Low}} (\gamma_{s_1}, \sigma_{s_1})$, and therefore, (S-Num) upholds Theorem 1.

## Unit

If (1) is found with

(S-Unit) $$\frac{}{\langle (), \gamma_{s_1}, \sigma_{s_1}\rangle \to \langle u, \gamma_{s_1}, \sigma_{s_1}\rangle}$$

Then (2) is found with

(S-Unit) $$\frac{}{\langle (), \gamma_{s_2}, \sigma_{s_2}\rangle \to \langle u, \gamma_{s_2}, \sigma_{s_2}\rangle}$$

And both are typed with:

(Unit) $$\frac{}{\Gamma, pc \vdash () : \mathsf{Low}@() \triangleright \Gamma}$$

Since there is no changes to the states, it is trivially true that for the final states, we have that $\Gamma \models (\gamma_{s_1}, \sigma_{s_1}) =_{\mathsf{Low}} (\gamma_{s_1}, \sigma_{s_1})$, and therefore, (S-Unit) upholds Theorem 1.

## Var

If (1) is found with

(S-Var) $$\frac{}{\langle x, \gamma_{s_1}, \sigma_{s_1}\rangle \to \langle v_{s_1}, \gamma_{s_1}, \sigma_{s_1}\rangle} \quad \gamma_{s_1}(x) = v_{s_1}$$

Then (2) is found with

(S-Var) $$\frac{}{\langle x, \gamma_{s_2}, \sigma_{s_2}\rangle \to \langle v_{s_2}, \gamma_{s_2}, \sigma_{s_2}\rangle} \quad \gamma_{s_2}(x) = v_{s_2}$$

And both are typed with:

(Var) $$\frac{}{\Gamma, pc \vdash x : t@() \triangleright \Gamma} \quad \Gamma(x) = t$$

Since there is no changes to the states, it is trivially true that for the final states, we have that $\Gamma \models (\gamma_{s_1}, \sigma_{s_1}) =_{\mathsf{Low}} (\gamma_{s_1}, \sigma_{s_1})$, and therefore, (S-Var) upholds Theorem 1.

## Bop

If (1) is found with

(S-Bop) $$\frac{\langle e_1, \gamma_{s_1}^1, \sigma_{s_1}^1\rangle \to \langle v_{s_1}^1, \gamma_{s_1}^2, \sigma_{s_1}^2\rangle \quad \langle e_2, \gamma_{s_1}^1, \sigma_{s_1}^2\rangle \to \langle v_{s_1}^2, \gamma_{s_1}^3, \sigma_{s_1}^3\rangle}{\langle e_1 \text{ bop } e_2, \gamma_{s_1}^1, \sigma_{s_1}^1\rangle \to \langle v_{s_1}^3, \gamma_{s_1}^3, \sigma_{s_1}^3\rangle} \quad v_{s_1}^3 = v_{s_1}^1 \text{ bop } v_{s_1}^2$$

Then (2) is also found with

(S-Bop) $$\frac{\langle e_1, \gamma_{s_2}^1, \sigma_{s_2}^1\rangle \to \langle v_{s_2}^1, \gamma_{s_2}^2, \sigma_{s_2}^2\rangle \quad \langle e_2, \gamma_{s_2}^1, \sigma_{s_2}^2\rangle \to \langle v_{s_2}^2, \gamma_{s_2}^3, \sigma_{s_2}^3\rangle}{\langle e_1 \text{ bop } e_2, \gamma_{s_2}^1, \sigma_{s_2}^1\rangle \to \langle v_{s_2}^3, \gamma_{s_2}^3, \sigma_{s_2}^3\rangle} \quad v_{s_2}^3 = v_{s_2}^1 \text{ bop } v_{s_2}^2$$

And both are typed with:

(Bop) $$\frac{\Gamma_1, pc_1 \vdash e_1 : t_1@t_2 \triangleright \Gamma_2 \quad \Gamma_1, pc_1 \vdash e_2 : t_3@t_4 \triangleright \Gamma_3}{\Gamma_1, pc_1 \vdash e_1 \text{ bop } e_2 : t_5@t_6 \triangleright \Gamma_1} \quad \begin{array}{l} t_1, t_3 \in \{\mathsf{Low}, \mathsf{High}\} \\ t_5 = \sqcup\{t_1, t_3\} \\ t_6 = \sqcap\{t_2, t_4\} \end{array}$$

Since $\langle e_1, \gamma_{s_1}^1, \sigma_{s_1}^1\rangle \to \langle v_{s_11}, \gamma_{s_1}^2, \sigma_{s_1}^2\rangle$, and $\langle e_1, \gamma_{s_2}^1, \sigma_{s_2}^1\rangle \to \langle v_{s_21}, \gamma_{s_2}^2, \sigma_{s_2}^2\rangle$, and $\Gamma_1, pc \vdash e_1 : t_1@t_2 \triangleright \Gamma_2$, we can use the inductive hypothesis, meaning that the transitions in the

A. 6

premise uphold Theorem 1, and we then know that $\Gamma_2 \models |(\gamma^2_{s_1}, \sigma^2_{s_1}) =_{\text{Low}} (\gamma^2_{s_2}, \sigma^2_{s_2})$

From Lemma 3, we then know that $\Gamma_1 \models |(\gamma^1_{s_1}, \sigma^2_{s_1}) =_{\text{Low}} (\gamma^1_{s_2}, \sigma^2_{s_2})$. Therefore, $\langle e_2, \gamma^1_{s_1}, \sigma^2_{s_1}\rangle \to \langle v_{s_1 2}, \gamma^3_{s_1}, \sigma^3_{s_1}\rangle$, and $\langle e_2, \gamma^1_{s_2}, \sigma^2_{s_2}\rangle \to \langle v_{s_2 2}, \gamma^3_{s_2}, \sigma^3_{s_2}\rangle$, and $\Gamma_1, pc \vdash e_2 : t_3 @ t_4 \triangleright \Gamma_3$ from the inductive hypothesis, we then know that $\Gamma_3 \models \langle \gamma^3_{s_1}, \sigma^3_{s_1}\rangle =_{\text{Low}} \langle \gamma^3_{s_2}, \sigma^3_{s_2}\rangle$ and from Lemma 3 we know that $\Gamma_1 \models \langle \gamma^1_{s_1}, \sigma^3_{s_1}\rangle =_{\text{Low}} \langle \gamma^1_{s_2}, \sigma^3_{s_2}\rangle$.

Therefore, (S-Bop) upholds Theorem 1.

## Let

If (1) is found with

$$\text{(S-Let)} \quad \frac{\langle e, \gamma^1_{s_1}, \sigma^1_{s_1}\rangle \to \langle v_{s_1}, \gamma^2_{s_1}, \sigma^2_{s_1}\rangle}{\langle \text{Let } x = e, \gamma^1_{s_1}, \sigma^1_{s_1}\rangle \to \langle u, \gamma^1_{s_1}[x \mapsto v_{s_1}], \sigma^2_{s_1}\rangle}$$

Then (2) is found with

$$\text{(S-Let)} \quad \frac{\langle e, \gamma^1_{s_2}, \sigma^1_{s_2}\rangle \to \langle v_{s_2}, \gamma^2_{s_2}, \sigma^2_{s_2}\rangle}{\langle \text{Let } x = e, \gamma^1_{s_2}, \sigma^1_{s_2}\rangle \to \langle u, \gamma^1_{s_2}[x \mapsto v_{s_2}], \sigma^2_{s_2}\rangle}$$

And both are typed with either:

$$\text{(Let-Base)} \quad \frac{\Gamma, pc \vdash e : t_1 @ t_2 \triangleright \Gamma_1}{\Gamma, pc \vdash \text{Let } x = e : \text{Low} @ t_3 \triangleright \Gamma[x \to t_1]} \quad \begin{array}{c} t_1 \in \{\text{Low}, \text{High}\} \\ t_1 \sqsupseteq pc \\ t_3 = \sqcap\{t_1, t_2\} \end{array}$$

or

$$\text{(Let-n)} \quad \frac{\Gamma, pc \vdash e : t_2 @ t_3 \triangleright \Gamma_1}{\Gamma, pc \vdash \text{Let } x_{t_1} = e : \text{Low} @ t_4 \triangleright \Gamma[x \to t_1]} \quad \begin{array}{c} t_1, t_2 \in \{\text{Low}, \text{High}\} \\ t_1 \sqsupseteq t_2 \\ t_1 \sqsupseteq pc \\ t_4 = \sqcap\{t_3, t_1\} \end{array}$$

or

$$\text{(Let-Base-Func)} \quad \frac{\Gamma, pc \vdash e : (t_1 \to t_2 @ t_3) @ t_4 \triangleright \Gamma_1}{\Gamma, pc \vdash \text{Let } x = e : \text{Low} @ \text{Low} \triangleright \Gamma[x \to (t_1 \to t_2 @ t_3)]} \quad \text{Low} \sqsupseteq pc$$

or

$$\text{(Let-Base-Ref)} \quad \frac{\Gamma, pc \vdash e : \text{ref } t_1 @ t_2 \triangleright \Gamma_1}{\Gamma, pc \vdash \text{Let } x = e : \text{Low} @ t_3 \triangleright \Gamma[x \to \text{ref } t_1]} \quad \begin{array}{c} t_3 = \sqcap\{t_2, t_1\} \\ t_3 \sqsupseteq pc \\ t_1 \in \{\text{Low}, \text{High}\} \end{array}$$

For all typing rules, we know that from the inductive hypothesis, since
$\Gamma \models (\gamma^1_{s_1}, \sigma^1_{s_1}) =_{\text{Low}} (\gamma^1_{s_2}, \sigma^1_{s_2})$,
and $\langle e_1, \gamma^1_{s_1}, \sigma^1_{s_1}\rangle \to \langle v_{s_1}, \gamma^2_{s_1}, \sigma^2_{s_1}\rangle$,
and $\langle e_1, \gamma^1_{s_2}, \sigma^1_{s_2}\rangle \to \langle v_{s_2}, \gamma^2_{s_2}, \sigma^2_{s_2}\rangle$,
then $\Gamma_1 \models (\gamma^2_{s_1}, \sigma^2_{s_1}) =_{\text{Low}} (\gamma^2_{s_2}, \sigma^2_{s_2})$.

From Lemma 3, we also know then that $\Gamma \models (\gamma^1_{s_1}, \sigma^2_{s_1}) =_{\text{Low}} (\gamma^1_{s_2}, \sigma^2_{s_2})$.

We will now look at each typing-rule separately.

A. 7

**Let-Base**

For (Let-Base) we have two cases, either $t_1 = \mathsf{Low}$ or $t_1 = \mathsf{High}$. We look at each case separately.

($t_1 = \mathsf{Low}$) In this case, we know from Lemma 1, that $v_{s_1} = v_{s_2}$. Therefore, the value stored in the $\mathsf{Low}$ variable $x$ is the same in both states, and therefore $\Gamma[x \to \mathsf{Low}] \models (\gamma^1_{s_1}[x \to v_{s_1}], \sigma^2_{s_1}) =_{\mathsf{Low}} (\gamma^1_{s_2}[x \to v_{s_2}], \sigma^2_{s_2})$.

($t_1 = \mathsf{High}$) In this case, we know that the resulting type-environment is $\Gamma[x \to \mathsf{High}]$, meaning that it does not change any low variable, and we therefore still upholds non-interference. Therefore, we know that $\Gamma[x \to \mathsf{High}] \models (\gamma^1_{s_1}[x \to v_{s_1}], \sigma^2_{s_1}) =_{\mathsf{Low}} (\gamma^1_{s_2}[x \to v_{s_2}], \sigma^2_{s_2})$.

Therefore, Theorem 1 holds for (S-Let), when it is typed with (Let-Base)

**Let-Base-Func**

From Lemma 1, since $\Gamma, pc \vdash e : (t_1 \to t_2 @ t_3) t_4 \triangleright \Gamma_1$ we know that $v_{s_1} = v_{s_2}$, meaning that the function will have the same inner expression, and the same variable. Therefore, $\Gamma[x \to t_1 \to t_2 @ t_3] \vdash (\gamma^1_{s_1}[x \to v_{s_1}], \sigma^2_{s_1}) =_{\mathsf{Low}} (\gamma^1_{s_2}[x \to v_{s_2}], \sigma^2_{s_2})$.

Therefore, Theorem 1 holds for (S-Let), when it is typed with (Let-Base-Func)

**Let-n**

For (Let-Base) we have two cases, either $t_1 = \mathsf{Low}$ or $t_1 = \mathsf{High}$. We look at each case separately.

($t_1 = \mathsf{Low}$) In this case, we know that $t_1 = \mathsf{Low}$, and since $t_1 \sqsubseteq t_2$, therefore $t_2 = \mathsf{Low}$. from Lemma 1, we then know that $v_{s_1} = v_{s_2}$. Therefore, the value stored in the $\mathsf{Low}$ variable is the same in both states, and therefore $\Gamma[x \to \mathsf{Low}] \models (\gamma^1_{s_1}[x \to v_{s_1}], \sigma^2_{s_1}) =_{\mathsf{Low}} (\gamma^1_{s_2}[x \to v_{s_2}], \sigma^2_{s_2})$.

($t_1 = \mathsf{High}$) In this case, we know that the resulting type-environment is $\Gamma[x \to \mathsf{High}]$, meaning that it does not change any low variable, and we therefore still upholds non-interference. Therefore, we know that $\Gamma[x \to \mathsf{High}] \models (\gamma^1_{s_1}[x \to v_{s_1}], \sigma^2_{s_1}) =_{\mathsf{Low}} (\gamma^1_{s_2}[x \to v_{s_2}], \sigma^2_{s_2})$.

Therefore, Theorem 1 holds for (S-Let), when it is typed with (Let-n)

**Let-Base-Ref**

For (Let-Base-Ref) we have two cases, either $t_3 = \mathsf{Low}$ or $t_3 = \mathsf{High}$. We look at each case separately.

($t_3 = \mathsf{Low}$) In this case, we know from Lemma 1, that the same value is stored in the location.

Since $\sigma^2_{s_1}(v_{s_1}) = \sigma^2_{s_2}(v_{s_2})$ $\mathsf{Low}$, therefore $\Gamma[x \to l_{\mathsf{Low}}] \models (\gamma^1_{s_1}[x \to v_{s_1}], \sigma^2_{s_1}) =_{\mathsf{Low}} (\gamma^1_{s_2}[x \to v_{s_2}], \sigma^2_{s_2})$.

($t_3 = \mathsf{High}$) In this case, we know that the resulting type-environment is $\Gamma[x \to \mathsf{ref\ High}]$, meaning that it does not change any variable of type $\mathsf{ref\ Low}$, and we therefore still upholds non-interference. Therefore, we know that $\Gamma[x \to \text{ref High}] \models (\gamma_{s_1}^1[x \to v_{s_1}], \sigma_{s_1}^2) =_{\text{Low}} (\gamma_{s_2}^1[x \to v_{s_2}], \sigma_{s_2}^2)$.

Therefore, Theorem 1 holds for (S-Let), when it is typed with (Let-Base-Ref). Since Theorem 1 holds, regardless of the typing rule, it holds for (S-Let)

## If-Else

In the case of the If-construct, we cannot be sure that (1) and (2) use the same rule. This comes from the fact that $\Gamma \models s_1 =_{\text{Low}} s_2$ does not guarantee that High variables have the same value. So if, for an example, the conditional expression $e_1$ was a High variable, then it might in one state evaluate to True, and in another state evaluate to False. For an example, the transition for the expression "if $h$ then 2 else 3", where $\Gamma(h) = \text{High}$ could be derived using the following 2 derivation trees:

$$\text{(S-If-Else-True)} \; \frac{\text{(S-Var)} \; \frac{}{\langle h, \gamma_{s_1}, \sigma_{s_1}\rangle \to \langle \text{True}, \gamma_{s_1}, \sigma_{s_1}\rangle} \; \gamma_{s_1}(h) = \text{True} \quad \text{(S-Num)} \; \frac{}{\langle 2, \gamma_{s_1}, \sigma_{s_1}\rangle \to \langle v_1, \gamma_{s_1}, \sigma_{s_1}\rangle} \; \mathbb{N}(2) = v_1}{\langle \text{if } h \text{ then 2 else 3}, \gamma_{s_1}, \sigma_{s_1}\rangle \to \langle v_1, \gamma_{s_1}, \sigma_{s_1}\rangle}$$

and

$$\text{(S-If-Else-False)} \; \frac{\text{(S-Var)} \; \frac{}{\langle h, \gamma_{s_2}, \sigma_{s_2}\rangle \to \langle \text{False}, \gamma_{s_2}, \sigma_{s_2}\rangle} \; \gamma_{s_2}(h) = \text{False} \quad \text{(S-Num)} \; \frac{}{\langle 3, \gamma_{s_2}, \sigma_{s_2}\rangle \to \langle v_2, \gamma_{s_2}, \sigma_{s_2}\rangle} \; \mathbb{N}(3) = v_2}{\langle \text{if } h \text{ then 2 else 3}, \gamma_{s_1}, \sigma_{s_1}\rangle \to \langle v_2, \gamma_{s_1}, \sigma_{s_1}\rangle}$$

Therefore, we need to show that regardless of which of the two rules are chosen for (2), non-interference still holds.

Therefore, (1) is concluded with:

$$\text{(S-If-Else-True)} \quad \frac{\langle e_1, \gamma_{s_1}^1, \sigma_{s_1}^1\rangle \to \langle \text{True}, \gamma_{s_1}^2, \sigma_{s_1}^2\rangle \quad \langle e_2, \gamma_{s_1}^1, \sigma_{s_1}^2\rangle \to \langle v_{s_1}, \gamma_{s_1}^3, \sigma_{s_1}^3\rangle}{\langle \text{If } e_1 \text{ then } e_2 \text{ else } e_3, \gamma_{s_1}^1, \sigma_{s_1}^1\rangle \to \langle v_{s_1}, \gamma_{s_1}^1, \sigma_{s_1}^3\rangle}$$

and (2) is concluded with either

$$\text{(S-If-Else-True)} \quad \frac{\langle e_1, \gamma_{s_2}^1, \sigma_{s_2}^1\rangle \to \langle \text{True}, \gamma_{s_2}^2, \sigma_{s_2}^2\rangle \quad \langle e_2, \gamma_{s_2}^1, \sigma_{s_2}^2\rangle \to \langle v_{s_2}, \gamma_{s_2}^3, \sigma_{s_2}^3\rangle}{\langle \text{If } e_1 \text{ then } e_2 \text{ else } e_3, \gamma_{s_2}^1, \sigma_{s_2}^1\rangle \to \langle v_{s_2}, \gamma_{s_2}^1, \sigma_{s_2}^3\rangle}$$

or

$$\text{(S-If-Else-False)} \quad \frac{\langle e_1, \gamma_{s_2}^1, \sigma_{s_2}^1\rangle \to \langle \text{False}, \gamma_{s_2}^2, \sigma_{s_2}^2\rangle \quad \langle e_3, \gamma_{s_2}^1, \sigma_{s_2}^2\rangle \to \langle v_{s_2}, \gamma_{s_2}^3, \sigma_{s_2}^3\rangle}{\langle \text{If } e_1 \text{ then } e_2 \text{ else } e_3, \gamma_{s_2}^1, \sigma_{s_2}^1\rangle \to \langle v_{s_2}, \gamma_{s_2}^1, \sigma_{s_2}^3\rangle}$$

However, regardless of what rule is chosen, a conditional-expression is always typed with:

$$\text{(If-Else)} \quad \frac{\begin{array}{l}\Gamma_1, pc_1 \vdash e_1 : t_1 @ t_2 \triangleright \Gamma_2 \\ \Gamma_1, pc_2 \vdash e_2 : t_3 @ t_4 \triangleright \Gamma_3 \\ \Gamma_1, pc_2 \vdash e_3 : t_5 @ t_6 \triangleright \Gamma_4\end{array}}{\Gamma_1, pc_1 \vdash \text{If } e_1 \text{ then } e_2 \text{ else } e_3 : t_7 @ t_8 \triangleright \Gamma_1} \quad \begin{array}{l}pc_2 = \sqcup\{pc_1, t_1\} \\ t_7 = \sqcup\{t_1, t_3, t_5\} \\ t_8 = \sqcap\{t_2, t_4, t_6\} \\ t_1 \in \{\text{Low}, \text{High}\}\end{array}$$

Now, from the fact that $t_1 \in \{\mathsf{Low}, \mathsf{High}\}$ we see that there are two cases, that we need to handle separately, namely that $t_1 = \mathsf{Low}$ and $t_1 = \mathsf{High}$.

We will begin with the case where $t_1 = \mathsf{Low}$. We will now use Lemma 1, which states that since $\Gamma_1, pc \vdash e_1 : \mathsf{Low}@t_2 \triangleright \Gamma_2$, and

$$\langle e_1, \gamma^1_{s_1}, \sigma^1_{s_1}\rangle \to \langle v_{s_1}, \gamma^2_{s_1}, \sigma^2_{s_1}\rangle$$

and

$$\langle e_1, \gamma^1_{s_2}, \sigma^1_{s_2}\rangle \to \langle v_{s_2}, \gamma^2_{s_2}, \sigma^2_{s_2}\rangle$$

then we have that $v_{s_1} = v_{s_2}$.

Therefore, the same transition rule is used, as if $v_{s_1} = v_{s_2} = \mathsf{True}$, then both (1) and (2) is concluded with (S-If-Else-True).

We will now use the inductive hypothesis, meaning that we assume the expressions in the precondition uphold Theorem 1. Then, since $e_1$ must be well-typed, for the expression "if $e_1$ then $e_2$ else $e_3$" to be well-typed, and the states it is evaluated in is $=_{\mathsf{Low}}$, we then know that $\Gamma_2 \models (\gamma^2_{s_1}, \sigma^2_{s_1}) =_{\mathsf{Low}} (\gamma^2_{s_2}, \sigma^2_{s_2})$.

We can then use lemma 4, to switch out the type environment and environments with earlier ones, so that we get $\Gamma_1 \models \langle\gamma^1_{s_1}, \sigma^2_{s_1}\rangle =_{\mathsf{Low}} \langle\gamma^1_{s_2}, \sigma^2_{s_2}\rangle$
Then, since the the states that the next expression is evaluated in is $=_{\mathsf{Low}}$, and $e_2$ and $e_3$ must also be well-typed, we have that

$$\langle e_2, \gamma^1_{s_1}, \sigma^2_{s_1}\rangle \to \langle v^1_{s_1}, \gamma^3_{s_1}, \sigma^3_{s_1}\rangle$$

and

$$\langle e_2, \gamma^1_{s_2}, \sigma^2_{s_2}\rangle \to \langle v^2_{s_2}, \gamma^3_{s_2}, \sigma^3_{s_2}\rangle$$

We then from the inductive hypothesis know, that $\Gamma_3 \models (\gamma^3_{s_1}, \sigma^3_{s_1}) =_{\mathsf{Low}} (\gamma^3_{s_2}, \sigma^3_{s_2})$

We can then use lemma 4, to once again, exit the local scope, and go back to $\Gamma_1$, $\gamma^1_1$ and $\gamma^1_2$, so that we have $\Gamma_1 \models \langle\gamma^1_{s_1}, \sigma^3_{s_1}\rangle =_{\mathsf{Low}} \langle\gamma^1_{s_2}, \sigma^3_{s_2}\rangle$.

As this is the final states and type environment, after the evaluation of an conditional-expression, this means that Theorem 1 holds for (S-If-Else-True) if $t_1 = \mathsf{Low}$. We will now look at the case where $t_1 = \mathsf{High}$

Since $t_1 = \mathsf{High}$, we then know that since $pc_2 = \sqcup\{pc_1, \mathsf{High}\}$, that $pc_2 = \mathsf{High}$. This can be seen by inspection of our inference rules found in Section 3.2, and the fact that $pc_2 \in \{\mathsf{Low}, \mathsf{High}\}$ by definition of our type system. We therefore have that:

$$\Gamma_1, \mathsf{High} \vdash e_2 : t_3@t_4 \triangleright \Gamma_2$$

and

$$\Gamma_1, \mathsf{High} \vdash e_3 : t_5@t_6 \triangleright \Gamma_2$$

A. 10

We can then use Lemma 2, to say that the state before evaluation of $e_2$ or $e_3$ in (2), will be low-equivalent to the new state after evaluation, meaning that

$$\Gamma_3 \models (\gamma_{s_1}^1, \sigma_{s_1}^2) =_{\mathsf{Low}} (\gamma_{s_1}^3, \sigma_{s_1}^3)$$

*and*

$$\Gamma_3 \models (\gamma_{s_2}^1, \sigma_{s_2}^2) =_{\mathsf{Low}} (\gamma_{s_2}^3, \sigma_{s_2}^3)$$

Here, we can now use Lemma 4, to exchange back to $\Gamma_1$, $\gamma_{s_1}^1$ and $\gamma_{s_2}^1$.

$$\Gamma_1 \models (\gamma_{s_1}^1, \sigma_{s_1}^2) =_{\mathsf{Low}} (\gamma_{s_1}^1, \sigma_{s_1}^3)$$

and

$$\Gamma_1 \models (\gamma_{s_1}^1, \sigma_{s_1}^2) =_{\mathsf{Low}} (\gamma_{s_1}^1, \sigma_{s_1}^3)$$

Now, we make an observation. The relation of $=_{\mathsf{Low}}$ with regards to a type environment $\Gamma$ is a transitive relation. This can be seen from the definition of Definition 2.2. Therefore, since we know that

$$\Gamma_1 \models (\gamma_{s_1}^1, \sigma_{s_1}^2) =_{\mathsf{Low}} (\gamma_{s_2}^1, \sigma_{s_2}^2)$$

and

$$\Gamma_1 \models (\gamma_{s_1}^1, \sigma_{s_1}^2) =_{\mathsf{Low}} (\gamma_{s_1}^1, \sigma_{s_1}^3)$$

and

$$\Gamma_1 \models (\gamma_{s_1}^1, \sigma_{s_1}^2) =_{\mathsf{Low}} (\gamma_{s_1}^1, \sigma_{s_1}^3)$$

We can then conclude that:

$$\Gamma_1 \models (\gamma_{s_1}^1, \sigma_{s_1}^3) =_{\mathsf{Low}} (\gamma_{s_2}^1, \sigma_{s_2}^3)$$

Since this is the final state and type environment after the evaluation, this means that Theorem 1 holds for (S-If-Else-True) if $t_1 = \mathsf{High}$. Since it hold for all possible typing's of $t_1$, where the conditional expression is well-typed, we know that Theorem 1 holds for (S-If-Else-True).

If (1) was concluded with

$$(\text{S-If-Else-False}) \quad \frac{\langle e_1, \gamma_{s_2}^1, \sigma_{s_2}^1 \rangle \to \langle \mathsf{False}, \gamma_{s_2}^2, \sigma_{s_2}^2 \rangle \quad \langle e_3, \gamma_{s_2}^1, \sigma_{s_2}^2 \rangle \to \langle v_{s_2}, \gamma_{s_2}^3, \sigma_{s_2}^3 \rangle}{\langle \mathsf{If}\ e_1\ \mathsf{then}\ e_2\ \mathsf{else}\ e_3, \gamma_{s_2}^1, \sigma_{s_2}^1 \rangle \to \langle v_{s_2}, \gamma_{s_2}^1, \sigma_{s_2}^3 \rangle}$$

Then the proof for it is also upholding Theorem 1, follows the exact same logic as for (S-If-Else-True), except that $e_2$ must be replaced with $e_3$.

## While

If (1) Is concluded with

$$\text{(S-While-False)} \quad \frac{\langle e_1, \gamma_{s_1}^1, \sigma_{s_1}^1\rangle \to \langle \mathsf{False}, \gamma_{s_1}^2, \sigma_{s_1}^2\rangle}{\langle \mathsf{While}\ e_1\ e_2, \gamma_{s_1}^1, \sigma_{s_1}^1\rangle \to \langle u, \gamma_{s_1}^1, \sigma_{s_1}^2\rangle}$$

then (2) is concluded with either

$$\text{(S-While-True)} \quad \frac{\begin{array}{c}\langle e_1, \gamma_{s_2}^1, \sigma_{s_2}^1\rangle \to \langle \mathsf{True}, \gamma_{s_2}^2, \sigma_{s_2}^2\rangle \\ \langle e_2, \gamma_{s_2}^1, \sigma_{s_2}^2\rangle \to \langle v_{s_2}^1, \gamma_{s_2}^3, \sigma_{s_2}^3\rangle \\ \langle \mathsf{While}\ e_1\ e_2, \gamma_{s_2}^1, \sigma_{s_2}^3\rangle \to \langle v_{s_2}^2, \gamma_{s_2}^4, \sigma_{s_2}^4\rangle \end{array}}{\langle \mathsf{While}\ e_1\ e_2, \gamma_{s_2}^1, \sigma_{s_2}^1\rangle \to \langle u, \gamma_{s_2}^1, \sigma_{s_2}^4\rangle}$$

or

$$\text{(S-While-False)} \quad \frac{\langle e_1, \gamma_{s_2}^1, \sigma_{s_2}^1\rangle \to \langle \mathsf{False}, \gamma_{s_2}^2, \sigma_{s_2}^2\rangle}{\langle \mathsf{While}\ e_1\ e_2, \gamma_{s_2}^1, \sigma_{s_2}^1\rangle \to \langle u, \gamma_{s_2}^1, \sigma_{s_2}^2\rangle}$$

We have that the typing of the While-statement is concluded with:

$$\text{(While)} \quad \frac{\begin{array}{c}\Gamma_1, pc_1 \vdash e_1 : t_1@t_2 \rhd \Gamma_2 \\ \Gamma_1, pc_2 \vdash e_2 : t_3@t_4 \rhd \Gamma_3 \\ \hline \Gamma_1, pc_1 \vdash \mathsf{While}\ e_1\ e_2 : \mathsf{Low}@t_6 \rhd \Gamma_1\end{array}} \quad \begin{array}{c} pc_2 = \sqcup\{pc_1, t_1\} \\ t_6 = \sqcap\{t_2, t_4\} \\ t_1 \in \{\mathsf{Low}, \mathsf{High}\}\end{array}$$

We have 2 cases, one where $t_1 = \mathsf{Low}$ and one where $t_1 = \mathsf{High}$.

$t_1 = \mathsf{Low}$

From Lemma 1, we then know when
$\langle e_1, \gamma_{s_1}^1, \sigma_{s_1}^1\rangle \to \langle v_{s_1}, \gamma_{s_1}^2, \sigma_{s_1}^2\rangle$
and
$\langle e_1, \gamma_{s_2}^1, \sigma_{s_2}^1\rangle \to \langle v_{s_2}, \gamma_{s_2}^2, \sigma_{s_2}^2\rangle$, then $v_{s_1} = v_{s_2}$.

Therefore, the same rule is chosen for (2). From the inductive hypothesis, we then know that since

$\Gamma_1 \models (\gamma_{s_1}^1, \sigma_{s_1}^1) =_{\mathsf{Low}} (\gamma_{s_2}^1, \sigma_{s_2}^1)$
and $\langle e_1, \gamma_{s_1}^1, \sigma_{s_1}^1\rangle \to \langle v_{s_1}, \gamma_{s_1}^2, \sigma_{s_1}^2\rangle$,
and $\langle e_1, \gamma_{s_2}^1, \sigma_{s_2}^1\rangle \to \langle v_{s_2}, \gamma_{s_2}^2, \sigma_{s_2}^2\rangle$,
Then

$\Gamma_2 \models (\gamma_{s_1}^2, \sigma_{s_1}^2) =_{\text{Low}} (\gamma_{s_2}^2, \sigma_{s_2}^2)$
From Lemma 3, we then know that
$\Gamma_1 \models (\gamma_{s_1}^1, \sigma_{s_1}^2) =_{\text{Low}} (\gamma_{s_2}^1, \sigma_{s_2}^2)$
We are now done, as we have already showed that
$\Gamma_1 \models \langle \gamma_{s_1}^1, \sigma_{s_1}^2 \rangle =_{\text{Low}} \langle \gamma_{s_2}^1, \sigma_{s_2}^2 \rangle$, which is the final states of (S-While-False) Therefore, Theorem 1 holds, if (1) is concluded with (S-While-False), and $t_1 = \text{Low}$.

As a short comment, if (1) instead is concluded with

$$\text{(S-While-True)} \quad \frac{\begin{array}{c} \langle e_1, \gamma_{s_1}^1, \sigma_{s_1}^1 \rangle \to \langle \text{True}, \gamma_{s_1}^2, \sigma_{s_1}^2 \rangle \\ \langle e_2, \gamma_{s_1}^1, \sigma_{s_1}^2 \rangle \to \langle v_{s_1}^1, \gamma_{s_1}^3, \sigma_{s_1}^3 \rangle \\ \langle \text{While } e_1 \ e_2, \gamma_{s_1}^1, \sigma_{s_1}^3 \rangle \to \langle v_{s_1}^2, \gamma_{s_1}^4, \sigma_{s_1}^4 \rangle \end{array}}{\langle \text{While } e_1 \ e_2, \gamma_{s_1}^1, \sigma_{s_1}^1 \rangle \to \langle u, \gamma_{s_1}^1, \sigma_{s_1}^4 \rangle}$$

we then know from Lemma 1, that (2) is concluded with

$$\text{(S-While-True)} \quad \frac{\begin{array}{c} \langle e_1, \gamma_{s_2}^1, \sigma_{s_2}^1 \rangle \to \langle \text{True}, \gamma_{s_2}^2, \sigma_{s_2}^2 \rangle \\ \langle e_2, \gamma_{s_2}^1, \sigma_{s_2}^2 \rangle \to \langle v_{s_2}^1, \gamma_{s_2}^3, \sigma_{s_2}^3 \rangle \\ \langle \text{While } e_1 \ e_2, \gamma_{s_2}^1, \sigma_{s_2}^3 \rangle \to \langle v_{s_2}^2, \gamma_{s_2}^4, \sigma_{s_2}^4 \rangle \end{array}}{\langle \text{While } e_1 \ e_2, \gamma_{s_2}^1, \sigma_{s_2}^1 \rangle \to \langle u, \gamma_{s_2}^1, \sigma_{s_2}^4 \rangle}$$

Since $v_{s_1} = v_{s_2} = \text{True}$. In this case, we need two more steps, to show that Theorem 1 holds, if (1) is concluded with (S-While-True), and $t_1 = \text{Low}$.

Since $\langle e_2, \gamma_{s_1}^1, \sigma_{s_1}^2 \rangle \to \langle v_{s_1}^1, \gamma_{s_1}^3, \sigma_{s_1}^3 \rangle$ and $\langle e_2, \gamma_{s_2}^1, \sigma_{s_2}^2 \rangle \to \langle v_{s_2}^1, \gamma_{s_2}^3, \sigma_{s_2}^3 \rangle$ Then, from the inductive hypothesis of non-interference, we have that:
$\Gamma_3 \models \langle \gamma_{s_1}^3, \sigma_{s_1}^3 \rangle =_{\text{Low}} \langle \gamma_{s_2}^3, \sigma_{s_2}^3 \rangle$
Then, using Lemma 3, we get that:
$\Gamma_1 \models (\gamma_{s_1}^1, \sigma_{s_1}^3) =_{\text{Low}} (\gamma_{s_2}^1, \sigma_{s_2}^3)$
Therefore
$\langle \text{While } e_1 \ e_2, \gamma_{s_1}^1, \sigma_{s_1}^3 \rangle \to \langle v_{s_1}^2, \gamma_{s_1}^4, \sigma_{s_1}^4 \rangle$ and $\langle \text{While } e_1 \ e_2, \gamma_{s_2}^1, \sigma_{s_2}^3 \rangle \to \langle v_{s_2}^2, \gamma_{s_2}^4, \sigma_{s_2}^4 \rangle$ And then, From the inductive hypothesis of non-interference, we have that
$\Gamma_4 \models (\gamma_{s_1}^4, \sigma_{s_1}^4) =_{\text{Low}} (\gamma_{s_2}^4, \sigma_{s_2}^4)$
From Lemma 3, we then have that:
$\Gamma_1 \models \langle \gamma_{s_1}^1, \sigma_{s_1}^4 \rangle =_{\text{Low}} \langle \gamma_{s_2}^1, \sigma_{s_2}^4 \rangle$

Therefore, Theorem 1 holds, if (1) is concluded with (S-While-True), and $t_1 = \text{Low}$.

$t_1 = \text{High}$

We go back to looking at (1) being typed with (S-While-False). From the typing rules, we know that $pc_2 = \sqcup\{t_1, pc_1\}$, meaning that $pc_2 \in \{(), \text{High}\}$. Therefore, if (2) is concluded with (S-While-True), then from Lemma 2, we know that since:
$\Gamma_1 \models (\gamma_{s_1}^1, \sigma_{s_1}^2) =_{\text{Low}} (\gamma_{s_2}^1, \sigma_{s_2}^2)$,
and $\langle e_2, \gamma_{s_2}^1, \sigma_{s_2}^2 \rangle \to \langle v_{s_1}, \gamma_{s_1}^3, \sigma_{s_1}^3 \rangle$,
Then
$\Gamma_3 \models (\gamma_{s_2}^1, \sigma_{s_2}^2) =_{\text{Low}} (\gamma_{s_2}^3, \sigma_{s_2}^3)$

A. 13

From Lemma 4, we then know that
$\Gamma_1 \models (\gamma^1_{s_2}, \sigma^2_{s_2}) =_{\text{Low}} (\gamma^1_{s_2}, \sigma^3_{s_2})$

We then know that since
$\langle \text{While } e_1\ e_2, \gamma^1_{s_2}, \sigma^3_{s_2} \rangle \rightarrow \langle v_{s_1}, \gamma^4_{s_2}, \sigma^4_{s_2} \rangle$,
Then from lemma 2
$\Gamma_4 \models (\gamma^1_{s_2}, \sigma^3_{s_2}) =_{\text{Low}} (\gamma^4_{s_2}, \sigma^4_{s_2})$
From lemma Lemma 4, we then know that
$\Gamma_1 \models (\gamma^1_{s_2}, \sigma^3_{s_2}) =_{\text{Low}} (\gamma^1_{s_2}, \sigma^4_{s_2})$

However, From the transitive property of $=_{\text{Low}}$, we then know that
$\Gamma_1 \models (\gamma^1_{s_2}, \sigma^4_{s_2}) =_{\text{Low}} (\gamma^1_{s_1}, \sigma^2_{s_1})$.

If (2) is concluded using (S-While-False), both use the same rule, and it is the same prove as if $t_1 = \text{Low}$. Therefore, since regardless of what rule (2) is concluded with, and what the type $t_1$ is, we know that (S-While-False) upholds Theorem 1

In the case that (1) is concluded with (S-While-True) and $t_1 = \text{High}$, we then still do not know if (2) is concluded with (S-While-True) or (S-While-False). In the case that it is concluded with (S-While-True), then the same prove as for $t_1 = \text{High}$ could be used. In the case where (S-While-False), the proof is identical with the proof for (S-While-False), except that (1) and (2) is switched around. We therefore know, that (S-While-True) also upholds Theorem 1

## For-Base

If (1) Is concluded with

(S-For-Base) $\dfrac{\langle e_1, \gamma^1_{s_1}, \sigma^1_{s_1} \rangle \rightarrow \langle v^1_{s_1}, \gamma^2_{s_1}, \sigma^2_{s_1} \rangle \quad \langle e_2, \gamma^1_{s_1}, \sigma^2_{s_1} \rangle \rightarrow \langle v^2_{s_1}, \gamma^3_{s_1}, \sigma^3_{s_1} \rangle \\ \langle e_3, \gamma^1_{s_1}[x \mapsto v^2_{s_1}], \sigma^3_{s_1} \rangle \rightarrow \langle v^3_{s_1}, \gamma^4_{s_1}, \sigma^4_{s_1} \rangle}{\langle \text{For } x\ e_1\ e_2\ e_3, \gamma^1_{s_1}, \sigma^1_{s_1} \rangle \rightarrow \langle u, \gamma^1_{s_1}, \sigma^4_{s_1} \rangle}$ $\quad v^1_{s_1} = v^2_{s_1}$

Then (2) Is concluded with either

(S-For-Base) $\dfrac{\langle e_1, \gamma^1_{s_2}, \sigma^1_{s_2} \rangle \rightarrow \langle v^1_{s_2}, \gamma^2_{s_2}, \sigma^2_{s_2} \rangle \quad \langle e_2, \gamma^1_{s_2}, \sigma^2_{s_2} \rangle \rightarrow \langle v^2_{s_2}, \gamma^3_{s_2}, \sigma^3_{s_2} \rangle \\ \langle e_3, \gamma^1_{s_2}[x \mapsto v^2_{s_2}], \sigma^3_{s_2} \rangle \rightarrow \langle v^3_{s_2}, \gamma^4_{s_2}, \sigma^4_{s_2} \rangle}{\langle \text{For } x\ e_1\ e_2\ e_3, \gamma^1_{s_2}, \sigma^1_{s_2} \rangle \rightarrow \langle u, \gamma^1_{s_2}, \sigma^4_{s_2} \rangle}$ $\quad v^1_{s_2} = v^2_{s_2}$

or

(S-For-Rec) $\dfrac{\langle e_1, \gamma^1_{s_2}, \sigma^1_{s_2} \rangle \rightarrow \langle v^1_{s_2}, \gamma^2_{s_2}, \sigma^2_{s_2} \rangle \quad \langle e_2, \gamma^1_{s_2}, \sigma^2_{s_2} \rangle \rightarrow \langle v^2_{s_2}, \gamma^3_{s_2}, \sigma^3_{s_2} \rangle \\ \langle \text{For } x\ n\ m, \gamma^1_{s_2}, \sigma^3_{s_2} \rangle \rightarrow \langle u, \gamma^4_{s_2}, \sigma^4_{s_2} \rangle \\ \langle e_3, \gamma^1_{s_2}[x \mapsto v^2_{s_2}], \sigma^4_{s_2} \rangle \rightarrow \langle v^4_{s_2}, \gamma^5_{s_2}, \sigma^5_{s_2} \rangle}{\langle \text{For } x\ e_1\ e_2\ e_3, \gamma^1_{s_2}, \sigma^1_{s_2} \rangle \rightarrow \langle u, \gamma^1_{s_2}, \sigma^5_{s_2} \rangle}$ $\quad \begin{matrix} n = \mathbb{N}^{-1}(v^1_{s_2}) \\ m = \mathbb{N}^{-1}(v^2_{s_2} - 1) \\ v^1_{s_2} < v^2_{s_2} \end{matrix}$

however, both are typed with:

(For) $\dfrac{\Gamma_1, pc_1 \vdash e_1 : t_1 @ t_2 \triangleright \Gamma_2 \\ \Gamma_1, pc_1 \vdash e_2 : t_3 @ t_4 \triangleright \Gamma_3 \\ \Gamma_1[x \rightarrow pc_2], pc_2 \vdash e_3 : t_5 @ t_6 \triangleright \Gamma_4}{\Gamma_1, pc_1 \vdash \text{For } x\ e_1\ e_2\ e_3 : \text{Low} @ t_8 \triangleright \Gamma_1}$ $\quad \begin{matrix} pc_2 = \sqcup\{pc_1, t_1, t_3\} \\ t_8 = \sqcap\{t_2, t_4, t_6\} \\ t_1, t_3 \in \{\text{Low}, \text{High}\} \end{matrix}$

From our inductive hypothesis, we know that since:
$\Gamma_1 \models \langle \gamma^1_{s_1}, \sigma^1_{s_1} \rangle =_{\mathsf{Low}} \langle \gamma^1_{s_2}, \sigma^1_{s_2} \rangle$
and $\langle e_1, \gamma^1_{s_1}, \sigma^1_{s_1} \rangle \to \langle v^1_{s_1}, \gamma^2_{s_1}, \sigma^2_{s_1} \rangle$
and $\langle e_1, \gamma^1_{s_2}, \sigma^1_{s_2} \rangle \to \langle v^1_{s_2}, \gamma^2_{s_2}, \sigma^2_{s_2} \rangle$
and $\Gamma_1, pc \vdash e_1 : t_1 @ t_2 \triangleright \Gamma_2$
then $\Gamma_2 \models \langle \gamma^2_{s_1}, \sigma^2_{s_1} \rangle =_{\mathsf{Low}} \langle \gamma^2_{s_2}, \sigma^2_{s_2} \rangle$

From Lemma 3, we then know that $\Gamma_1 \models \langle \gamma^1_{s_1}, \sigma^2_{s_1} \rangle =_{\mathsf{Low}} \langle \gamma^1_{s_2}, \sigma^2_{s_2} \rangle$
From inductive hypothesis, we can then say that
$\Gamma_1 \models \langle \gamma^1_{s_1}, \sigma^2_{s_1} \rangle =_{\mathsf{Low}} \langle \gamma^1_{s_2}, \sigma^2_{s_2} \rangle$
and $\langle e_2, \gamma^1_{s_1}, \sigma^2_{s_1} \rangle \to \langle v^2_{s_1}, \gamma^3_{s_1}, \sigma^3_{s_1} \rangle$
and $\langle e_2, \gamma^1_{s_2}, \sigma^2_{s_2} \rangle \to \langle v^2_{s_2}, \gamma^3_{s_2}, \sigma^3_{s_2} \rangle$
and since $\Gamma_1, pc \vdash e_2 : t_3 @ t_4 \triangleright \Gamma_3$
then $\Gamma_1 \models \langle \gamma^3_{s_1}, \sigma^3_{s_1} \rangle =_{\mathsf{Low}} \langle \gamma^3_{s_2}, \sigma^3_{s_2} \rangle$

From Lemma 3, we then know that $\Gamma_1 \models \langle \gamma^1_{s_1}, \sigma^3_{s_1} \rangle =_{\mathsf{Low}} \langle \gamma^1_{s_2}, \sigma^3_{s_2} \rangle$
Now, if $pc_2 = \mathsf{Low}$, then both $e_1$ and $e_2$ are typed as $\mathsf{Low}$. Therefore, from Lemma 1, we know that $v^1_{s_1} = v^1_{s_2}$ and $v^2_{s_1} = v^2_{s_2}$. Therefore, (2) must also use S-For-Base. We then know that from the inductive hypothesis that
$\Gamma_1[x \mapsto pc_2] \models (\gamma^1_{s_1}[x \mapsto v^2_{s_1}], \sigma^3_{s_1}) =_{\mathsf{Low}} (\gamma^1_{s_2}[x \mapsto v^2_{s_2}], \sigma^3_{s_2})$
and $\langle e_3, \gamma^1_{s_1}[x \mapsto v^2_{s_1}], \sigma^3_{s_1} \rangle \to \langle v^3_{s_1}, \gamma^4_{s_1}, \sigma^4_{s_1} \rangle$
and $\langle e_3, \gamma^1_{s_2}[x \mapsto v^2_{s_2}], \sigma^3_{s_2} \rangle \to \langle v^3_{s_2}, \gamma^4_{s_2}, \sigma^4_{s_2} \rangle$
and $\Gamma_1[x \mapsto pc_2], pc \vdash e_3 : t_5 @ t_6 \triangleright \Gamma_4$
then $\Gamma_4 \models (\gamma^4_{s_1}, \sigma^4_{s_1}) =_{\mathsf{Low}} (\gamma^4_{s_2}, \sigma^4_{s_2})$

From Lemma 3, we then know that $\Gamma_1 \models \langle \gamma^1_{s_1}, \sigma^4_{s_1} \rangle =_{\mathsf{Low}} \langle \gamma^1_{s_2}, \sigma^4_{s_2} \rangle$

This means that theorem 1 holds for (S-For-Base) when $pc_2 = \mathsf{Low}$.

We will also show that it holds if $pc_2 = \mathsf{High}$ Because $pc_2 = \mathsf{High}$, then $t_1, t_3 = \mathsf{High}$. From Lemma 2, we then know that, if (2) is typed with (S-For-Rec), it will maintain non-interference, and therefore:
$\Gamma_1 \models (\gamma^1_{s_1}, \sigma^4_{s_1}) =_{\mathsf{Low}} (\gamma^1_{s_2}, \sigma^5_{s_2})$

If (2) is typed with (S-For-Base), the proof is identical to the proof when $pc_2 = \mathsf{Low}$. This means that 1 holds for (S-For-Base) when $pc_2 = \mathsf{High}$. Therefore, since Theorem 1 holds for (S-For-Base), when both $pc_2 = \mathsf{High}$ and $pc_2 = \mathsf{Low}$, we know it holds for all cases of (S-For-Base).

### For-Rec

If (1) Is concluded with

(S-For-Rec) $$\dfrac{\begin{array}{c}\langle e_1, \gamma^1_{s_1}, \sigma^1_{s_1}\rangle \to \langle v^1_{s_1}, \gamma^2_{s_1}, \sigma^2_{s_1}\rangle \quad \langle e_2, \gamma^1_{s_1}, \sigma^2_{s_1}\rangle \to \langle v^2_{s_1}, \gamma^3_{s_1}, \sigma^3_{s_1}\rangle \\ \langle \mathsf{For}\ x\ n\ m, \gamma^1_{s_1}, \sigma^3_{s_1}\rangle \to (u, \gamma^4_{s_1}, \sigma^4_{s_1}) \\ \langle e_3, \gamma^1_{s_1}[x \mapsto v^2_{s_1}], \sigma^4_{s_1}\rangle \to (v^4_{s_1}, \gamma^5_{s_1}, \sigma^5_{s_1}) \end{array}}{\langle \mathsf{For}\ x\ e_1\ e_2\ e_3, \gamma^1_{s_1}, \sigma^1_{s_1}\rangle \to \langle u, \gamma^1_{s_1}, \sigma^5_{s_1}\rangle} \quad \begin{array}{l} n = \mathbb{N}^{-1}(v^1_{s_1}) \\ m = \mathbb{N}^{-1}(v^2_{s_1} - 1) \\ v^1_{s_1} < v^2_{s_1} \end{array}$$

Then (2) Is concluded with either

(S-For-Base) $$\dfrac{\langle e_1, \gamma_{s_2}^1, \sigma_{s_2}^1 \rangle \to \langle v_{s_2}^1, \gamma_{s_2}^2, \sigma_{s_2}^2 \rangle \quad \langle e_2, \gamma_{s_2}^1, \sigma_{s_2}^2 \rangle \to \langle v_{s_2}^2, \gamma_{s_2}^3, \sigma_{s_2}^3 \rangle \\ \langle e_3, \gamma_{s_2}^1[x \mapsto v_{s_2}^2], \sigma_{s_2}^3 \rangle \to \langle v_{s_2}^3, \gamma_{s_2}^4, \sigma_{s_2}^4 \rangle}{\langle \mathsf{For}\ x\ e_1\ e_2\ e_3, \gamma_{s_2}^1, \sigma_{s_2}^1 \rangle \to \langle u, \gamma_{s_2}^1, \sigma_{s_2}^4 \rangle} \quad v_{s_2}^1 = v_{s_2}^2$$

or

(S-For-Rec) $$\dfrac{\langle e_1, \gamma_{s_2}^1, \sigma_{s_2}^1 \rangle \to \langle v_{s_2}^1, \gamma_{s_2}^2, \sigma_{s_2}^2 \rangle \quad \langle e_2, \gamma_{s_2}^1, \sigma_{s_2}^2 \rangle \to \langle v_{s_2}^2, \gamma_{s_2}^3, \sigma_{s_2}^3 \rangle \\ \langle \mathsf{For}\ x\ n\ m, \gamma_{s_2}^1, \sigma_{s_2}^3 \rangle \to (u, \gamma_{s_2}^4, \sigma_{s_2}^4) \\ \langle e_3, \gamma_{s_2}^1[x \mapsto v_{s_2}^2], \sigma_{s_2}^4 \rangle \to (v_{s_2}^4, \gamma_{s_2}^5, \sigma_{s_2}^5)}{\langle \mathsf{For}\ x\ e_1\ e_2\ e_3, \gamma_{s_2}^1, \sigma_{s_2}^1 \rangle \to \langle u, \gamma_{s_2}^1, \sigma_{s_2}^5 \rangle} \quad \begin{array}{l} n = \mathbb{N}^{-1}(v_{s_2}^1) \\ m = \mathbb{N}^{-1}(v_{s_2}^2 - 1) \\ v_{s_2}^1 < v_{s_2}^2 \end{array}$$

however, both are typed with:

(For) $$\dfrac{\Gamma_1, pc_1 \vdash e_1 : t_1 @ t_2 \rhd \Gamma_2 \\ \Gamma_1, pc_1 \vdash e_2 : t_3 @ t_4 \rhd \Gamma_3 \\ \Gamma_1[x \to pc_2], pc_2 \vdash e_3 : t_5 @ t_6 \rhd \Gamma_4}{\Gamma_1, pc_1 \vdash \mathsf{For}\ x\ e_1\ e_2\ e_3 : \mathsf{Low} @ t_8 \rhd \Gamma_1} \quad \begin{array}{l} pc_2 = \sqcup \{pc_1, t_1, t_3\} \\ t_8 = \sqcap \{t_2, t_4, t_6\} \\ t_1, t_3 \in \{\mathsf{Low}, \mathsf{High}\} \end{array}$$

From our inductive hypothesis, we know that since :
$\Gamma_1 \models \langle \gamma_{s_1}^1, \sigma_{s_1}^1 \rangle =_{\mathsf{Low}} \langle \gamma_{s_2}^1, \sigma_{s_2}^1 \rangle$
and $\langle e_1, \gamma_{s_1}^1, \sigma_{s_1}^1 \rangle \to \langle v_{s_1}^1, \gamma_{s_1}^2, \sigma_{s_1}^2 \rangle$
and $\langle e_1, \gamma_{s_2}^1, \sigma_{s_2}^1 \rangle \to \langle v_{s_2}^1, \gamma_{s_2}^2, \sigma_{s_2}^2 \rangle$
and $\Gamma_1, pc \vdash e_1 : t_1 @ t_2 \rhd \Gamma_2$
then $\Gamma_2 \models \langle \gamma_{s_1}^2, \sigma_{s_1}^2 \rangle =_{\mathsf{Low}} \langle \gamma_{s_2}^2, \sigma_{s_2}^2 \rangle$

From Lemma 3, we then know that $\Gamma_1 \models \langle \gamma_{s_1}^1, \sigma_{s_1}^2 \rangle =_{\mathsf{Low}} \langle \gamma_{s_2}^1, \sigma_{s_2}^2 \rangle$
From inductive hypothesis, we can then say that

$\Gamma_1 \models \langle \gamma_{s_1}^1, \sigma_{s_1}^2 \rangle =_{\mathsf{Low}} \langle \gamma_{s_2}^1, \sigma_{s_2}^2 \rangle$
and $\langle e_2, \gamma_{s_1}^1, \sigma_{s_1}^2 \rangle \to \langle v_{s_1}^2, \gamma_{s_1}^3, \sigma_{s_1}^3 \rangle$
and $\langle e_2, \gamma_{s_2}^1, \sigma_{s_2}^2 \rangle \to \langle v_{s_2}^2, \gamma_{s_2}^3, \sigma_{s_2}^3 \rangle$
and since $\Gamma_1, pc \vdash e_2 : t_4 @ t_5 \rhd \Gamma_3$
then $\Gamma_1 \models \langle \gamma_{s_1}^3, \sigma_{s_1}^3 \rangle =_{\mathsf{Low}} \langle \gamma_{s_2}^3, \sigma_{s_2}^3 \rangle$

From Lemma 3, we then know that $\Gamma_1 \models \langle \gamma_{s_1}^1, \sigma_{s_1}^3 \rangle =_{\mathsf{Low}} \langle \gamma_{s_2}^1, \sigma_{s_2}^3 \rangle$
Now, if $pc_2 = \mathsf{Low}$, then both $e_1$ and $e_2$ are typed as $\mathsf{Low}$. Therefore, from Lemma 1, we know that $v_{s_1}^1 = v_{s_2}^1$ and $v_{s_1}^2 = v_{s_2}^2$. Therefore, (2) must also use S-For-Base. We then know that from the inductive hypothesis that
if $\Gamma_1 \models (\gamma_{s_1}^1, \sigma_{s_1}^3) =_{\mathsf{Low}} (\gamma_{s_2}^1, \sigma_{s_2}^3)$
and $\langle \mathsf{For}\ x\ e_1\ e_2\ e_3, \gamma_{s_1}^1, \sigma_{s_1}^3 \rangle \to \langle v_{s_1}^3, \gamma_{s_1}^4, \sigma_{s_1}^4 \rangle$
and $\langle \mathsf{For}\ x\ e_1\ e_2\ e_3, \gamma_{s_2}^1, \sigma_{s_2}^3 \rangle \to \langle v_{s_2}^3, \gamma_{s_2}^4, \sigma_{s_2}^4 \rangle$
then $\Gamma_1 \models (\gamma_{s_1}^4, \sigma_{s_1}^4) =_{\mathsf{Low}} (\gamma_{s_2}^4, \sigma_{s_4}^3)$

From Lemma 3, we then know that $\Gamma_1 \models \langle \gamma_{s_1}^1, \sigma_{s_1}^4 \rangle =_{\mathsf{Low}} \langle \gamma_{s_2}^1, \sigma_{s_2}^4 \rangle$

From the inductive hypothesis, we then know that:
$\Gamma_1[x \mapsto pc_2] \models (\gamma_{s_1}^1[x \mapsto v_{s_1}^2], \sigma_{s_1}^4) =_{\text{Low}} (\gamma_{s_2}^1[x \mapsto v_{s_2}^2], \sigma_{s_2}^4)$
and $\langle e_3, \gamma_{s_1}^1[x \mapsto v_{s_1}^2], \sigma_{s_1}^4 \rangle \to \langle v_{s_1}^3, \gamma_{s_1}^4, \sigma_{s_1}^5 \rangle$
and $\langle e_3, \gamma_{s_2}^1[x \mapsto v_{s_2}^2], \sigma_{s_2}^4 \rangle \to \langle v_{s_2}^3, \gamma_{s_2}^4, \sigma_{s_2}^5 \rangle$
and $\Gamma_1[x \mapsto pc_2], pc \vdash e_3 : t_5@t_6 \triangleright \Gamma_4$
then $\Gamma_4 \models (\gamma_{s_1}^5, \sigma_{s_1}^5) =_{\text{Low}} (\gamma_{s_2}^5, \sigma_{s_2}^5)$

From Lemma 3, we then know that $\Gamma_1 \models \langle \gamma_{s_1}^1, \sigma_{s_1}^5 \rangle =_{\text{Low}} \langle \gamma_{s_2}^1, \sigma_{s_2}^5 \rangle$
Which is the final state. Therefore,

Therefore, Theorem 1 holds for (S-For-Rec), when $pc_2 = $ Low.

For the case where $pc_2 = $ High, if (2) also is concluded with (S-For-Rec), then the proof is identical to when $pc_2 = $ Low. If (2) is concluded with (S-For-Base), then the proof is identical to the proof when $pc_2 = $ High for (S-For-Base), except that (1) and (2) are swithced. Therefore, since Theorem 1 holds for (S-For-Rec), when both $pc_2 = $ High and $pc_2 = $ Low, we know it holds for all cases of (S-For-Rec).

## Function

If (1) is found with
(S-Func) $\dfrac{}{\langle \lambda x.e, \gamma_{s_1}, \sigma_{s_1} \rangle \to \langle v_{s_1}, \gamma_{s_1}, \sigma_{s_1} \rangle} \quad v_{s_1} = (e, x, \gamma_{s_1})$

Then (2) is found with
(S-Func) $\dfrac{}{\langle \lambda x.e, \gamma_{s_2}, \sigma_{s_2} \rangle \to \langle v_{s_2}, \gamma_{s_2}, \sigma_{s_2} \rangle} \quad v_{s_2} = (e, x, \gamma_{s_2})$

And both are typed with:
(Func) $\dfrac{\Gamma_1[x \mapsto t_1], pc \vdash e : t_2@t_3 \triangleright \Gamma_2}{\Gamma_1, pc \vdash \lambda x_{t_1}.e : (t_1 \to t_2@t_3)@() \triangleright \Gamma_1} \quad t_3 \sqsupseteq pc$

Since the state is not changed, it is trivially true that $\Gamma_1 \models (\gamma_{s_1}^1, \sigma_{s_1}^1) =_{\text{Low}} (\gamma_{s_2}^1, \sigma_{s_2}^1)$

## Application

The (App) rule is interesting to look as it requires that functions work similarly in different states, in order to prove that we maintain non-interference. Essentially, we need to know that if an expression can evaluate to a function, then applying that function will not change the non-interference-property of the states. This is why it is required in definition 2.3, that the functions in the states are $\sim_{t_1 \to t_2}$

Now if (1) is found with
(S-App) $\dfrac{\langle e_1, \gamma_{s_1}^1, \sigma_{s_1}^1 \rangle \to \langle v_{s_1}^1, \gamma_{s_1}^2, \sigma_{s_1}^2 \rangle \quad \langle e_2, \gamma_{s_1}^1, \sigma_{s_1}^2 \rangle \to \langle v_{s_1}^2, \gamma_{s_1}^3, \sigma_{s_1}^3 \rangle \quad \langle e_3, \gamma_{s_1}^5[x \mapsto v_{s_1}^2], \sigma_{s_1}^3 \rangle \to \langle v_{s_1}^3, \gamma_{s_1}^4, \sigma_{s_1}^4 \rangle}{\langle e_1\ e_2, \gamma_{s_1}^1, \sigma_{s_1}^1 \rangle \to \langle v_{s_1}^3, \gamma_{s_1}^1, \sigma_{s_1}^4 \rangle} \quad v_{s_1}^1 = (e_3, x, \gamma_{s_1}^5)$

Then (2) is found with
(S-App) $\dfrac{\langle e_1, \gamma_{s_2}^1, \sigma_{s_2}^1 \rangle \to \langle v_{s_2}^1, \gamma_{s_2}^2, \sigma_{s_2}^2 \rangle \quad \langle e_2, \gamma_{s_2}^1, \sigma_{s_2}^2 \rangle \to \langle v_{s_2}^2, \gamma_{s_2}^3, \sigma_{s_2}^3 \rangle \quad \langle e_3, \gamma_{s_2}^5[x \mapsto v_{s_2}^2], \sigma_{s_2}^3 \rangle \to \langle v_{s_2}^3, \gamma_{s_2}^4, \sigma_{s_2}^4 \rangle}{\langle e_1\ e_2, \gamma_{s_2}^1, \sigma_{s_2}^1 \rangle \to \langle v_{s_2}^3, \gamma_{s_2}^1, \sigma_{s_2}^4 \rangle} \quad v_{s_2}^1 = (e_3, x, \gamma_{s_2}^5)$

And both are typed with:

$$\text{(App)} \quad \frac{\begin{array}{c} \Gamma_1, pc_1 \vdash e_2 : t_1@t_2 \triangleright \Gamma_2 \\ \Gamma_1, pc_1 \vdash e_1 : (t_1 \to t_3@t_4)@t_5 \triangleright \Gamma_3 \end{array} \quad \begin{array}{c} t_6 = \sqcap\{t_2, t_4, t_5\} \\ \end{array}}{\Gamma_1, pc_1 \vdash e_1 \; e_2 : t_3@t_6 \triangleright \Gamma_1 \quad t_6 \sqsupseteq pc_1}$$

Since we know that $\langle e_1, \gamma_{s_1}^1, \sigma_{s_1}^1 \rangle \to \langle v_{s_1}^1, \gamma_{s_1}^2, \sigma_{s_1}^2 \rangle$, and $\langle e_1, \gamma_{s_2}^1, \sigma_{s_2}^1 \rangle \to \langle v_{s_2}^1, \gamma_{s_2}^2, \sigma_{s_2}^2 \rangle$, we can use the inductive hypothesis, to say that $\Gamma_2 \models (\gamma_{s_2}^2, \sigma_{s_2}^2) =_{\text{Low}} \gamma_{s_1}^2, \sigma_{s_1}^2$.

We can then use Lemma 3, to say that $\Gamma_1 \models (\gamma_{s_2}^1, \sigma_{s_2}^2) =_{\text{Low}} \gamma_{s_1}^1, \sigma_{s_1}^2$.

Since we then know that $\langle e_2, \gamma_{s_1}^1, \sigma_{s_1}^2 \rangle \to \langle v_{s_1}^2, \gamma_{s_1}^3, \sigma_{s_1}^3 \rangle$, and $\langle e_2, \gamma_{s_2}^1, \sigma_{s_2}^2 \rangle \to \langle v_{s_2}^2, \gamma_{s_2}^3, \sigma_{s_2}^3 \rangle$, we can use the inductive hypothesis, to say that $\Gamma_3 \models (\gamma_{s_1}^3, \sigma_{s_1}^3) =_{\text{Low}} \gamma_{s_2}^3, \sigma_{s_2}^3$.

We can then use Lemma 3, to say that $\Gamma_1 \models (\gamma_{s_2}^1, \sigma_{s_2}^3) =_{\text{Low}} \gamma_{s_1}^1, \sigma_{s_1}^3$.

From Lemma 1, we have that since $e_1$ is typed as a $(t_1 \to t_2@t_3)$, we then know that $(v_{s_1}^1, \sigma_{s_1}^2) \sim_{t_1 \to t_2} (v_{s_2}^1, \sigma_{s_1}^2)$. Since locations always are new with regards to the store, we cannot overwrite old locations, and therefore, we can exchange $\sigma_{s_1}^2$ and $\sigma_{s_2}^2$, with the newer stores $\sigma_{s_1}^3$ and $\sigma_{s_2}^3$, and we have that $(v_{s_1}^1, \sigma_{s_1}^3) \sim_{t_1 \to t_2} (v_{s_2}^1, \sigma_{s_1}^3)$. Therefore, we know that regardless of what value the functions are applied to, they will maintain non-interference of the states. Therefore, because of $\langle e_3, \gamma_{s_1}^5[x \to v_{s_1}^2], \sigma_{s_1}^3 \rangle \to \langle v_{s_1}^3, \gamma_{s_1}^4, \sigma_{s_1}^4 \rangle$ and $\langle e_3, \gamma_{s_2}^5[x \to v_{s_2}^2], \sigma_{s_2}^3 \rangle \to \langle v_{s_2}^3, \gamma_{s_2}^4, \sigma_{s_2}^4 \rangle$, then we know that, for some $\Gamma'$, that was the type-environment when the functions was defined, then $\Gamma' \models (\gamma_{s_2}^4, \sigma_{s_2}^4) =_{\text{Low}} \gamma_{s_1}^4, \sigma_{s_1}^4$.

To show how to find $\Gamma'$, we simply inspect the function typing-rule. This typing rule must have been used where the function was initially defined:

$$\text{(Func)} \quad \frac{\Gamma_1[x \mapsto t_1], pc \vdash e : t_2@t_3 \triangleright \Gamma_2}{\Gamma_1, pc \vdash \lambda x_{t_1}.e : (t_1 \to t_2@t_3)@() \triangleright \Gamma_1} \quad t_3 \sqsupseteq pc$$

It is the typing environment $\Gamma_1$ in the (Func)-rule, we refer to when we say "the type-environment $\Gamma'$ when the functions was defined".

The expression $e$ in the (Func) rule, is the inner expression found in a function value, as can be seen by the (S-Func) rule:

$$\text{(S-Func)} \quad \frac{}{\langle \lambda x.e, \gamma, \sigma \rangle \to \langle v, \gamma, \sigma \rangle} \quad v = (e, x, \gamma)$$

Now, we switch back to the type-environment $\Gamma_1$, and environments $\gamma_{s_1}^1$ and $\gamma_{s_2}^1$, using Lemma 3, and we get that:

$\Gamma_1 \models (\gamma_{s_1}^1, \sigma_{s_1}^4) =_{\text{Low}} (\gamma_{s_2}^1, \sigma_{s_2}^4)$.

## Seq

If (1) is found with

$$\text{(S-Seq)} \quad \frac{\langle e_1, \gamma_{s_1}^1, \sigma_{s_1}^1 \rangle \to \langle v_{s_1}^1, \gamma_{s_1}^2, \sigma_{s_1}^2 \rangle \quad \langle e_2, \gamma_{s_1}^2, \sigma_{s_1}^2 \rangle \to \langle v_{s_1}^2, \gamma_{s_1}^3, \sigma_{s_1}^3 \rangle}{\langle e_1; e_2, \gamma_{s_1}^1, \sigma_{s_1}^1 \rangle \to \langle v_{s_1}^2, \gamma_{s_1}^3, \sigma_{s_1}^3 \rangle}$$

Then (2) is found with

$$\text{(S-Seq)} \quad \frac{\langle e_1, \gamma_{s_2}^1, \sigma_{s_2}^1 \rangle \to \langle v_{s_2}^1, \gamma_{s_2}^2, \sigma_{s_2}^2 \rangle \quad \langle e_2, \gamma_{s_2}^2, \sigma_{s_2}^2 \rangle \to \langle v_{s_2}^2, \gamma_{s_2}^3, \sigma_{s_2}^3 \rangle}{\langle e_1; e_2, \gamma_{s_2}^1, \sigma_{s_2}^1 \rangle \to \langle v_{s_2}^2, \gamma_{s_2}^3, \sigma_{s_2}^3 \rangle}$$

And both are typed with:

$$\text{(Ref)} \quad \frac{\Gamma, pc \vdash e : t_1 @ t_2 \triangleright \Gamma'}{\Gamma, pc \vdash \mathsf{Ref}(e) : \mathsf{ref}\ t_1 @ t_2 \triangleright \Gamma} \quad t_1 \in \{\mathsf{Low}, \mathsf{High}\}$$

Because of $\langle e_1, \gamma_{s_1}^1, \sigma_{s_1}^1 \rangle \to \langle v_{s_1}^1, \gamma_{s_1}^2, \sigma_{s_1}^2 \rangle$
and $\langle e_1, \gamma_{s_2}^1, \sigma_{s_2}^1 \rangle \to \langle v_{s_2}^1, \gamma_{s_2}^2, \sigma_{s_2}^2 \rangle$
From the inductive hypothesis, we then know that
$\Gamma_2 \models (\gamma_{s_1}^2, \sigma_{s_1}^2) =_{\mathsf{Low}} (\gamma_{s_2}^2, \sigma_{s_2}^2)$.
and because of
$\langle e_2, \gamma_{s_1}^2, \sigma_{s_1}^2 \rangle \to \langle v_{s_1}^2, \gamma_{s_1}^3, \sigma_{s_1}^3 \rangle$
and
$\langle e_2, \gamma_{s_2}^2, \sigma_{s_2}^2 \rangle \to \langle v_{s_2}^2, \gamma_{s_2}^3, \sigma_{s_2}^3 \rangle$

From inductive hypothesis, we then know that
$\Gamma_3 \models (\gamma_{s_1}^3, \sigma_{s_1}^3) =_{\mathsf{Low}} (\gamma_{s_2}^3, \sigma_{s_2}^3)$.
Therefore, theorem Theorem 1 holds when (1) is typed with (S-Ref).

## Ref

If (1) is found with

$$\text{(S-Ref)} \quad \frac{\langle e, \gamma_{s_1}^1, \sigma_{s_1}^1 \rangle \to \langle v_{s_1}, \gamma_{s_1}^2, \sigma_{s_1}^2 \rangle}{\langle \mathsf{Ref}(e), \gamma_{s_1}^1, \sigma_{s_1}^1 \rangle \to \langle \ell_{s_1}, \gamma_{s_1}^1, \sigma_{s_1}^2[\ell_{s_1} \to v_{s_1}] \rangle} \quad \ell_{s_1} = new(\sigma_{s_1}^2)$$

Then (2) is found with

$$\text{(S-Ref)} \quad \frac{\langle e, \gamma_{s_2}^1, \sigma_{s_2}^1 \rangle \to \langle v_{s_2}, \gamma_{s_2}^2, \sigma_{s_2}^2 \rangle}{\langle \mathsf{Ref}(e), \gamma_{s_2}^1, \sigma_{s_2}^1 \rangle \to \langle \ell_{s_2}, \gamma_{s_2}^1, \sigma_{s_2}^2[\ell_{s_2} \to v_{s_2}] \rangle} \quad \ell_{s_2} = new(\sigma_{s_2}^2)$$

And both are typed with:

$$\text{(Ref)} \quad \frac{\Gamma, pc \vdash e : t_1 @ t_2 \triangleright \Gamma'}{\Gamma, pc \vdash \mathsf{Ref}(e) : \mathsf{ref}\ t_1 @ t_2 \triangleright \Gamma} \quad t_1 \in \{\mathsf{Low}, \mathsf{High}\}$$

From the inductive hypothesis, we know that since
$\langle e_1, \gamma_{s_1}^1, \sigma_{s_1}^1 \rangle \to \langle v_{s_1}^1, \gamma_{s_1}^2, \sigma_{s_1}^2 \rangle$
and
$\langle e_1, \gamma_{s_2}^1, \sigma_{s_2}^1 \rangle \to \langle v_{s_2}^1, \gamma_{s_2}^2, \sigma_{s_2}^2 \rangle$
then
$\Gamma' \models (\gamma_{s_1}^2, \sigma_{s_1}^2) =_{\mathsf{Low}} (\gamma_{s_2}^2, \sigma_{s_2}^2)$.

From Lemma 3, we then know that $\Gamma \models (\gamma_{s_1}^1, \sigma_{s_1}^2) =_{\mathsf{Low}} (\gamma_{s_2}^1, \sigma_{s_2}^2)$.

Since that the location is a new-one, with regards to the store, there cannot already be a variable referencing it. Therefore, the change to the store, does not change the non-interference properties of the states, and therefore, we have that: $\Gamma \models (\gamma_{s_1}^1, \sigma_{s_1}^2[\ell_{s_1}^1 \to v_{s_1}^1]) =_{\mathsf{Low}} (\gamma_{s_2}^1, \sigma_{s_2}^2[\ell_{s_2}^1 \to v_{s_2}^1])$

Furthermore, if $\Gamma, pc \vdash e : \mathsf{Ref}(\mathsf{Low}) @ t_2 \triangleright \Gamma'$, then we know from Lemma 1, that it is the same value that is stored. This we used when proving that theorem 1 holds for (S-Let), when typed with (Let-Base-Ref).

## Deref

If (1) is found with

(S-Deref) $\dfrac{}{\langle !x, \gamma^1_{s_1}, \sigma^1_{s_1} \rangle \to \langle v^2_{s_1}, \gamma^1_{s_1}, \sigma^1_{s_1} \rangle}$ $\quad \gamma^1_{s_1}(x) = \ell^1_{s_1}$
$v^2_{s_1} = \sigma^1_{s_1}(\ell^1_{s_1})$

Then (2) is found with

(S-Deref) $\dfrac{}{\langle !x, \gamma^1_{s_2}, \sigma^1_{s_2} \rangle \to \langle v^2_{s_2}, \gamma^1_{s_2}, \sigma^1_{s_2} \rangle}$ $\quad \gamma^1_{s_2}(x) = \ell^1_{s_2}$
$v^2_{s_2} = \sigma^1_{s_2}(\ell^1_{s_2})$

And both are typed with:

(Deref) $\dfrac{}{\Gamma, pc \vdash !x : t_1@() \triangleright \Gamma} \quad \Gamma(x) = \mathsf{ref}\ t_1$

Since there are no changes to the states, it is still true that $\Gamma \models (\gamma^1_{s_1}, \sigma^1_{s_1}) =_{\mathsf{Low}} (\gamma^1_{s_2}, \sigma^1_{s_2})$.

## Reassign

If (1) is found with

(S-Reassign) $\dfrac{\langle e, \gamma^1_{s_1}, \sigma^1_{s_1} \rangle \to \langle v^1_{s_1}, \gamma^2_{s_1}, \sigma^2_{s_1} \rangle}{\langle x := e, \gamma^1_{s_1}, \sigma^1_{s_1} \rangle \to \langle u, \gamma^1_{s_1}, \sigma^2_{s_1}[\ell_{s_1} \to v^1_{s_1}] \rangle}$ $\quad \gamma^1_{s_1}(x) = \ell_{s_1}$

then (2) is found with

(S-Reassign) $\dfrac{\langle e, \gamma^1_{s_2}, \sigma^1_{s_2} \rangle \to \langle v^1_{s_2}, \gamma^2_{s_2}, \sigma^2_{s_2} \rangle}{\langle x := e, \gamma^1_{s_2}, \sigma^1_{s_2} \rangle \to \langle u, \gamma^1_{s_2}, \sigma^2_{s_2}[\ell_{s_2} \to v^1_{s_2}] \rangle}$ $\quad \gamma^1_{s_2}(x) = \ell_{s_2}$

And both are typed with:

(Reassign) $\dfrac{\Gamma, pc \vdash e : t_1@t_2 \triangleright \Gamma_1}{\Gamma, pc \vdash x := e : \mathsf{Low}@t_4 \triangleright \Gamma}$ $\quad \begin{aligned} \Gamma(x) &= \mathsf{ref}\ t_3 \\ t_3 &\in \{\mathsf{Low}, \mathsf{High}\} \\ t_3 &\sqsupseteq t_1 \\ t_3 &\sqsupseteq pc \\ t_4 &= \sqcap\{t_3, t_2\} \end{aligned}$

From the inductive hypothesis, we know that since
$\langle e_1, \gamma^1_{s_1}, \sigma^1_{s_1} \rangle \to \langle v^1_{s_1}, \gamma^2_{s_1}, \sigma^2_{s_1} \rangle$
and
$\langle e_1, \gamma^1_{s_2}, \sigma^1_{s_2} \rangle \to \langle v^1_{s_2}, \gamma^2_{s_2}, \sigma^2_{s_2} \rangle$
then
$\Gamma' \models (\gamma^2_{s_1}, \sigma^2_{s_1}) =_{\mathsf{Low}} (\gamma^2_{s_2}, \sigma^2_{s_2})$.

From Lemma 3, we then know that $\Gamma \models (\gamma^1_{s_1}, \sigma^2_{s_1}) =_{\mathsf{Low}} (\gamma^1_{s_2}, \sigma^2_{s_2})$.

we have two cases, one where $\Gamma(x) = \mathsf{Ref}(\mathsf{Low})$, and one where $\Gamma(x) = \mathsf{ref}\ \mathsf{High}$. We will look at each case separately.

$\Gamma(x) = \mathsf{ref}\ \mathsf{Low}$

Since we have that $t_3 \sqsupseteq t_1$, then it must be that $t_1 = \mathsf{Low}$. Therefore, from Lemma 1, we know that $v_{s_1} = v_{s_2}$. Therefore, we still have that $\Gamma \models (\gamma^1_{s_1}, \sigma^2_{s_1}[\ell^1_{s_1} \to v^1_{s_2}]) =_{\mathsf{Low}} (\gamma^1_{s_2}, \sigma^2_{\sigma_2}[\ell^1_{s_1} \to v^1_{s_2}])$

A. 20

$\Gamma(x) = \mathsf{ref\ High}$

Since $\Gamma(x) = \ell_{\mathsf{High}}$, then changes to it, does not change the non-interference-property, as we only care about variables that are typed as $\mathsf{Ref(Low)}$. Therefore, it is still the case that $\Gamma \models (\gamma_{s_1}^1, \sigma_{s_1}^2[\ell_{s_1}^1 \to v_{s_2}^1]) =_{\mathsf{Low}} (\gamma_{s_2}^1, \sigma_{\sigma_2}^2[\ell_{s_1}^1 \to v_{s_2}^1])$

Since it holds in both cases, the Theorem 1 holds for (Reassign).

∎

# D  Proof of Lemma 1

*Proof.*

**Lemma 1** (Low Evaluation). *Let*

- $\Gamma, pc \vdash e : t_1 @ t_2$
- $t_1 \in \{\mathsf{Low}, \mathbf{Func}, l_{\mathsf{Low}}\}$
- $\forall s_1, s_2 \in \mathbf{States}$ *where* $s_1 =_{\mathsf{Low}} s_2$

*if*

$$\langle e, \gamma_{s_1}, \sigma_{s_1} \rangle \to \langle v_{s_1}, \gamma'_{s_1}, \sigma'_{s_1} \rangle \quad (3)$$
$$\langle e, \gamma_{s_2}, \sigma_{s_2} \rangle \to \langle v_{s_2}, \gamma'_{s_2}, \sigma'_{s_2} \rangle \quad (4)$$

*then if* $t_1 \in \{\mathsf{Low}, l_{\mathsf{Low}}\}$

$$v_{s_1} \sim_{\mathsf{Low}} v_{s_2}$$

*else if* $t_1 = (t_1 \to t_2 @ t_3)$

$$v_{s_1} \sim_{t_1 \to t_2} v_{s_2}$$

We will prove this, by using induction on the rules for the derivation of (3).

### Unit

If (3) is found with:  (S-Unit)

$$\frac{}{\langle (), \gamma_{s_1}, \sigma_{s_1} \rangle \to \langle u, \gamma_{s_1}, \sigma_{s_1} \rangle}$$

Then (4) is found with:  (S-Unit)

$$\frac{}{\langle (), \gamma_{s_2}, \sigma_{s_2} \rangle \to \langle u, \gamma_{s_2}, \sigma_{s_2} \rangle}$$

And they are both typed with:

(Unit)  $$\frac{}{\Gamma, pc \vdash () : \mathsf{Low}@() \triangleright \Gamma}$$

Since the value $u$ does not depend on the state, it can easily be seen that for both states, the same value is reached. Therefore, Lemma 1 holds for (Bool)

## Bool

If (3) is found with:    (S-Bool)

Then (4) is found with:    (S-Bool)

$$\frac{\langle b, \gamma_{s_1}, \sigma_{s_1}\rangle \to \langle v_{s_1}, \gamma_{s_1}, \sigma_{s_1}\rangle \quad v_{s_1} = \mathbb{B}(b)}{\langle b, \gamma_{s_2}, \sigma_{s_2}\rangle \to \langle v_{s_2}, \gamma_{s_2}, \sigma_{s_2}\rangle \quad v_{s_2} = \mathbb{B}(b)}$$

And they are both typed with:

(Bool)    $\dfrac{}{\Gamma, pc \vdash b : \mathsf{Low}@() \triangleright \Gamma}$

Since the value $v$ does not depend on the state, it can easily be seen that for both states, the same value is reached. Therefore, Lemma 1 holds for (Bool)

## Num

If (3) is found with:    (S-Num)

Then (4) is found with:    (S-Num)

$$\frac{\langle n, \gamma_{s_1}, \sigma_{s_1}\rangle \to \langle v_{s_1}, \gamma_{s_1}, \sigma_{s_1}\rangle \quad v_{s_1} = \mathbb{N}(n)}{\langle n, \gamma_{s_2}, \sigma_{s_2}\rangle \to \langle v_{s_2}, \gamma_{s_2}, \sigma_{s_2}\rangle \quad v_{s_2} = \mathbb{N}(n)}$$

And they are both typed with:

(Num)    $\dfrac{}{\Gamma, pc \vdash n : \mathsf{Low}@() \triangleright \Gamma}$

Since the value $v$ does not depend on the state, it can easily be seen that for both states, the same value is reached. Therefore, Lemma 1 holds for (Num)

## Var

If (3) is found with:    (S-Var)

Then (4) is found with:    (S-Var)

$$\frac{\langle x, \gamma_{s_1}, \sigma_{s_1}\rangle \to \langle v_{s_1}, \gamma_{s_1}, \sigma_{s_1}\rangle \quad \gamma_{s_1}(x) = v_{s_1}}{\langle x, \gamma_{s_2}, \sigma_{s_2}\rangle \to \langle v_{s_2}, \gamma_{s_2}, \sigma_{s_2}\rangle \quad \gamma_{s_2}(x) = v_{s_2}}$$

And they are both typed with:

(Var)    $\dfrac{\Gamma(x) = t}{\Gamma, pc \vdash x : t@() \triangleright \Gamma}$

We use the fact, that for Lemma 1 to hold, the type of $x$ must be in either **Func**, ref Low or Low, and the states before evaluation must be $=_{\mathsf{Low}}$. Since we know that if $\Gamma(x) \in \{\mathsf{ref\ Low}, \mathsf{Low}\}$, then $(v_{s_1}, \sigma_{s_1}) \sim_{\mathsf{Low}} (v_{s_2}, \sigma_{s_2})$, and if $\Gamma(x) = (t_1 \to t_2@t_3)$, then we know that $(v_{s_1}, \sigma_{s_1}) \sim_{t_1 \to t_2} (v_{s_2}, \sigma_{s_2})$. Since this is also the requirements of Lemma 1, it therefore holds for (S-Var).

## Let-Base

If (3) is found with:    (S-Let)

Then (4) is found with:    (S-Let)

$$\frac{\langle e, \gamma^1_{s_1}, \sigma^1_{s_1}\rangle \to \langle v_{s_1}, \gamma^2_{s_1}, \sigma^2_{s_1}\rangle}{\langle \mathsf{Let}\ x =\ e, \gamma^1_{s_1}, \sigma^1_{s_1}\rangle \to \langle u, \gamma^1_{s_1}[x \mapsto v_{s_1}], \sigma^2_{s_1}\rangle}$$

$$\frac{\langle e, \gamma^1_{s_2}, \sigma^1_{s_2}\rangle \to \langle v_{s_2}, \gamma^2_{s_2}, \sigma^2_{s_2}\rangle}{\langle \mathsf{Let}\ x =\ e, \gamma^1_{s_2}, \sigma^1_{s_2}\rangle \to \langle u, \gamma^1_{s_2}[x \mapsto v_{s_2}], \sigma^2_{s_2}\rangle}$$

And they are both typed with either:

(Let-Base) $$\frac{\Gamma, pc \vdash e : t_1@t_2 \triangleright \Gamma_1}{\Gamma, pc \vdash \mathsf{Let}\ x =\ e\ : \mathsf{Low}@t_3 \triangleright \Gamma[x \to t_1]} \quad \begin{array}{l} t_1 \in \{\mathsf{Low}, \mathsf{High}\} \\ t_1 \sqsupseteq pc \\ t_3 = \sqcap\{t_1, t_2\} \end{array}$$

or

(Let-Base-Func) $$\frac{\Gamma, pc \vdash e : (t_1 \to t_2@t_3)@t_4 \triangleright \Gamma_1}{\Gamma, pc \vdash \mathsf{Let}\ x =\ e\ : \mathsf{Low}@\mathsf{Low} \triangleright \Gamma[x \to (t_1 \to t_2@t_3)]} \quad \mathsf{Low} \sqsupseteq pc$$

or

(Let-n) $$\frac{\Gamma, pc \vdash e : t_2@t_3 \triangleright \Gamma_1}{\Gamma, pc \vdash \mathsf{Let}\ x_{t_1} =\ e\ : \mathsf{Low}@t_4 \triangleright \Gamma[x \to t_1]} \quad \begin{array}{l} t_1, t_2 \in \{\mathsf{Low}, \mathsf{High}\} \\ t_1 \sqsupseteq t_2 \\ t_1 \sqsupseteq pc \\ t_4 = \sqcap\{t_3, t_1\} \end{array}$$

or

(Let-Base-Ref) $$\frac{\Gamma, pc \vdash e : \mathsf{ref}\ t_1@t_2 \triangleright \Gamma_1}{\Gamma, pc \vdash \mathsf{Let}\ x =\ e\ : \mathsf{Low}@t_3 \triangleright \Gamma[x \to \mathsf{ref}\ t_1]} \quad \begin{array}{l} t_3 = \sqcap\{t_2, t_1\} \\ t_3 \sqsupseteq pc \\ t_1 \in \{\mathsf{Low}, \mathsf{High}\} \end{array}$$

Since with all 4 typing-rules, the value will become $u$ for both (3) and (4), it is also true that when $e$ is typed as Low, it will still evaluate to $u$. Therefore, Lemma 1 holds for (S-Let)

### Func

If (3) is found with:

(S-Func) $$\frac{}{\langle \lambda x.e, \gamma_{s_1}, \sigma_{s_1} \rangle \to \langle v_{s_1}, \gamma_{s_1}, \sigma_{s_1} \rangle} \quad v_{s_1} = (e, x, \gamma_{s_1})$$

Then (4) is found with:

(S-Func) $$\frac{}{\langle \lambda x.e, \gamma_{s_2}, \sigma_{s_2} \rangle \to \langle v_{s_2}, \gamma_{s_2}, \sigma_{s_2} \rangle} \quad v_{s_2} = (e, x, \gamma_{s_2})$$

And they are both typed with:

(Func) $$\frac{\Gamma_1[x \mapsto t_1], pc \vdash e : t_2@t_3 \triangleright \Gamma_2}{\Gamma_1, pc \vdash \lambda x_{t_1}.e : (t_1 \to t_2@t_3)@() \triangleright \Gamma_1} \quad t_3 \sqsupseteq pc$$

Since we assume that $\Gamma_1 \models (\gamma_{s_1}^1, \sigma_{s_1}^1) =_{\mathsf{Low}} (\gamma_{s_2}^1, \sigma_{s_2}^1)$, and we know that $e$ is well-typed with regards to $\Gamma_1[x \to t_1]$, we can, using Theorem 1 say that, for any values $v_1, v_2$ where $(v_1, \sigma_{s_1}^1) \sim_{t_1} (v_1, \sigma_{s_1}^1)$, or if $(v_1, \sigma_{s_1}^1) \sim_{t_a \to t_b} (v_2, \sigma_{s_1}^1)$:
if $\langle e, \gamma_{s_1}^1[x \to v_1]\sigma_{s_1}^1 \rangle \to \langle v_3, \gamma_{s_1}^2 \sigma_{s_1}^2 \rangle$
and
$\langle e, \gamma_{s_2}^1[x \to v_2]\sigma_{s_2}^1 \rangle \to \langle v_4, \gamma_{s_2}^2 \sigma_{s_2}^2 \rangle$
means that
$\Gamma_2 \models (\gamma_{s_1}^2, \sigma_{s_1}^2) =_{\mathsf{Low}} (\gamma_{s_2}^2, \sigma_{s_2}^2)$

From Lemma 3, we can then say that:
$\Gamma_1 \models (\gamma_{s_1}^1, \sigma_{s_1}^2) =_{\mathsf{Low}} (\gamma_{s_2}^1, \sigma_{s_2}^2)$.

At last, if $t_2 \in \mathsf{Low}$, then we know from Lemma 1, that $v_3 \sim_{\mathsf{Low}} v_4$, and if $t_2 = (t_c \to t_d@t_e)$, then $v_3 \sim_{t_c \to t_d} v_4$

Since these are the requirements for $(v_{s_1}, \sigma_{s_1}) \sim_{t_1 \to t_2} (v_{s_2}, \sigma_{s_2})$, we then know that Lemma 1 holds for (Func)

## App

If (3) is found with:

(S-App) $\dfrac{\langle e_1, \gamma_{s_1}^1, \sigma_{s_1}^1\rangle \rightarrow \langle v_{s_1}^1, \gamma_{s_1}^2, \sigma_{s_1}^2\rangle \quad \langle e_2, \gamma_{s_1}^1, \sigma_{s_1}^2\rangle \rightarrow \langle v_{s_1}^2, \gamma_{s_1}^3, \sigma_{s_1}^3\rangle \\ \langle e_3, \gamma_{s_1}^5[x \mapsto v_{s_1}^2], \sigma_{s_1}^3\rangle \rightarrow \langle v_{s_1}^3, \gamma_{s_1}^4, \sigma_{s_1}^4\rangle}{\langle e_1\ e_2, \gamma_{s_1}^1, \sigma_{s_1}^1\rangle \rightarrow \langle v_{s_1}^3, \gamma_{s_1}^1, \sigma_{s_1}^4\rangle}$ $\quad v_{s_1}^1 = (e_3, x, \gamma_{s_1}^5)$

Then (4) is found with:

(S-App) $\dfrac{\langle e_1, \gamma_{s_2}^1, \sigma_{s_2}^1\rangle \rightarrow \langle v_{s_2}^1, \gamma_{s_2}^2, \sigma_{s_2}^2\rangle \quad \langle e_2, \gamma_{s_2}^1, \sigma_{s_2}^2\rangle \rightarrow \langle v_{s_2}^2, \gamma_{s_2}^3, \sigma_{s_2}^3\rangle \\ \langle e_3, \gamma_{s_2}^5[x \mapsto v_{s_2}^2], \sigma_{s_2}^3\rangle \rightarrow \langle v_{s_2}^3, \gamma_{s_2}^4, \sigma_{s_2}^4\rangle}{\langle e_1\ e_2, \gamma_{s_2}^1, \sigma_{s_2}^1\rangle \rightarrow \langle v_{s_2}^3, \gamma_{s_2}^1, \sigma_{s_2}^4\rangle}$ $\quad v_{s_2}^1 = (e_3, x, \gamma_{s_2}^5)$

And they are both typed with:

(App) $\dfrac{\Gamma_1, pc_1 \vdash e_2 : t_1@t_2 \triangleright \Gamma_2 \\ \Gamma_1, pc_1 \vdash e_1 : (t_1 \rightarrow t_3@t_4)@t_5 \triangleright \Gamma_3 \quad t_6 = \sqcap\{t_2, t_4, t_5\}}{\Gamma_1, pc_1 \vdash e_1\ e_2 : t_3@t_6 \triangleright \Gamma_1 \quad t_6 \sqsupseteq pc_1}$

From the typing rule, we know that the type of $e_1$ must be $(t_1 \rightarrow t_3@t_4)@t_5$. Therefore, from the inductive hypothesis, we know that that $v_{s_1}^1 \sim_{t_1 \rightarrow t_3} v_{s_2}^1$.

We then know, as shown in C, that we can show that $\Gamma' \models (\gamma_{s_1}^5, \sigma_{s_1}^3) =_{\mathsf{Low}} (\gamma_{s_2}^5, \sigma_{s_2}^3)$, for some $\Gamma'$, which was the type environment when the function was defined. Now, from inductive hypothesis, we know that if $t_1 \in \{\mathsf{Low}, \mathsf{ref\ Low}\}$, then $(v_{s_1}^2, \sigma_{s_1}^3) \sim_{\mathsf{Low}} (v_{s_2}^2, \sigma_{s_2}^3)$. If $t_1 = (t_a \rightarrow t_b@t_c)$, then we also from the inductive hypothesis know that $(v_{s_1}^2, \sigma_{s_1}^3) \sim_{t_a \rightarrow t_b} (v_{s_2}^2, \sigma_{s_2}^3)$. If $t_1 \in \{\mathsf{High}, \mathsf{ref\ High}\}$, then it easily shown that $(v_{s_1}^2, \sigma_{s_1}^3) \sim_{\mathsf{High}} (v_{s_2}^2, \sigma_{s_2}^3)$, since we dont actually care about the values they have. Since all the requirements of the input values for $(v_{s_1}^2, \sigma_{s_1}^3) \sim_{t_1 \rightarrow t_3} (v_{s_2}^2, \sigma_{s_2}^3)$ is fulfilled, we then know that the for resulting values, $(v_{s_1}^3, \sigma_{s_1}^4) \sim_{t_3} (v_{s_2}^3, \sigma_{s_2}^4)$ Therefore, Lemma 1 holds for $e_3$, because regardless of the type, as the values are equivalent with regards to their type.

## Bop

If (3) is found with:

(S-Bop) $\dfrac{\langle e_1, \gamma_{s_1}^1, \sigma_{s_1}^1\rangle \rightarrow \langle v_{s_1}^1, \gamma_{s_1}^2, \sigma_{s_1}^2\rangle \quad \langle e_2, \gamma_{s_1}^1, \sigma_{s_1}^2\rangle \rightarrow \langle v_{s_1}^2, \gamma_{s_1}^3, \sigma_{s_1}^3\rangle}{\langle e_1\ \mathsf{bop}\ e_2, \gamma_{s_1}^1, \sigma_{s_1}^1\rangle \rightarrow \langle v_{s_1}^3, \gamma_{s_1}^1, \sigma_{s_1}^3\rangle}$ $\quad v_{s_1}^3 = v_{s_1}^1\ \mathsf{bop}\ v_{s_1}^2$

Then (4) is found with:

(S-Bop) $\dfrac{\langle e_1, \gamma_{s_2}^1, \sigma_{s_2}^1\rangle \rightarrow \langle v_{s_2}^1, \gamma_{s_2}^2, \sigma_{s_2}^2\rangle \quad \langle e_2, \gamma_{s_2}^1, \sigma_{s_2}^2\rangle \rightarrow \langle v_{s_2}^2, \gamma_{s_2}^3, \sigma_{s_2}^3\rangle}{\langle e_1\ \mathsf{bop}\ e_2, \gamma_{s_2}^1, \sigma_{s_2}^1\rangle \rightarrow \langle v_{s_2}^3, \gamma_{s_2}^1, \sigma_{s_2}^3\rangle}$ $\quad v_{s_2}^3 = v_{s_2}^1\ \mathsf{bop}\ v_{s_2}^2$

And they are both typed with:

(Bop) $\dfrac{\Gamma_1, pc_1 \vdash e_1 : t_1@t_2 \triangleright \Gamma_2 \\ \Gamma_1, pc_1 \vdash e_2 : t_3@t_4 \triangleright \Gamma_3}{\Gamma_1, pc_1 \vdash e_1\ bop\ e_2 : t_5@t_6 \triangleright \Gamma_1}$ $\quad \begin{array}{l} t_1, t_3 \in \{\mathsf{Low, High}\} \\ t_5 = \sqcup\{t_1, t_3\} \\ t_6 = \sqcap\{t_2, t_4\} \end{array}$

From the type rule, we know that for $\Gamma_1, pc \vdash e_1\ \mathsf{bop}\ e_2 : \mathsf{Low} \triangleright \Gamma$ then both $e_1$ and $e_2$ must be typed as $\mathsf{Low}$, as $t_1, t_3 \in \{\mathsf{Low, High}\}$, and $t_5 = \sqcap t_1, t_3$

From the inductive hypothesis, we therefore know that $v_{s_1}^1 = v_{s_2}^1$, and $v_{s_1}^2 = v_{s_2}^2$.

Therefore, $v_3 = v_{s_1}^1\ \mathsf{bop}\ v_{s_1}^2 = v_{s_2}^1\ \mathsf{bop}\ v_{s_2}^2$

Which means that Lemma Lemma 1 holds for (S-Bop)

## Seq

If (3) is found with:

(S-Seq) $\dfrac{\langle e_1, \gamma_{s_1}^1, \sigma_{s_1}^1 \rangle \to \langle v_{s_1}^1, \gamma_{s_1}^2, \sigma_{s_1}^2 \rangle \quad \langle e_2, \gamma_{s_1}^2, \sigma_{s_1}^2 \rangle \to \langle v_{s_1}^2, \gamma_{s_1}^3, \sigma_{s_1}^3 \rangle}{\langle e_1; e_2, \gamma_{s_1}^1, \sigma_{s_1}^1 \rangle \to \langle v_{s_1}^2, \gamma_{s_1}^3, \sigma_{s_1}^3 \rangle}$

Then (4) is found with:

(S-Seq) $\dfrac{\langle e_1, \gamma_{s_2}^1, \sigma_{s_2}^1 \rangle \to \langle v_{s_2}^1, \gamma_{s_2}^2, \sigma_{s_2}^2 \rangle \quad \langle e_2, \gamma_{s_2}^2, \sigma_{s_2}^2 \rangle \to \langle v_{s_2}^2, \gamma_{s_2}^3, \sigma_{s_2}^3 \rangle}{\langle e_1; e_2, \gamma_{s_2}^1, \sigma_{s_2}^1 \rangle \to \langle v_{s_2}^2, \gamma_{s_2}^3, \sigma_{s_2}^3 \rangle}$

And they are both typed with: (Seq) $\dfrac{\Gamma_1, pc \vdash e_1 : t_1@t_2 \triangleright \Gamma_2 \qquad \Gamma_2, pc \vdash e_2 : t_3@t_4 \triangleright \Gamma_3}{\Gamma_1, pc \vdash e_1; e_2 : t_3@t_5 \triangleright \Gamma_3} \quad t_5 = \sqcap\{t_2, t_4\}$

From the non-interference property, we know that $\Gamma_2 \vdash (\gamma_{s_1}^2, \sigma_{s_1}^2) =_{\mathsf{Low}} (\gamma_{s_1}^2, \sigma_{s_1}^2)$. Therefore, from the inductive hypothesis, we know that $v_{s_1}^2$ and $v_{s_2}^2$ and the stores uphold Lemma 1, meaning that the rule (S-Seq) also upholds it.

## If

If (3) are found with:

(S-If-Else-True) $\dfrac{\langle e_1, \gamma_{s_1}^1, \sigma_{s_1}^1 \rangle \to \langle \mathsf{True}, \gamma_{s_1}^2, \sigma_{s_1}^2 \rangle \quad \langle e_2, \gamma_{s_1}^1, \sigma_{s_1}^2 \rangle \to \langle v_{s_1}, \gamma_{s_1}^3, \sigma_{s_1}^3 \rangle}{\langle \mathsf{If}\ e_1\ \mathsf{then}\ e_2\ \mathsf{else}\ e_3, \gamma_{s_1}^1, \sigma_{s_1}^1 \rangle \to \langle v_{s_1}, \gamma_{s_1}^1, \sigma_{s_1}^3 \rangle}$

Then (4) are also found with:

(S-If-Else-True) $\dfrac{\langle e_1, \gamma_{s_2}^1, \sigma_{s_2}^1 \rangle \to \langle \mathsf{True}, \gamma_{s_2}^2, \sigma_{s_2}^2 \rangle \quad \langle e_2, \gamma_{s_2}^1, \sigma_{s_2}^2 \rangle \to \langle v_{s_2}, \gamma_{s_2}^3, \sigma_{s_2}^3 \rangle}{\langle \mathsf{If}\ e_1\ \mathsf{then}\ e_2\ \mathsf{else}\ e_3, \gamma_{s_2}^1, \sigma_{s_2}^1 \rangle \to \langle v_{s_2}, \gamma_{s_2}^1, \sigma_{s_2}^3 \rangle}$

and both are typed with

(If-Else) $\dfrac{\begin{array}{l}\Gamma_1, pc_1 \vdash e_1 : t_1@t_2 \triangleright \Gamma_2 \\ \Gamma_1, pc_2 \vdash e_2 : t_3@t_4 \triangleright \Gamma_3 \\ \Gamma_1, pc_2 \vdash e_3 : t_5@t_6 \triangleright \Gamma_4\end{array}}{\Gamma_1, pc_1 \vdash \mathsf{If}\ e_1\ \mathsf{then}\ e_2\ \mathsf{else}\ e_3 : t_7@t_8 \triangleright \Gamma_1} \quad \begin{array}{l}pc_2 = \sqcup\{pc_1, t_1\} \\ t_7 = \sqcup\{t_1, t_3, t_5\} \\ t_8 = \sqcap\{t_2, t_4, t_6\} \\ t_1 \in \{\mathsf{Low}, \mathsf{High}\}\end{array}$

This is because, from inspection of the partial order of types, then the only type that the (S-If-Else-True) can be typed with, that is relevant to our lemma, is Low. From the inductive hypothesis, we then know that $e_1$ will evaluate to the same Low value, as $t_7 = \sqcup\{t_1, t_3, t_5\}$ means that $e_1$ must be typed as Low.

Furthermore, from Lemma 3 and the non-interference property we know that $\Gamma \models (\gamma_{s_1}^1, \sigma_{s_1}^2) =_{\mathsf{Low}} (\gamma_{s_2}^1, \sigma_{s_2}^2)$, and therefore, from the inductive hypothesis, we know that $v_{s_1}^1 = v_{s_2}^1$, since $t_3 = t_5 = \mathsf{Low}$, for $t_7 = \mathsf{Low}$. Therefore, Lemma 1 holds for (S-If-Else-True).

If (3) are found with:

(S-If-Else-False) $\dfrac{\langle e_1, \gamma_{s_1}^1, \sigma_{s_1}^1 \rangle \to \langle \mathsf{False}, \gamma_{s_1}^2, \sigma_{s_1}^2 \rangle \quad \langle e_3, \gamma_{s_1}^1, \sigma_{s_1}^2 \rangle \to \langle v_{s_1}, \gamma_{s_1}^3, \sigma_{s_1}^3 \rangle}{\langle \mathsf{If}\ e_1\ \mathsf{then}\ e_2\ \mathsf{else}\ e_3, \gamma_{s_1}^1, \sigma_{s_1}^1 \rangle \to \langle v_{s_1}, \gamma_{s_1}^1, \sigma_{s_1}^3 \rangle}$

Then the prove follows the same logic, but with $e_2$ replaced with $e_3$.

## While

If (3) are found with either:

(S-While-True)
$$\frac{\langle e_1, \gamma_{s_1}^1, \sigma_{s_1}^1 \rangle \to \langle \mathsf{True}, \gamma_{s_1}^2, \sigma_{s_1}^2 \rangle \quad \langle e_2, \gamma_{s_1}^1, \sigma_{s_1}^2 \rangle \to \langle v_{s_1}^1, \gamma_{s_1}^3, \sigma_{s_1}^3 \rangle \quad \langle \mathsf{While}\ e_1\ e_2, \gamma_{s_1}^1, \sigma_{s_1}^3 \rangle \to \langle v_{s_1}^2, \gamma_{s_1}^4, \sigma_{s_1}^4 \rangle}{\langle \mathsf{While}\ e_1\ e_2, \gamma_{s_1}^1, \sigma_{s_1}^1 \rangle \to \langle u, \gamma_{s_1}^1, \sigma_{s_1}^4 \rangle}$$

or

(S-While-False)
$$\frac{\langle e_1, \gamma_{s_1}^1, \sigma_{s_1}^1 \rangle \to \langle \mathsf{False}, \gamma_{s_1}^2, \sigma_{s_1}^2 \rangle}{\langle \mathsf{While}\ e_1\ e_2, \gamma_{s_1}^1, \sigma_{s_1}^1 \rangle \to \langle u, \gamma_{s_1}^1, \sigma_{s_1}^2 \rangle}$$

and if (4) are found with either:

(S-While-True)
$$\frac{\langle e_1, \gamma_{s_2}^1, \sigma_{s_2}^1 \rangle \to \langle \mathsf{True}, \gamma_{s_2}^2, \sigma_{s_2}^2 \rangle \quad \langle e_2, \gamma_{s_2}^1, \sigma_{s_2}^2 \rangle \to \langle v_{s_2}^1, \gamma_{s_2}^3, \sigma_{s_2}^3 \rangle \quad \langle \mathsf{While}\ e_1\ e_2, \gamma_{s_2}^1, \sigma_{s_2}^3 \rangle \to \langle v_{s_2}^2, \gamma_{s_2}^4, \sigma_{s_2}^4 \rangle}{\langle \mathsf{While}\ e_1\ e_2, \gamma_{s_2}^1, \sigma_{s_2}^1 \rangle \to \langle u, \gamma_{s_2}^1, \sigma_{s_2}^4 \rangle}$$

or

(S-While-False)
$$\frac{\langle e_1, \gamma_{s_2}^1, \sigma_{s_2}^1 \rangle \to \langle \mathsf{False}, \gamma_{s_2}^2, \sigma_{s_2}^2 \rangle}{\langle \mathsf{While}\ e_1\ e_2, \gamma_{s_2}^1, \sigma_{s_2}^1 \rangle \to \langle u, \gamma_{s_2}^1, \sigma_{s_2}^2 \rangle}$$

and both are typed with

(While)
$$\frac{\Gamma_1, pc_1 \vdash e_1 : t_1 @ t_2 \triangleright \Gamma_2 \quad \Gamma_1, pc_2 \vdash e_2 : t_3 @ t_4 \triangleright \Gamma_3}{\Gamma_1, pc_1 \vdash \mathsf{While}\ e_1\ e_2 : \mathsf{Low} @ t_6 \triangleright \Gamma_1} \quad \begin{array}{l} pc_2 = \sqcup\{pc_1, t_1\} \\ t_6 = \sqcap\{t_2, t_4\} \\ t_1 \in \{\mathsf{Low}, \mathsf{High}\} \end{array}$$

As regardless of what rules are chosen, both (3) and (4) will evaluate to $u$, and if the while-construct is well-typed, then the type will be Low. therefore, it can easily be seen, that the while-construct upholds Lemma 1.

## For

Regardless of if (3) are found with either:

(S-For-Base)
$$\frac{\langle e_1, \gamma_{s_1}^1, \sigma_{s_1}^1 \rangle \to \langle v_{s_1}^1, \gamma_{s_1}^2, \sigma_{s_1}^2 \rangle \quad \langle e_2, \gamma_{s_1}^1, \sigma_{s_1}^2 \rangle \to \langle v_{s_1}^2, \gamma_{s_1}^3, \sigma_{s_1}^3 \rangle \quad \langle e_3, \gamma_{s_1}^1[x \mapsto v_{s_1}^2], \sigma_{s_1}^3 \rangle \to \langle v_{s_1}^3, \gamma_{s_1}^4, \sigma_{s_1}^4 \rangle}{\langle \mathsf{For}\ x\ e_1\ e_2\ e_3, \gamma_{s_1}^1, \sigma_{s_1}^1 \rangle \to \langle u, \gamma_{s_1}^1, \sigma_{s_1}^4 \rangle} \quad v_{s_1}^1 = v_{s_1}^2$$

or

(S-For-Rec)
$$\frac{\langle e_1, \gamma_{s_1}^1, \sigma_{s_1}^1 \rangle \to \langle v_{s_1}^1, \gamma_{s_1}^2, \sigma_{s_1}^2 \rangle \quad \langle e_2, \gamma_{s_1}^1, \sigma_{s_1}^2 \rangle \to \langle v_{s_1}^2, \gamma_{s_1}^3, \sigma_{s_1}^3 \rangle \quad \langle \mathsf{For}\ x\ n\ m, \gamma_{s_1}^1, \sigma_{s_1}^3 \rangle \to (u, \gamma_{s_1}^4, \sigma_{s_1}^4) \quad \langle e_3, \gamma_{s_1}^1[x \mapsto v_{s_1}^2], \sigma_{s_1}^4 \rangle \to \langle v_{s_1}^4, \gamma_{s_1}^5, \sigma_{s_1}^5 \rangle}{\langle \mathsf{For}\ x\ e_1\ e_2\ e_3, \gamma_{s_1}^1, \sigma_{s_1}^1 \rangle \to \langle u, \gamma_{s_1}^1, \sigma_{s_1}^5 \rangle} \quad \begin{array}{l} n = \mathbb{N}^{-1}(v_{s_1}^1) \\ m = \mathbb{N}^{-1}(v_{s_1}^2 - 1) \\ v_{s_1}^1 < v_{s_1}^2 \end{array}$$

and if (4) are found with either:

(S-For-Base)
$$\frac{\langle e_1, \gamma_{s_2}^1, \sigma_{s_2}^1 \rangle \to \langle v_{s_2}^1, \gamma_{s_2}^2, \sigma_{s_2}^2 \rangle \quad \langle e_2, \gamma_{s_2}^1, \sigma_{s_2}^2 \rangle \to \langle v_{s_2}^2, \gamma_{s_2}^3, \sigma_{s_2}^3 \rangle \quad \langle e_3, \gamma_{s_2}^1[x \mapsto v_{s_2}^2], \sigma_{s_2}^3 \rangle \to \langle v_{s_2}^3, \gamma_{s_2}^4, \sigma_{s_2}^4 \rangle}{\langle \mathsf{For}\ x\ e_1\ e_2\ e_3, \gamma_{s_2}^1, \sigma_{s_2}^1 \rangle \to \langle u, \gamma_{s_2}^1, \sigma_{s_2}^4 \rangle} \quad v_{s_2}^1 = v_{s_2}^2$$

A. 26

or

$$(\text{S-For-Rec}) \quad \frac{\begin{array}{c}\langle e_1, \gamma_{s_2}^1, \sigma_{s_2}^1\rangle \to \langle v_{s_2}^1, \gamma_{s_2}^2, \sigma_{s_2}^2\rangle \quad \langle e_2, \gamma_{s_2}^1, \sigma_{s_2}^2\rangle \to \langle v_{s_2}^2, \gamma_{s_2}^3, \sigma_{s_2}^3\rangle \\ \langle \text{For } x \ n \ m, \gamma_{s_2}^1, \sigma_{s_2}^3\rangle \to (u, \gamma_{s_2}^4, \sigma_{s_2}^4) \\ \langle e_3, \gamma_{s_2}^1[x \mapsto v_{s_2}^2], \sigma_{s_2}^4\rangle \to (v_{s_2}^4, \gamma_{s_2}^5, \sigma_{s_2}^5) \end{array}}{\langle \text{For } x \ e_1 \ e_2 \ e_3, \gamma_{s_2}^1, \sigma_{s_2}^1\rangle \to \langle u, \gamma_{s_2}^1, \sigma_{s_2}^5\rangle} \quad \begin{array}{l} n = \mathbb{N}^{-1}(v_{s_2}^1) \\ m = \mathbb{N}^{-1}(v_{s_2}^2 - 1) \\ v_{s_2}^1 < v_{s_2}^2 \end{array}$$

and both are typed with

$$(\text{For}) \quad \frac{\begin{array}{c}\Gamma_1, pc_1 \vdash e_1 : t_1@t_2 \triangleright \Gamma_2 \\ \Gamma_1, pc_1 \vdash e_2 : t_3@t_4 \triangleright \Gamma_3 \\ \Gamma_1[x \to pc_2], pc_2 \vdash e_3 : t_5@t_6 \triangleright \Gamma_4 \end{array}}{\Gamma_1, pc_1 \vdash \text{For } x \ e_1 \ e_2 \ e_3 : \text{Low}@t_8 \triangleright \Gamma_1} \quad \begin{array}{l} pc_2 = \sqcup\{pc_1, t_1, t_3\} \\ t_8 = \sqcap\{t_2, t_4, t_6\} \\ t_1, t_3 \in \{\text{Low}, \text{High}\} \end{array}$$

As regardless of what rules are chosen, both (3) and (4) will evaluate to $u$, and if the for-construct is well-typed, then the type will be Low, it can easily be seen, that the for-construct upholds Lemma 1.

## Deref

If (3) is found with:

$$(\text{S-Deref}) \quad \frac{}{\langle !x, \gamma_{s_1}^1, \sigma_{s_1}^1\rangle \to \langle v_{s_1}^2, \gamma_{s_1}^1, \sigma_{s_1}^1\rangle} \quad \begin{array}{l}\gamma_{s_1}^1(x) = \ell_{s_1}^1 \\ v_{s_1}^2 = \sigma_{s_1}^1(\ell_{s_1}^1)\end{array}$$

Then (4) is found with:

$$(\text{S-Deref}) \quad \frac{}{\langle !x, \gamma_{s_2}^1, \sigma_{s_2}^1\rangle \to \langle v_{s_2}^2, \gamma_{s_2}^1, \sigma_{s_2}^1\rangle} \quad \begin{array}{l}\gamma_{s_2}^1(x) = \ell_{s_2}^1 \\ v_{s_2}^2 = \sigma_{s_2}^1(\ell_{s_2}^1)\end{array}$$

And they are both typed with: $\quad (\text{Deref}) \quad \dfrac{}{\Gamma, pc \vdash !x : t_1@() \triangleright \Gamma} \quad \Gamma(x) = \text{ref } t_1$

Since the only type that this rule can be typed as, that is relevant to Lemma 1 is Low, then $\Gamma(x) = \text{ref Low}$. This means that, since we know that $G \vdash (\gamma_{s_1}^1, \sigma_{s_1}^1) =_{\text{Low}} (\gamma_{s_1}^1, \sigma_{s_1}^1)$, we then know that the $\sigma_{s_1}^1(\gamma_{s_1}^1(x) = \sigma_{s_2}^1(\gamma_{s_2}^1(x))$. Therefore, (Deref) also upholds Lemma 1.

## Ref

If (3) is found with:

$$(\text{S-Ref}) \quad \frac{\langle e, \gamma_{s_1}^1, \sigma_{s_1}^1\rangle \to \langle v_{s_1}, \gamma_{s_1}^2, \sigma_{s_1}^2\rangle}{\langle \text{Ref}(e), \gamma_{s_1}^1, \sigma_{s_1}^1\rangle \to \langle \ell_{s_1}, \gamma_{s_1}^1, \sigma_{s_1}^2[\ell_{s_1} \to v_{s_1}]\rangle} \quad \ell_{s_1} = new(\sigma_{s_1}^2)$$

Then (4) is found with:

$$(\text{S-Ref}) \quad \frac{\langle e, \gamma_{s_2}^1, \sigma_{s_2}^1\rangle \to \langle v_{s_2}, \gamma_{s_2}^2, \sigma_{s_2}^2\rangle}{\langle \text{Ref}(e), \gamma_{s_2}^1, \sigma_{s_2}^1\rangle \to \langle \ell_{s_2}, \gamma_{s_2}^1, \sigma_{s_2}^2[\ell_{s_2} \to v_{s_2}]\rangle} \quad \ell_{s_2} = new(\sigma_{s_2}^2)$$

And they are both typed with: $\quad (\text{Ref}) \quad \dfrac{\Gamma, pc \vdash e : t_1@t_2 \triangleright \Gamma'}{\Gamma, pc \vdash \text{Ref}(e) : \text{ref } t_1@t_2 \triangleright \Gamma} \quad t_1 \in \{\text{Low}, \text{High}\}$

The only relevant type for Lemma 1 that this rule can evaluate to is ref Low. For this to happen, then $t_1 = \text{Low}$. From the inductive hypothesis, we then know that $v_{s_1} = v_{s_2}$.

A. 27

Therefore, $\sigma_{s_1}[l_{s_1} \to v_{s_1}](l_{s_1}) = \sigma_{s_2}[l_{s_2} \to v_{s_2}](l_{s_2})$, which is the requirement, when the type is ref Low, for the expression to uphold Lemma 1. Therefore, (S-Ref) upholds Lemma 1.

### Reassign

(1) is found with:

$$\text{(S-Reassign)} \quad \frac{\langle e, \gamma^1_{s_1}, \sigma^1_{s_1} \rangle \to \langle v^1_{s_1}, \gamma^2_{s_1}, \sigma^2_{s_1} \rangle}{\langle x := e, \gamma^1_{s_1}, \sigma^1_{s_1} \rangle \to \langle u, \gamma^1_{s_1}, \sigma^2_{s_1}[\ell_{s_1} \to v^1_{s_1}] \rangle} \quad \gamma^1_{s_1}(x) = \ell_{s_1}$$

(2) is found with:

$$\text{(S-Reassign)} \quad \frac{\langle e, \gamma^1_{s_2}, \sigma^1_{s_2} \rangle \to \langle v^1_{s_2}, \gamma^2_{s_2}, \sigma^2_{s_2} \rangle}{\langle x := e, \gamma^1_{s_2}, \sigma^1_{s_2} \rangle \to \langle u, \gamma^1_{s_2}, \sigma^2_{s_2}[\ell_{s_2} \to v^1_{s_2}] \rangle} \quad \gamma^1_{s_2}(x) = \ell_{s_2}$$

And they are both typed with:

$$\text{(Reassign)} \quad \frac{\Gamma, pc \vdash e : t_1 @ t_2 \triangleright \Gamma_1}{\Gamma, pc \vdash x := e : \text{Low} @ t_4 \triangleright \Gamma} \quad \begin{array}{l} \Gamma(x) = \text{ref } t_3 \\ t_3 \in \{\text{Low}, \text{High}\} \\ t_3 \sqsupseteq t_1 \\ t_3 \sqsupseteq pc \\ t_4 = \sqcap\{t_3, t_2\} \end{array}$$

As (Reassign) always evaluates to $u$, it also always upholds Lemma 1.

∎

## E  Proof of Lemma 2 and 5

Lemmas we use in our 2-level semantics. We will prove both lemmas simultaneously, as they are very similar, and the inductive hypothesis, results in the same thing, namely that $\Gamma_1 \models s =_{\text{Low}} s'$, and we only need Lemma 5 to prove Lemma 2 in the application rule.

**Lemma 2** (Side-effect Equivalent High PC). *Given*

$$\Gamma, \text{High} \vdash e : t_1 @ t_2 \triangleright \Gamma_1$$

*if*

$$\langle e, s \rangle \longrightarrow \langle v, s' \rangle \tag{5}$$

*then*

$$\Gamma_1 \models s =_{\text{Low}} s'$$

*and*

**Lemma 5** (Side-effect Equivalent $t_2 \in \{(), \text{High}\}$). *Given*

$$\Gamma, pc \vdash e : t_1 @ t_2 \triangleright \Gamma_1$$

*if*

$$\langle e, s \rangle \longrightarrow \langle v, s' \rangle \tag{9}$$

$$t_2 \in \{(), \mathsf{High}\}$$

*then*

$$\Gamma_1 \models s =_{\mathsf{Low}} s'$$

## Unit

(5) and (9) is found with: (S-Unit) $\quad \dfrac{}{\langle (), \gamma, \sigma \rangle \to \langle u, \gamma, \sigma \rangle}$

And it is typed with:

(Unit) $\quad \dfrac{}{\Gamma, pc \vdash () : \mathsf{Low}@() \triangleright \Gamma}$

As the state before and after are the same, it is trivially true, that Lemma 2 holds for (S-Bool). Since the side-effect is always typed as (), Lemma 5 is also trivially true.

## Bool

(5) and (9) is found with: (S-Bool) $\quad \dfrac{}{\langle b, \gamma, \sigma \rangle \to \langle v, \gamma, \sigma \rangle} \quad v = \mathbb{B}(b)$

And it is typed with:

(Bool) $\quad \dfrac{}{\Gamma, pc \vdash b : \mathsf{Low}@() \triangleright \Gamma}$

As the state before and after are the same, it is trivially true, that Lemma 2 holds for (S-Bool). Since the side-effect is always typed as (), Lemma 5 is also trivially true.

## Num

(5) and (9) is found with: (S-Num) $\quad \dfrac{}{\langle n, \gamma, \sigma \rangle \to \langle v, \gamma, \sigma \rangle} \quad v = \mathbb{N}(n)$

And it is typed with:
(Num) $\quad \dfrac{}{\Gamma, pc \vdash n : \mathsf{Low}@() \triangleright \Gamma}$

As the state before and after are the same, it is trivially true, that Lemma 2 holds for (S-Num). Since the side-effect is always typed as (), Lemma 5 is also trivially true.

## Var

(5) and (9) is found with:
(S-Var) $\quad \dfrac{}{\langle x, \gamma, \sigma \rangle \to \langle v, \gamma, \sigma \rangle} \quad \gamma(x) = v$

And it is typed with:
(Var) $\quad \dfrac{}{\Gamma, pc \vdash x : t@() \triangleright \Gamma} \quad \Gamma(x) = t$

As the state before and after are the same, it is trivially true, that Lemma 2 holds for (S-Var). Since the side-effect is always typed as (), Lemma 5 is also trivially true.

## Bop

(5) and (9) is found with:

(S-Bop) $$\frac{\langle e_1, \gamma^1, \sigma^1 \rangle \to \langle v^1, \gamma^2, \sigma^2 \rangle \quad \langle e_2, \gamma^1, \sigma^2 \rangle \to \langle v^2, \gamma^3, \sigma^3 \rangle}{\langle e_1 \text{ bop } e_2, \gamma^1, \sigma^1 \rangle \to \langle v^3, \gamma^1, \sigma^3 \rangle} \quad v^3 = v^1 \text{ bop } v^2$$

And it is typed with:

(Bop) $$\frac{\Gamma_1, pc_1 \vdash e_1 : t_1@t_2 \triangleright \Gamma_2 \qquad \Gamma_1, pc_1 \vdash e_2 : t_3@t_4 \triangleright \Gamma_3}{\Gamma_1, pc_1 \vdash e_1 \text{ bop } e_2 : t_5@t_6 \triangleright \Gamma_1} \quad \begin{array}{l} t_1, t_3 \in \{\mathsf{Low}, \mathsf{High}\} \\ t_5 = \sqcup\{t_1, t_3\} \\ t_6 = \sqcap\{t_2, t_4\} \end{array}$$

From the inductive hypothesis of Lemma 2, since we know that $\langle e_1, \gamma^1, \sigma^1 \rangle \to \langle v_1, \gamma^2, \sigma^2 \rangle$, which means that $\Gamma_2 \models (\gamma^1, \sigma^1) =_{\mathsf{Low}} (\gamma^2, \sigma^2)$.

We then know from Lemma 4 that $\Gamma_1 \models (\gamma^1, \sigma^1) =_{\mathsf{Low}} (\gamma^1, \sigma^2)$.

Therefore, with $\langle e_2, \gamma^1, \sigma^2 \rangle \to \langle v_2, \gamma^3, \sigma^3 \rangle$, which with the inductive hypothesis of Lemma 2 means that $\Gamma_3 \models (\gamma^1, \sigma^2) =_{\mathsf{Low}} (\gamma^3, \sigma^3)$, which with Lemma 4 means that $\Gamma_1 \models (\gamma^1, \sigma^2) =_{\mathsf{Low}} (\gamma^1, \sigma^3)$.

Which means that from the transitive property of $=_{\mathsf{Low}}$, we know that $\Gamma_1 \models (\gamma^1, \sigma^1) =_{\mathsf{Low}} (\gamma^1, \sigma^3)$. Therefore, (S-Bop) upholds Lemma 2.

with regards to Lemma 5, since $t_6 = \sqcap\{t_2, t_4\}$, for $t_6 \in \{(), \mathsf{High}\}$, then $t_2, t_4 \in \{(), \mathsf{High}\}$. Therefore, the inductive hypothesis of Lemma 5 holds for $e_1$ and $e_2$. Therefore, the same derivation as above can be derived for Lemma 5, so that we see that $\Gamma_1 \models (\gamma^1, \sigma^1) =_{\mathsf{Low}} (\gamma^1, \sigma^3)$.

## Let

(5) and (9) is found with:

(S-Let) $$\frac{\langle e, \gamma^1, \sigma^1 \rangle \to \langle v, \gamma^2, \sigma^2 \rangle}{\langle \mathsf{Let}\ x =\ e, \gamma^1, \sigma^1 \rangle \to \langle u, \gamma^1[x \mapsto v], \sigma^2 \rangle}$$

And it is typed with either:

(Let-Base) $$\frac{\Gamma, pc \vdash e : t_1@t_2 \triangleright \Gamma_1}{\Gamma, pc \vdash \mathsf{Let}\ x =\ e\ : \mathsf{Low}@t_3 \triangleright \Gamma[x \to t_1]} \quad \begin{array}{l} t_1 \in \{\mathsf{Low}, \mathsf{High}\} \\ t_1 \sqsupseteq pc \\ t_3 = \sqcap\{t_1, t_2\} \end{array}$$

or

(Let-n) $$\frac{\Gamma, pc \vdash e : t_2@t_3 \triangleright \Gamma_1}{\Gamma, pc \vdash \mathsf{Let}\ x_{t_1} =\ e\ : \mathsf{Low}@t_4 \triangleright \Gamma[x \to t_1]} \quad \begin{array}{l} t_1, t_2 \in \{\mathsf{Low}, \mathsf{High}\} \\ t_1 \sqsupseteq t_2 \\ t_1 \sqsupseteq pc \\ t_4 = \sqcap\{t_3, t_1\} \end{array}$$

or

(Let-Base-Ref) $$\frac{\Gamma, pc \vdash e : \mathsf{ref}\ t_1@t_2 \triangleright \Gamma_1}{\Gamma, pc \vdash \mathsf{Let}\ x =\ e\ : \mathsf{Low}@t_3 \triangleright \Gamma[x \to \mathsf{ref}\ t_1]} \quad \begin{array}{l} t_3 = \sqcap\{t_2, t_1\} \\ t_3 \sqsupseteq pc \\ t_1 \in \{\mathsf{Low}, \mathsf{High}\} \end{array}$$

It cannot be typed with (Let-Base-Func) with regards to Lemma 2, As it requires that $\mathsf{Low} \sqsupseteq pc$, or with Lemma 5, as the side-effect is always $\mathsf{Low}$:

$$\text{(Let-Base-Func)} \quad \frac{\Gamma, pc \vdash e : (t_1 \to t_2@t_3)@t_4 \triangleright \Gamma_1}{\Gamma, pc \vdash \text{Let } x = e : \text{Low@Low} \triangleright \Gamma[x \to (t_1 \to t_2@t_3)]} \quad \text{Low} \sqsupseteq pc$$

We have from the inductive hypothesis of Lemma 2 and Lemma 4, that $\langle e, \gamma^1, \sigma^1 \rangle \to \langle v_1, \gamma^2, \sigma^2 \rangle$ means that $\Gamma_1 \models (\gamma^1, \sigma^1) =_{\text{Low}} (\gamma^1, \sigma^2)$.

We then look at the consequences of the changed environment, for each typing rule:

**Let-Base**

With regards to Lemma 2, since we know that $pc = \text{High}$, we know that $t_1 \sqsupseteq \text{High}$, and since $t_1 \in \{\text{Low}, \text{High}\}$, it must mean that $t_1 = \text{High}$.

Therefore, $\Gamma[x \to \text{High}] \models (\gamma^1, \sigma^1) =_{\text{Low}} (\gamma^1[x \to v], \sigma^2)$, as $\Gamma(x) = \text{High}$ means that we do not care about changes to it. Therefore, Lemma 2 holds for rule (S-Let), if it is typed with (Let-Base).

With regards to Lemma 5, since we know that $t_3 \in \{\text{High}, ()\}$, and we know that $t_3 = \sqcap\{t_1, t_2\}$, it must mean that $t_2 \in \{\text{High}, ()\}$. Therefore, with Lemma 4, $\Gamma_1[x \to \text{High}] \models (\gamma^1, \sigma^1) =_{\text{Low}} (\gamma^1[x \to v], \sigma^2)$. Since it must also be the case that $t_1 \in \{\text{Low}, \text{High}\}$ and $t_1 \in \{(), \text{High}\}$, it must be the case that $t_1 = \text{High}$. Therefore, changes to it does not change non-interference, and therefore, Lemma 5 holds for rule (S-Let), if it is typed with (Let-Base).

**Let-n**

With regards to Lemma 2, since we know that $pc = \text{High}$, we know that $t_1 \sqsupseteq \text{High}$, and since $t_1 \in \{\text{Low}, \text{High}\}$, it must mean that $t_1 = \text{High}$.

Therefore, $\Gamma[x \to \text{High}] \models (\gamma^1, \sigma^1) =_{\text{Low}} (\gamma^1[x \to v], \sigma^2)$, as $\Gamma(x) = \text{High}$, and it therefore does not change low-variables.

As in both cases, Lemma 2 holds, Lemma 2 holds for the (S-Let)-construct, when typed with (Let-n).

With regards to Lemma 5, since we know that $t_4 \in \{\text{High}, ()\}$, and we know that $t_4 = \sqcap\{t_1, t_3\}$, it must mean that $t_3 \in \{\text{High}, ()\}$. Therefore, with Lemma 4, $\Gamma_1[x \to \text{High}] \models (\gamma^1, \sigma^1) =_{\text{Low}} (\gamma^1[x \to v], \sigma^2)$, since it must also be the case that $t_1 \in \{\text{Low}, \text{High}\}$ and $t_1 \in \{(), \text{High}\}$, it must be the case that $t_1 = \text{High}$. Therefore, changes to it does not change non-interference, and therefore, Lemma 5 holds for rule (S-Let), if it is typed with (Let-n).

**Let-Base-Ref**

With regards to Lemma 2, since we know that $pc = \text{High}$, we then know that $t_3 \sqsupseteq \text{High}$, and since $t_3 \sqcap \{t_1, t_2\}$, and $t_1 \in \{\text{Low}, \text{High}\}$ it must mean that $t_1 = \text{High}$.

Therefore, $\Gamma[x \to \text{ref High}] \models (\gamma^1, \sigma^1) =_{\text{Low}} (\gamma^1[x \to v], \sigma^2)$, as $\Gamma(x) = \text{ref High}$, and it therefore does not change ref Low variables, meaning it does not change non-interference. Therefore, Lemma 2 holds for rule (S-Let), if it is typed with (Let-Base-Ref).

With regards to Lemma 5, since we know that $t_3 \in \{\text{High}, ()\}$, and we know that $t_3 = \sqcap\{t_1, t_2\}$, it must mean that $t_2 \in \{\text{High}, ()\}$. Therefore, with Lemma 4, $\Gamma_1 \models (\gamma^1, \sigma^1) =_{\text{Low}}$ $(\gamma^1, \sigma^2)$, since it must also be the case that $t_1 \in \{\mathsf{Low}, \mathsf{High}\}$ and $t_1 \in \{(), \mathsf{High}\}$, it must be the case that $t_1 = \mathsf{Low}$. Therefore $\Gamma(x) = \mathsf{ref\ High}$, and changes to it does not change non-interference, and therefore, Lemma 5 holds for rule (S-Let), if it is typed with (Let-Base-Ref).

We have now shown, that regardless of what typing rule is used, Lemma 2 and Lemma 5 holds for rule (S-Let).

### If-Else

(5) and (9) is found with either:

(S-If-Else-True) $\dfrac{\langle e_1, \gamma^1, \sigma^1 \rangle \to \langle \mathsf{True}, \gamma^2, \sigma^2 \rangle \quad \langle e_2, \gamma^1, \sigma^2 \rangle \to \langle v, \gamma^3, \sigma^3 \rangle}{\langle \mathsf{If}\ e_1\ \mathsf{then}\ e_2\ \mathsf{else}\ e_3, \gamma^1, \sigma^1 \rangle \to \langle v, \gamma^1, \sigma^3 \rangle}$

or

(S-If-Else-False) $\dfrac{\langle e_1, \gamma^1, \sigma^1 \rangle \to \langle \mathsf{False}, \gamma^2, \sigma^2 \rangle \quad \langle e_3, \gamma^1, \sigma^2 \rangle \to \langle v, \gamma^3, \sigma^3 \rangle}{\langle \mathsf{If}\ e_1\ \mathsf{then}\ e_2\ \mathsf{else}\ e_3, \gamma^1, \sigma^1 \rangle \to \langle v, \gamma^1, \sigma^3 \rangle}$

And it is typed with:

(If-Else) $\dfrac{\begin{array}{l}\Gamma_1, pc_1 \vdash e_1 : t_1@t_2 \triangleright \Gamma_2 \\ \Gamma_1, pc_2 \vdash e_2 : t_3@t_4 \triangleright \Gamma_3 \\ \Gamma_1, pc_2 \vdash e_3 : t_5@t_6 \triangleright \Gamma_4\end{array}}{\Gamma_1, pc_1 \vdash \mathsf{If}\ e_1\ \mathsf{then}\ e_2\ \mathsf{else}\ e_3 : t_7@t_8 \triangleright \Gamma_1} \quad \begin{array}{l}pc_2 = \sqcup\{pc_1, t_1\} \\ t_7 = \sqcup\{t_1, t_3, t_5\} \\ t_8 = \sqcap\{t_2, t_4, t_6\} \\ t_1 \in \{\mathsf{Low}, \mathsf{High}\}\end{array}$

With regards to Lemma 2, since $pc_2 = \sqcup\{pc_1, t_1\}$ and $pc_1 = \mathsf{High}$, then $pc_2 = \mathsf{High}$. Therefore, the inductive hypothesis can be used for all inner expressions. For Lemma 5, since $t_8 = \sqcap\{t_2, t_4, t_6\}$, and $t_8 \in \{(), \mathsf{High}\}$, then $t_2, t_4, t_6 \in \{(), \mathsf{High}\}$, and we can also use the inductive hypothesis for all inner expressions.

From the inductive hypothesis, since we know that $\langle e_1, \gamma^1, \sigma^1 \rangle \to \langle v_1, \gamma^2, \sigma^2 \rangle$, which means that $\Gamma_2 \models (\gamma^1, \sigma^1) =_{\mathsf{Low}} (\gamma^2, \sigma^2)$. We then know from Lemma 4 that $\Gamma_1 \models (\gamma^1, \sigma^1) =_{\mathsf{Low}} (\gamma^1, \sigma^2)$.

We could look at each derivation-rule individually, however, the method is identically in each, so we will just use the general case, $e_n$, where $e_n \in \{2, 3\}$.

We know that $\langle e_n, \gamma^1, \sigma^2 \rangle \to \langle v_2, \gamma^3, \sigma^3 \rangle$, which means that $\Gamma_3 \models (\gamma^1, \sigma^1) =_{\mathsf{Low}} (\gamma^3, \sigma^3)$. From Lemma 4, we then know that $\Gamma_1 \models (\gamma^1, \sigma^1) =_{\mathsf{Low}} (\gamma^1, \sigma^3)$. Which proves, that Lemma 2 and Lemma 5 holds for the if-construct.

### While

(5) and (9) is found with either:

(S-While-True) $\dfrac{\begin{array}{c}\langle e_1, \gamma^1, \sigma^1 \rangle \to \langle \mathsf{True}, \gamma^2, \sigma^2 \rangle \\ \langle e_2, \gamma^1, \sigma^2 \rangle \to \langle v^1, \gamma^3, \sigma^3 \rangle \\ \langle \mathsf{While}\ e_1\ e_2, \gamma^1, \sigma^3 \rangle \to \langle v^2, \gamma^4, \sigma^4 \rangle\end{array}}{\langle \mathsf{While}\ e_1\ e_2, \gamma^1, \sigma^1 \rangle \to \langle u, \gamma^1, \sigma^4 \rangle}$

or

(S-While-False) $\dfrac{\langle e_1, \gamma^1, \sigma^1 \rangle \to \langle \mathsf{False}, \gamma^2, \sigma^2 \rangle}{\langle \mathsf{While}\ e_1\ e_2, \gamma^1, \sigma^1 \rangle \to \langle u, \gamma^1, \sigma^2 \rangle}$

And it is typed with:

(While) $$\frac{\Gamma_1, pc_1 \vdash e_1 : t_1@t_2 \triangleright \Gamma_2 \quad \Gamma_1, pc_2 \vdash e_2 : t_3@t_4 \triangleright \Gamma_3}{\Gamma_1, pc_1 \vdash \text{While } e_1 \ e_2 : \text{Low}@t_6 \triangleright \Gamma_1}$$
$pc_2 = \sqcup\{pc_1, t_1\}$
$t_6 = \sqcap\{t_2, t_4\}$
$t_1 \in \{\text{Low}, \text{High}\}$

We look at (S-While-False) first.

**S-While-False**

With regards to Lemma 2, since $pc_2 \sqcup \{pc_1, t_1\}$ and $pc_1 = \text{High}$, then $pc_2 = \text{High}$. Therefore, the inductive hypothesis can be used for all inner expressions. For Lemma 5, since $t_6 = \sqcap\{t_2, t_4\}$, and $t_6 \in \{(), \text{High}\}$, then $t_2, t_4 \in \{(), \text{High}\}$, and we can also use the inductive hypothesis for all inner expressions.

From the inductive hypothesis, since we know that $\langle e_1, \gamma^1, \sigma^1 \rangle \to \langle \text{False}, \gamma^2, \sigma^2 \rangle$, which means that $\Gamma_2 \models (\gamma^1, \sigma^1) =_{\text{Low}} (\gamma^2, \sigma^2)$. We then know from Lemma 4 that $\Gamma_1 \models (\gamma^1, \sigma^1) =_{\text{Low}} (\gamma^1, \sigma^2)$, which proves Lemma 2 and Lemma 5 holds for (S-While-False).

**S-While-true**

From the inductive hypothesis, we know that $\langle e_1, \gamma^1, \sigma^1 \rangle \to \langle \text{True}, \gamma^2, \sigma^2 \rangle$, which means that $\Gamma_2 \models (\gamma^1, \sigma^1) =_{\text{Low}} (\gamma^2, \sigma^2)$. We then know from Lemma 4 that $\Gamma_1 \models (\gamma^1, \sigma^1) =_{\text{Low}} (\gamma^1, \sigma^2)$.

Therefore, from the inductive hypothesis, we know that $\langle e_2, \gamma^1, \sigma^2 \rangle \to \langle v^1, \gamma^3, \sigma^3 \rangle$, which means that $\Gamma_3 \models (\gamma^1, \sigma^2) =_{\text{Low}} (\gamma^3, \sigma^3)$. From Lemma 4, we then know that $\Gamma_1 \models (\gamma^1, \sigma^2) =_{\text{Low}} (\gamma^1, \sigma^3)$.

Therefore, from the inductive hypothesis, we know that $\langle \text{While } e_1 e_2, \gamma^1, \sigma^3 \rangle \to \langle v^2, \gamma^4, \sigma^4 \rangle$, which means that, $\Gamma_1 \models (\gamma^1, \sigma^2) =_{\text{Low}} (\gamma^4, \sigma^4)$, as the resulting type-environment after the typing of a while, is the same as the initial one. From Lemma 4, we then know that $\Gamma_1 \models (\gamma^1, \sigma^1) =_{\text{Low}} (\gamma^1, \sigma^4)$, from the transitive property of $=_{\text{Low}}$.

**For**

(5) and (9) is found with either:

(S-For-Base) $$\frac{\langle e_1, \gamma^1, \sigma^1 \rangle \to \langle v^1, \gamma^2, \sigma^2 \rangle \quad \langle e_2, \gamma^1, \sigma^2 \rangle \to \langle v^2, \gamma^3, \sigma^3 \rangle \quad \langle e_3, \gamma^1[x \mapsto v^2], \sigma^3 \rangle \to \langle v^3, \gamma^4, \sigma^4 \rangle}{\langle \text{For } x \ e_1 \ e_2 \ e_3, \gamma^1, \sigma^1 \rangle \to \langle u, \gamma^1, \sigma^4 \rangle} \quad v^1 = v^2$$

or

(S-For-Rec) $$\frac{\langle e_1, \gamma^1, \sigma^1 \rangle \to \langle v^1, \gamma^2, \sigma^2 \rangle \quad \langle e_2, \gamma^1, \sigma^2 \rangle \to \langle v^2, \gamma^3, \sigma^3 \rangle \quad \langle \text{For } x \ n \ m, \gamma^1, \sigma^3 \rangle \to (u, \gamma^4, \sigma^4) \quad \langle e_3, \gamma^1[x \mapsto v^2], \sigma^4 \rangle \to (v^4, \gamma^5, \sigma^5)}{\langle \text{For } x \ e_1 \ e_2 \ e_3, \gamma^1, \sigma^1 \rangle \to \langle u, \gamma^1, \sigma^5 \rangle}$$
$n = \mathbb{N}^{-1}(v^1)$
$m = \mathbb{N}^{-1}(v^2 - 1)$
$v^1 < v^2$

And it is typed with:

(For) $$\frac{\Gamma_1, pc_1 \vdash e_1 : t_1@t_2 \triangleright \Gamma_2 \quad \Gamma_1, pc_1 \vdash e_2 : t_3@t_4 \triangleright \Gamma_3 \quad \Gamma_1[x \to pc_2], pc_2 \vdash e_3 : t_5@t_6 \triangleright \Gamma_4}{\Gamma_1, pc_1 \vdash \text{For } x \ e_1 \ e_2 \ e_3 : \text{Low}@t_8 \triangleright \Gamma_1}$$
$pc_2 = \sqcup\{pc_1, t_1, t_3\}$
$t_8 = \sqcap\{t_2, t_4, t_6\}$
$t_1, t_3 \in \{\text{Low}, \text{High}\}$

With regards to Lemma 2, since $pc_2 \sqcup \{pc_1, t_1, t_3\}$ and $pc_1 = \mathsf{High}$, then $pc_2 = \mathsf{High}$. Therefore, the inductive hypothesis can be used for all inner expressions. For Lemma 5, since $t_8 = \sqcap \{t_2, t_4, t_6\}$, and $t_8 \in \{(), \mathsf{High}\}$, then $t_2, t_4, t_6 \in \{(), \mathsf{High}\}$, and we can also use the inductive hypothesis for all inner expressions.

We look at (S-For-Base) first.

**S-For-Base**

From the inductive hypothesis, we know that since $\langle e_1, \gamma^1, \sigma^1 \rangle \to \langle v^1, \gamma^2, \sigma^2 \rangle$, then $\Gamma_2 \models (\gamma^1, \sigma^1) =_{\mathsf{Low}} (\gamma^2, \sigma^2)$. We then know from Lemma 4 that $\Gamma_1 \models (\gamma^1, \sigma^1) =_{\mathsf{Low}} (\gamma^1, \sigma^2)$.

Therefore, from the inductive hypothesis, we know that $\langle e_2, \gamma^1, \sigma^2 \rangle \to \langle v^2, \gamma^3, \sigma^3 \rangle$, which means that $\Gamma_3 \models (\gamma^1, \sigma^2) =_{\mathsf{Low}} (\gamma^3, \sigma^3)$. From Lemma 4, we then know that $\Gamma_1 \models (\gamma^1, \sigma^2) =_{\mathsf{Low}} (\gamma^1, \sigma^3)$.

Now we make an observation. For Lemma 2, as $pc_1 = \mathsf{High}$, we know that $pc_2 = \mathsf{High}$, as $pc_2 = \sqcup \{pc_1, t_1, t_3\}$ and $t_1, t_3 \in \{\mathsf{Low}, \mathsf{High}\}$. Therefore, changes to $x$ will not influence the $=_{\mathsf{Low}}$ of the states. For Lemma 5, we cannot be certain that $pc_2 = \mathsf{High}$, but because for $pc_2 = \mathsf{Low}$, then $t_1, t_3 = \mathsf{Low}$, and from Lemma 1, we then know that the same value is stored there, and therefore, it does not affect $=_{\mathsf{Low}}$. Therefore, from the inductive hypothesis, since we know that $\langle e_3, \gamma^1[x \to v^2], \sigma^3 \rangle \to \langle v^3, \gamma^4, \sigma^4 \rangle$, which means that, $\Gamma_4 \models (\gamma^1, \sigma^3) =_{\mathsf{Low}} (\gamma^4, \sigma^4)$. From Lemma 4, we then know that $\Gamma_1 \models (\gamma^1, \sigma^3) =_{\mathsf{Low}} (\gamma^1, \sigma^4)$. From the transitive property of $=_{\mathsf{Low}}$, we then know that $\Gamma_1 \models (\gamma^1, \sigma^1) =_{\mathsf{Low}} (\gamma^1, \sigma^4)$. Therefore, Lemma 2 and Lemma 5 holds for (S-For-Base)

**S-For-Rec**

From the inductive hypothesis, we know that since $\langle e_1, \gamma^1, \sigma^1 \rangle \to \langle v^1, \gamma^2, \sigma^2 \rangle$, then $\Gamma_2 \models (\gamma^1, \sigma^1) =_{\mathsf{Low}} (\gamma^2, \sigma^2)$. We then know from Lemma 4 that $\Gamma_1 \models (\gamma^1, \sigma^1) =_{\mathsf{Low}} (\gamma^1, \sigma^2)$.

Therefore, from the inductive hypothesis, we know that $\langle e_2, \gamma^1, \sigma^2 \rangle \to \langle v^2, \gamma^3, \sigma^3 \rangle$, which means that $\Gamma_3 \models (\gamma^1, \sigma^2) =_{\mathsf{Low}} (\gamma^3, \sigma^3)$. From Lemma 4, we then know that $\Gamma_1 \models (\gamma^1, \sigma^2) =_{\mathsf{Low}} (\gamma^1, \sigma^3)$.

Therefore, from the inductive hypothesis, we know that $\langle For x n m, \gamma^1, \sigma^3 \rangle \to \langle u, \gamma^4, \sigma^4 \rangle$, which means that $\Gamma_1 \models (\gamma^1, \sigma^3) =_{\mathsf{Low}} (\gamma^4, \sigma^4)$, since the type-environment after a *for*-construct is the same.

From Lemma 4, we then know that $\Gamma_1 \models (\gamma^1, \sigma^3) =_{\mathsf{Low}} (\gamma^1, \sigma^4)$.

Now we make an observation. For Lemma 2, as $pc_1 = \mathsf{High}$, we know that $pc_2 = \mathsf{High}$, as $pc_2 = \sqcup \{pc_1, t_1, t_3\}$ and $t_1, t_3 \in \{\mathsf{Low}, \mathsf{High}\}$. Therefore, changes to $x$ will not influence the $=_{\mathsf{Low}}$ of the states. For Lemma 5, we cannot be certain that $pc_2 = \mathsf{High}$, but because for $pc_2 = \mathsf{Low}$, then $t_1, t_3 = \mathsf{Low}$, and from Lemma 1, we then know that the same value is stored there, and therefore, it does not affect $=_{\mathsf{Low}}$. Therefore, $\Gamma_1 \models (\gamma^1, \sigma^4) =_{\mathsf{Low}} (\gamma^5, \sigma^5)$.

From Lemma 4, we then know that $\Gamma_1 \models (\gamma^1, \sigma^4) =_{\mathsf{Low}} (\gamma^1, \sigma^5)$, and because of the transitive property of $=_{\mathsf{Low}}$ we then know that $\Gamma_1 \models (\gamma^1, \sigma^1) =_{\mathsf{Low}} (\gamma^1, \sigma^5)$. Therefore, Lemma 2 and Lemma 5 holds for (S-For-Rec)

## Func

(5) and (9) is found with:

(S-Func) $$\frac{}{\langle \lambda x.e, \gamma, \sigma \rangle \rightarrow \langle v, \gamma, \sigma \rangle} \quad v = (e, x, \gamma)$$

And it is typed with with:

(Func) $$\frac{\Gamma_1[x \mapsto t_1], pc \vdash e : t_2@t_3 \triangleright \Gamma_2}{\Gamma_1, pc \vdash \lambda x_{t_1}.e : (t_1 \rightarrow t_2@t_3)@() \triangleright \Gamma_1} \quad t_3 \sqsupseteq pc$$

Since the state is not changed, it is trivially true, that the state before and after are the same. Therefore, Lemma 2 and Lemma 5 holds for (S-For-Rec)

Furthermore, given an expression $e_1$ with type $(t_1 \rightarrow t_2@t_3)@t_4$, and that can evaluate to a function $v = (x, e_2, \gamma)$, $t_2@t_3$ is the type of the inner expression $e_2$. Therefore, from Lemma 5, if $t_3 \in \{\mathsf{High}, ()\}$, we know that the state after executing it, the state will be low-equivalent to before the execution. This we will use in the proof for rule (App).

## App

(5) and (9) is found with:

(S-App) $$\frac{\langle e_1, \gamma^1, \sigma^1 \rangle \rightarrow \langle v^1, \gamma^2, \sigma^2 \rangle \quad \langle e_2, \gamma^1, \sigma^2 \rangle \rightarrow \langle v^2, \gamma^3, \sigma^3 \rangle}{\langle e_3, \gamma^5[x \mapsto v^2], \sigma^3 \rangle \rightarrow \langle v^3, \gamma^4, \sigma^4 \rangle} \quad v^1 = (e_3, x, \gamma^5)$$
$$\langle e_1 \ e_2, \gamma^1, \sigma^1 \rangle \rightarrow \langle v^3, \gamma^1, \sigma^4 \rangle$$

And it is typed with:

(App) $$\frac{\Gamma_1, pc_1 \vdash e_2 : t_1@t_2 \triangleright \Gamma_2 \quad \Gamma_1, pc_1 \vdash e_1 : (t_1 \rightarrow t_3@t_4)@t_5 \triangleright \Gamma_3}{\Gamma_1, pc_1 \vdash e_1 \ e_2 : t_3@t_6 \triangleright \Gamma_1} \quad \begin{array}{l} t_6 = \sqcap\{t_2, t_4, t_5\} \\ t_6 \sqsupseteq pc_1 \end{array}$$

With regards to Lemma 2, since $pc_1 = \mathsf{High}$, the inductive hypothesis can be used for $e_1$ and $e_2$. For Lemma 5, since $t_6 = \sqcap\{t_2, t_4, t_5\}$, and $t_6 \in \{(), \mathsf{High}\}$, then $t_2, t_4, t_5 \in \{(), \mathsf{High}\}$, and we can use the inductive hypothesis for all inner expressions.

From the inductive hypothesis, we know that since $\langle e_1, \gamma^1, \sigma^1 \rangle \rightarrow \langle v_1, \gamma^2, \sigma^2 \rangle$, then $\Gamma_3 \vdash (\gamma_1, \sigma_1) =_{\mathsf{Low}} (\gamma_2, \sigma_2)$.

From the inductive hypothesis of Lemma 2, and Lemma 4, we then know that since $\langle e_2, \gamma^1, \sigma^2 \rangle \rightarrow \langle v_1, \gamma^3, \sigma^3 \rangle$, then $\Gamma_2 \vdash (\gamma^1, \sigma^2) =_{\mathsf{Low}} (\gamma^3, \sigma^3)$.

From lemma 4, we then know that $\Gamma_1 \vdash (\gamma^1, \sigma^2) =_{\mathsf{Low}} (\gamma^1, \sigma^3)$

We now use the fact that $t_3@t_4$ is the type of the inner expression, under the environment that it was declared under. From Lemma 5 we know that if $t_4 \in \{(), \mathsf{High}\}$, then the state after evaluating $e_3$ will be $=_{\mathsf{Low}}$ with the state before. Since $t_6 = \sqcap\{t_2, t_4, t_6\}$, and $t_6 \sqsubseteq \mathsf{High}$, since $t_6$ must be higher or equal security level to $pc$ then it must be that $t_4 \in \{(), \mathsf{High}\}$. Therefore, the new state must be $=_{\mathsf{Low}}$ with the previous state, with regards to the proof of Lemma 2.

This essentially means, that the only changes that $e_3$ can make to the state, are to the values stored at $\mathsf{ref\ High}$. Therefore, no values where $\Gamma_1(x) = \mathsf{ref\ Low}$ are changed, so we can go back to $\gamma_1$ and $\Gamma_1$, and therefore have, that $\Gamma_1 \vdash (\gamma_1, \sigma_1) =_{\mathsf{Low}} ((\gamma_1, \sigma_4)$. Therefore, Lemma 2 and Lemma 5 holds for (S-App).

**Seq**

*Proof.* (5) and (9) are found with

$$(\text{S-Seq}) \quad \frac{\langle e_1, \gamma^1, \sigma^1 \rangle \to \langle v^1, \gamma^2, \sigma^2 \rangle \quad \langle e_2, \gamma^2, \sigma^2 \rangle \to \langle v^2, \gamma^3, \sigma^3 \rangle}{\langle e_1; e_2, \gamma^1, \sigma^1 \rangle \to \langle v^2, \gamma^3, \sigma^3 \rangle}$$

and are typed with

$$(\text{Seq}) \quad \frac{\Gamma_1, pc \vdash e_1 : t_1 @ t_2 \triangleright \Gamma_2 \quad \Gamma_2, pc \vdash e_2 : t_3 @ t_4 \triangleright \Gamma_3}{\Gamma_1, pc \vdash e_1; e_2 : t_3 @ t_5 \triangleright \Gamma_3} \quad t_5 = \sqcap \{t_2, t_4\}$$

From the inductive hypothesis of Lemma 2, we get that, since $pc = \mathsf{High}$:

$\Gamma_2 \models s_1 =_{\mathsf{Low}} s_2$, and, $\Gamma_3 \models s_2 =_{\mathsf{Low}} s_3$ If $pc = \mathsf{High}$.

With regards to Lemma 5, we know that for $t_5 \in \{(), \mathsf{High}\}$, then $t_2, t_4 \in \{(), \mathsf{High}\}$, as $t_5 \sqcap \{t_2, t_4\}$.

Therefore, from the inductive hypothesis of Lemma 5, we know that $\Gamma_2 \models s_1 =_{\mathsf{Low}} s_2$, and, $\Gamma_3 \models s_2 =_{\mathsf{Low}} s_3$.

Here we make an observation. Since there has not been any changes, or assignments to new variables $x \in \{\mathsf{Low}, \mathbf{Func}, \mathsf{ref}\ \mathsf{Low}$, in either $e_1$ or $e_2$, it must be the case that $\Gamma_1 \models s_3 =_{\mathsf{Low}} s_3$, since the only variables that can have been added to $\mathbf{dom}(s_3)$, will be $\mathsf{High}$ or $\mathsf{ref}(\mathsf{High})$, and therefore, not change non-interference. If there were added a variable $x \in \{\mathsf{Low}, \mathbf{Func}, \mathsf{ref}\ \mathsf{Low}\}$, then $s_1, s_2$ and $s_2, s_3$ would not be $=_{\mathsf{Low}}$ Therefore, for both Lemma 2 and Lemma 5, it must be that $\Gamma_3 \models s_1 =_{\mathsf{Low}} s_3$.

**Ref**

(5) and (9) is found with:

$$(\text{S-Ref}) \quad \frac{\langle e, \gamma^1, \sigma^1 \rangle \to \langle v, \gamma^2, \sigma^2 \rangle}{\langle \mathsf{Ref}(e), \gamma^1, \sigma^1 \rangle \to \langle \ell, \gamma^1, \sigma^2[\ell \to v] \rangle} \quad \ell = new(\sigma^2)$$

And it is typed with:

$$(\text{Ref}) \quad \frac{\Gamma, pc \vdash e : t_1 @ t_2 \triangleright \Gamma'}{\Gamma, pc \vdash \mathsf{Ref}(e) : \mathsf{ref}\ t_1 @ t_2 \triangleright \Gamma} \quad t_1 \in \{\mathsf{Low}, \mathsf{High}\}$$

For Lemma 2, since we assume that $pc = \mathsf{High}$, we then from the inductive hypothesis know, that $\Gamma' \models (\gamma^1, \sigma^1) =_{\mathsf{Low}} (\gamma^2, \sigma^2)$. From Lemma 4, we then know that $\Gamma \models (\gamma^1, \sigma^1) =_{\mathsf{Low}} (\gamma^1, \sigma^2)$. Therefore, Lemma 2 holds for (Ref), as no variables are yet pointing to the new location.

For Lemma 5, since we assume that $t_2 \in \{(), \mathsf{High}\}$, we then from the inductive hypothesis know, that $\Gamma \models (\gamma^1, \sigma^1) =_{\mathsf{Low}} (\gamma^2, \sigma^2)$. From Lemma 4, we then know that $\Gamma \models (\gamma^1, \sigma^1) =_{\mathsf{Low}} (\gamma^1, \sigma^2)$. Therefore, Lemma 5 holds for (Ref), as no variables are yet pointing to the new location.

## Deref

(5) and (9) is found with:

(S-Deref) $$\frac{}{\langle !x, \gamma^1, \sigma^1 \rangle \to \langle v^2, \gamma^1, \sigma^1 \rangle} \quad \begin{array}{l} \gamma^1(x) = \ell^1 \\ v^2 = \sigma^1(\ell^1) \end{array}$$

And it is typed with:

(Deref) $$\frac{}{\Gamma, pc \vdash !x : t_1@() \triangleright \Gamma} \quad \Gamma(x) = \text{ref } t_1$$

Since there happen no changes to the state, it is trivially true, that Lemma 2 and Lemma 5 holds for (S-Deref), since $\Gamma \models (\gamma^1, \sigma^1) = \text{Low}(\gamma^1, \sigma^1)$.

## Reassign

(S-Reassign) $$\frac{\langle e, \gamma^1, \sigma^1 \rangle \to \langle v^1, \gamma^2, \sigma^2 \rangle}{\langle x := e, \gamma^1, \sigma^1 \rangle \to \langle u, \gamma^1, \sigma^2[\ell \to v^1] \rangle} \quad \gamma^1(x) = \ell$$

And it is typed with:

(Reassign) $$\frac{\Gamma, pc \vdash e : t_1@t_2 \triangleright \Gamma_1}{\Gamma, pc \vdash x := e : \text{Low}@t_4 \triangleright \Gamma} \quad \begin{array}{l} \Gamma(x) = \text{ref } t_3 \\ t_3 \in \{\text{Low}, \text{High}\} \\ t_3 \sqsupseteq t_1 \\ t_3 \sqsupseteq pc \\ t_4 = \sqcap\{t_3, t_2\} \end{array}$$

For Lemma 2, first, we use the inductive hypothesis on $e$, and since $pc = \text{High}$, and $\langle e, \gamma^1, \sigma^1 \rangle \to \langle v^1, \gamma^2, \sigma^2 \rangle$, we then know that $\Gamma_1 \vdash (\gamma^1, \sigma^1) =_{\text{Low}} (\gamma^2, \sigma^2)$. Using Lemma 4, we can then say that $\Gamma \vdash (\gamma^1, \sigma^1) =_{\text{Low}} (\gamma^1, \sigma^2)$. because $pc = \text{High}$, we know that $t_3 = \text{High}$, since $t_3 \sqsupseteq pc$, and $t_3 \in \{\text{Low}, \text{High}\}$. Therefore, the changes will not change the $=_{\text{Low}}$, and therefore $\Gamma \vdash (\gamma^1, \sigma^1) =_{\text{Low}} (\gamma^1, \sigma^2[\ell^1 \to v^1])$.

For Lemma 5, first, we use the inductive hypothesis on $e$, and since $t_4 \in \{(), \text{High}\}$, and $t_4 = \sqcap\{t_3, t_2\}$ and therefore, $t_2 \in \{(), \text{High}\}$, and $\langle e, \gamma^1, \sigma^1 \rangle \to \langle v^1, \gamma^2, \sigma^2 \rangle$, we then know that $\Gamma_1 \vdash (\gamma^1, \sigma^1) =_{\text{Low}} (\gamma^2, \sigma^2)$. Using Lemma 4, we can then say that $\Gamma \vdash (\gamma^1, \sigma^1) =_{\text{Low}} (\gamma^1, \sigma^2)$. Since $t_4 \in \{(), \text{High}\}$, and $t_4 = \sqcap\{t_3, t_2\}$, it is also the case that $t_3 \in \{(), \text{High}\}$. Therefore, the changes will not change the $=_{\text{Low}}$, as $\Gamma(x) = \text{ref High}$ and therefore $\Gamma \vdash (\gamma^1, \sigma^1) =_{\text{Low}} (\gamma^1, \sigma^2[\ell^1 \to v^1])$.

∎

# F Proof of Lemma 3 and 4

Lemma 3 states the following: Through inspection of the type-rules, and semantic rules, we see that through our rules, there cannot be removed variables from either the type-environment, or the environment through a transition. From this follows, that we can always replace the type-environment and the environments with earlier ones, and keep all non-interference properties. Therefore, we have if a program ends up with the following states:

$$\Gamma[x \to t] \models (\gamma_{s_1}[x \to v_1], \sigma_{s_1}) =_{\text{Low}} (\gamma_{s_2}[x \to v_2], \sigma_{s_2})$$

Then we also have
$$\Gamma \models (\gamma_{s_1}, \sigma_{s_1}) =_{\text{Low}} (\gamma_{s_2}, \sigma_{s_2})$$

From this follows Lemma 3:

**Lemma 3** (Exiting scope non-interference). *If*
$$\Gamma \models (\gamma^1_{s_1}, \sigma^1_{s_1}) =_{\text{Low}} (\gamma^1_{s_2}, \sigma^1_{s_2})$$

*and*

$$\langle e, \gamma^1_{s_1}, \sigma^1_{s_1} \rangle \rightarrow \langle v_{s_1}, \gamma^2_{s_1}, \sigma^2_{s_1} \rangle \quad (6)$$
$$\langle e, \gamma^1_{s_2}, \sigma^1_{s_2} \rangle \rightarrow \langle v_{s_2}, \gamma^2_{s_2}, \sigma^2_{s_2} \rangle \quad (7)$$
$$\Gamma, pc \vdash e : t \triangleright \Gamma'$$

*then*

$$\Gamma \models (\gamma^1_{s_1}, \sigma^2_{s_1}) =_{\text{Low}} (\gamma^1_{s_2}, \sigma^2_{s_2})$$

Sometimes, we also need to be able to back in environments and type-enviroments, when we have 2 states, where only one of them has gotten a variable change:

$$\Gamma[x \rightarrow t] \vdash (\gamma_{s_1}, \sigma_{s_1}) =_{\text{Low}} (\gamma_{s_2}[x \rightarrow v_2], \sigma_{s_2})$$

Then we also have
$$\Gamma \vdash (\gamma_{s_1}, \sigma_{s_1}) =_{\text{Low}} (\gamma_{s_2}, \sigma_{s_2})$$

From this follows Lemma 4.

**Lemma 4** (Exiting scope non-interference Extended). *If*
$$\Gamma' \models (\gamma^1_{s_1}, \sigma^1_{s_1}) =_{\text{Low}} (\gamma^2_{s_1}, \sigma^2_{s1})$$

*and*

$$\langle e, \gamma^1_{s_1}, \sigma^1_{s_1} \rangle \rightarrow \langle v_{s_1}, \gamma^2_{s_1}, \sigma^2_{s_1} \rangle \quad (8)$$
$$\Gamma, pc \vdash e : t \triangleright \Gamma'$$

*then*

$$\Gamma \models (\gamma^1_{s_1}, \sigma^1_{s_1}) =_{\text{Low}} (\gamma^1_{s_1}, \sigma^2_{s_1})$$

We have not proven these two lemmas in this paper, as we deemed it outside of the scope of our project. However, through inspection of the rules, it can be seen that as long as both type-environment, and environments are set back to the same point, then non-interference should be maintained. This comes from the fact, that non-interference must persist all the way through a program, and the Low values stored in the store must always be the same regardless of the state.